\documentclass[
amsmath,amssymb,
aps, physrev,
notitlepage
]{revtex4-2}

\usepackage{graphicx}% Include figure files
\usepackage{dcolumn}% Align table columns on decimal point
\usepackage{bm}% bold math
\usepackage{setspace}
\usepackage{hyperref}
\usepackage{adjustbox}
\usepackage{array}
\usepackage{booktabs}
\usepackage{array}
\usepackage{multirow}
\usepackage{ltablex}
\usepackage{amsmath} % for advanced math typesetting
\usepackage{titlesec}

\usepackage{esvect}  % for fancy vector arrows
\usepackage{lineno}
\usepackage{soul}
\usepackage{float}
\usepackage{natbib} % G: for citation style
\usepackage{nomencl}
\usepackage{xcolor} % fore revision
\makenomenclature
\begin{document}
\setstretch{1.12}  % or 1.3
%\preprint{APS/123-QED}

\titlespacing\section{0pt}{8pt plus 4pt minus 2pt}{0pt plus 2pt minus 2pt}
\titlespacing\subsection{0pt}{7pt plus 4pt minus 2pt}{0pt plus 2pt minus 2pt}
\titlespacing\subsubsection{0pt}{6pt plus 4pt minus 2pt}{0pt plus 2pt minus 2pt}

\title{Scaling Laws for Caudal Fin Swimmers Incorporating Hydrodynamics, Kinematics, Morphology, and Scale Effects 
 %\\with Forced Linebreak
}% Force line breaks with \\

\author{Jung Hee Seo}
% \altaffiliation{Mechanical Engineering, Johns Hopkins University University.}%Lines break automatically or can be forced with \\
\author{Ji Zhou}
\author{Rajat Mittal}%
 \email{mittal@jhu.edu}
\affiliation{%
 Mechanical Engineering, Johns Hopkins University \\
 Baltimore, MD 21218
}%
\date{\today}% It is always \today, today,
\begin{abstract}
Many species of fish, as well as biorobotic underwater vehicles (BUVs), employ body–caudal fin (BCF) propulsion, in which a wave-like body motion culminates in high-amplitude caudal fin oscillations to generate thrust. This study uses high-fidelity simulations of a mackerel-inspired caudal fin swimmer across a wide range of Reynolds and Strouhal numbers to analyze the relationship between swimming kinematics and hydrodynamic forces. Central to this work is the derivation and use of a model for the leading-edge vortex (LEV) on the caudal fin. This vortex dominates the thrust production from the fin and the LEV model forms the basis for the derivation of scaling laws grounded in flow physics. Scaling laws are derived for thrust, power, efficiency, cost-of-transport, and swimming speed, and are parameterized using data from high-fidelity simulations. These laws are validated against published simulation and experimental data, revealing several new kinematic and morphometric parameters that critically influence hydrodynamic performance. The results provide a mechanistic framework for understanding thrust generation, optimizing swimming performance, and assessing the effects of scale and morphology in aquatic locomotion of both fish and BUVs.
\end{abstract}


\maketitle

\section{\label{sec:background}Introduction}
Most fish have multiple fins that are used for propulsion and maneuvering, but caudal fin driven propulsion in the carangiform, sub-carangiform, and thunniform body-caudal fin (BCF) modes are employed by many of these animals, especially for rectilinear swimming. In this mode of propulsion, fish employ a wave-like motion that increases in amplitude as it propagates towards the tail. The maximum amplitude is reached at the caudal fin, thereby imparting a relatively high lateral velocity to the propulsion surface of the fin. Given that the pressure on a surface roughly scales with the square of its velocity relative to the flow, the caudal fin can generate a large pressure-induced thrust force from its fin. 

These caudal fin swimmers exist on scales ranging from O(1 cm) (such as juvenile Zebrafish) to O(10 m) (such as many cetaceans and whale sharks), but the effect of scale on the swimming hydrodynamics of these types of swimmers has not been examined in detail. 
One question of fundamental importance to fish (or fish-like) swimming is the scaling relationship between the morphology (shape and size) of the body and caudal fin of the fish as well as the swimming kinematics of the fish, and the swimming performance of the fish, which is characterized by the swimming speed and the efficiency.
One measure of scale is the Reynolds number based on the body length ($L$) and swimming speed ($U$), which is defined as $\textrm{Re}_U = UL/\nu$. Another important parameter for these swimmers is the Strouhal number for the caudal fin, $\textrm{St}_A=f A_F/U $ where $f$ is the frequency of the tail beat, and $A_F$ is the peak-to-peak amplitude of the tail, which is taken as an estimate of the width of the wake. Based on the fluid dynamics principle, the Strouhal number should be a function of the Reynolds number, but the relation between these two non-dimensional numbers may depend on the swimming kinematics and the morphology of a fish.
Thus, the relation between the Strouhal and Reynolds numbers may provide insights into the role of kinematics and morphology on swimming performance. Investigation of the relation requires detailed analysis of the hydrodynamics of a caudal fin swimmer that may be characterized by the following elements: (a) swimming kinematics and swimming speed; (b) hydrodynamic forces on the swimmer and swimming efficiency; (c) details of the flow velocity and pressure over the body of the swimmer; and (d) vortex topologies and flow features over the body and in the wake of the swimmer.

The classic work of Bainbridge\cite{bainbridge1958speed} examined the scaling of swimming velocity ($U$) with tail beat frequency ($f$), and body-length ($L$) for trout (\emph{Salmo irideus}), dace (\emph{Leuciscus leuciscus}) and goldfish (\emph{Carassius auratus}), and proposed the following relationship between these variables: $U=L \left( \tfrac{3}{4}f - 1 [\textrm{Hz}] \right)$,
where $f$ is the tail-beat frequency in Hz and the formulation was derived for $f>5 [\textrm{Hz}]$. The fish in their experiments ranged in length from about 4 cm to 30 cm, and swimming speed ranged from 0.5 $L/s$ (body length per second) to over 10 $L/s$. They also found that the peak-to-peak amplitude at the distal end of the caudal fin (designated here as $A_F$) was well approximated by 0.18$L$ for the higher speeds for most of the fish in their experiments. We estimate that in their experiments the Reynolds numbers based on body-length ($\textrm{Re}_U = UL/\nu$) covered a wide range from about 20,000 for the smaller fish to nearly $10^6$ for the larger (or faster)  fish. The above formula can be rewritten to give $\textrm{St}_A=f A_F/U = \tfrac{4}{3}(A_F/L) \left[ 1 + f_{0}L/U \right]$, where $f_0=1 [\textrm{Hz}]$. The Strouhal number therefore reduces with increasing swimming speed, and for large swimming velocities ($U$ is much larger than $1 L/s$), where $(A_F/L) \approx 0.18$, the Strouhal number would approach a value of 0.24. The above study did not examine the flow characteristics, thrust, drag, lateral forces, mechanical power, or the cost-of-transport (COT) for these fish, and therefore did not provide any reasoning for this scaling based on the fluid dynamics of swimming nor any indication of the effect of scale (and Reynolds number) on these quantities related to swimming performance.

Flow simulations have been employed to examine the effect of the Reynolds number on the hydrodynamic characteristics of carangiform swimmers. Borazjani and Sotiropoulos\cite{borazjani2008numerical} examined a carangiform swimmer based on the kinematics measured by Hess and Videler\cite{videler1984fast} at Reynolds numbers ranging from 300 to 4000 and Strouhal numbers from 0.0 to 1.2. Using simulations with ``tethered" fish, they found that for Reynolds numbers of 300 and 4000, terminal swimming velocity (where drag matched thrust) was reached at Strouhal numbers of 1.1 and 0.6, respectively. 
The results indicated that the fish with a low Reynolds number may swim at a high Strouhal number.
Li et al.\cite{li2021fishes} performed flow simulations for anguilliform and carangiform swimmers as well as larval zebrafish models with various tail-beat frequencies and amplitudes. The simulations covered the Reynolds numbers ranging from 1 to 6000. Based on the simulation results, they suggested that fish may change their swimming speed by changing their tail-beat frequency rather than amplitude to minimize the cost of transport, and this may be the reason why the fish swim within a narrow range of Strouhal numbers.

Triantafyllou et al.\cite{triantafyllou1993optimal} conducted a comprehensive survey of data on carangiform fish and cetaceans and concluded that most of these animals swim with a Strouhal number ranging (based on the tail amplitude) from 0.25 to 0.35, which was shown to be optimal from flapping foil experiments and wake stability analysis. Taylor et al.\cite{taylor2003flying} also showed that most swimming and flying animals operate within a narrow range of Strouhal number from 0.2 to 0.4.

Based on the optimization of Lighthill's elongated body theory, Eloy\cite{eloy2012optimal} proposed a relation between the optimal Strouhal number and the Lighthill number, $\textrm{Li}$, which is defined by $\textrm{Li}=S_bC_d/h^2$, where $S_b$ is the body surface area, $h$ is the height of a fish (or tail), and $C_d=F_D/(\frac{1}{2}\rho U^2 S_b)$ is the drag coefficient based on the total surface area. It was shown that the optimal Strouhal number increased with the Lighthill number. The drag coefficient, however, depends on the body shape, flow condition, and flow Reynolds number, and thus, the Lighthill number is not easy to obtain from observations, especially at high Reynolds numbers. Since the drag coefficient decreases for the higher Reynolds numbers in general, the relation implies that the optimal Strouhal number may be lower at a higher Reynolds number. 

Gazzola et al.\cite{gazzola2014scaling} introduced the swimming number, $\textrm{Sw}=2\pi fAL/\nu$, where $A$ is the tail-beat amplitude (not peak-to-peak), which is in fact the Reynolds number based on the lateral velocity of the tail, and proposed a scaling law: $\textrm{Re}_U\sim \textrm{Sw}^{4/3}$ by assuming a laminar Blasius flow over the fish body (i.e. $C_d\sim 1/\textrm{Re}_U^{1/2}$). 
By definition, $\textrm{Sw}=\pi\textrm{St}_A\textrm{Re}_U$, and thus the scaling yields $\textrm{St}_A\sim \textrm{Re}_U^{-1/4}$ for the laminar, Blasius flow. This scaling law requires the expression for the drag coefficient, which again depends on the body shape and flow conditions. 
Recently, Vent`ejou et al.\cite{ventejou2025universal} proposed a similar scaling law by introducing the thrust number, $\textrm{Th}=\rho f_T L^3/\nu^2$, where $f_T$ is the thrust force density. For the laminar Blasius flow, they obtained a scaling law: $\textrm{Re}_U\sim \textrm{Th}^{2/3}$. In the work of Gazzola et al., they scaled the thrust with $\rho(2\pi fA)^2L$, and thus, $\textrm{Th}\sim \textrm{Sw}^2$. This leads to the same scaling law of $\textrm{St}_A\sim \textrm{Re}_U^{-1/4}$ for the laminar, Blasius flow. 
Das et al.\cite{das2022contrasting} proposed a similar relation, $\textrm{St}_A\sim\textrm{Re}_U^{-0.375}$ for a self-propelling pitching and heaving foil at $\textrm{Re}_U \le 1000$.
While these scaling laws may represent an overall relationship between the Strouhal and Reynolds numbers for swimming animals, the above studies showed that detailed analysis of the flow physics of thrust and drag generation would be required to derive a more comprehensive relationship between morphology, kinematics, and scale.

In this regard, although several different scaling laws have been proposed for swimming fish previously, their connection to the force generation mechanism by the caudal fin, which is key to carangiform propulsion, is missing.
A motivation of the current work, therefore, is the application of the leading-edge vortex (LEV) based model to derive scaling laws for a swimming fish. Seo and Mittal conducted simulations of carangiform swimming at Re=5000\citep{seo2022improved} and showed that the LEV that forms over the caudal fin is a dominant contributor to the thrust. Raut et al.\cite{raut2024hydrodynamic} applied the LEV-based model to a pitching and heaving foil and derived a functional relation between the thrust and kinematic parameters. Recently, Zhou et al.\cite{Zhou2025hydrodynamical} applied the LEV-based model to investigate the hydrodynamic interaction in schooling fish. Since the caudal fin of body-caudal-fin (BCF) swimmers can be considered as a pitching and heaving foil, the application of the LEV-based model to the caudal fin may provide a functional relationship between the forces generated by the caudal fin and kinematic parameters. 
\begin{figure}
    \begin{center}
    \includegraphics[width=0.55\textwidth]{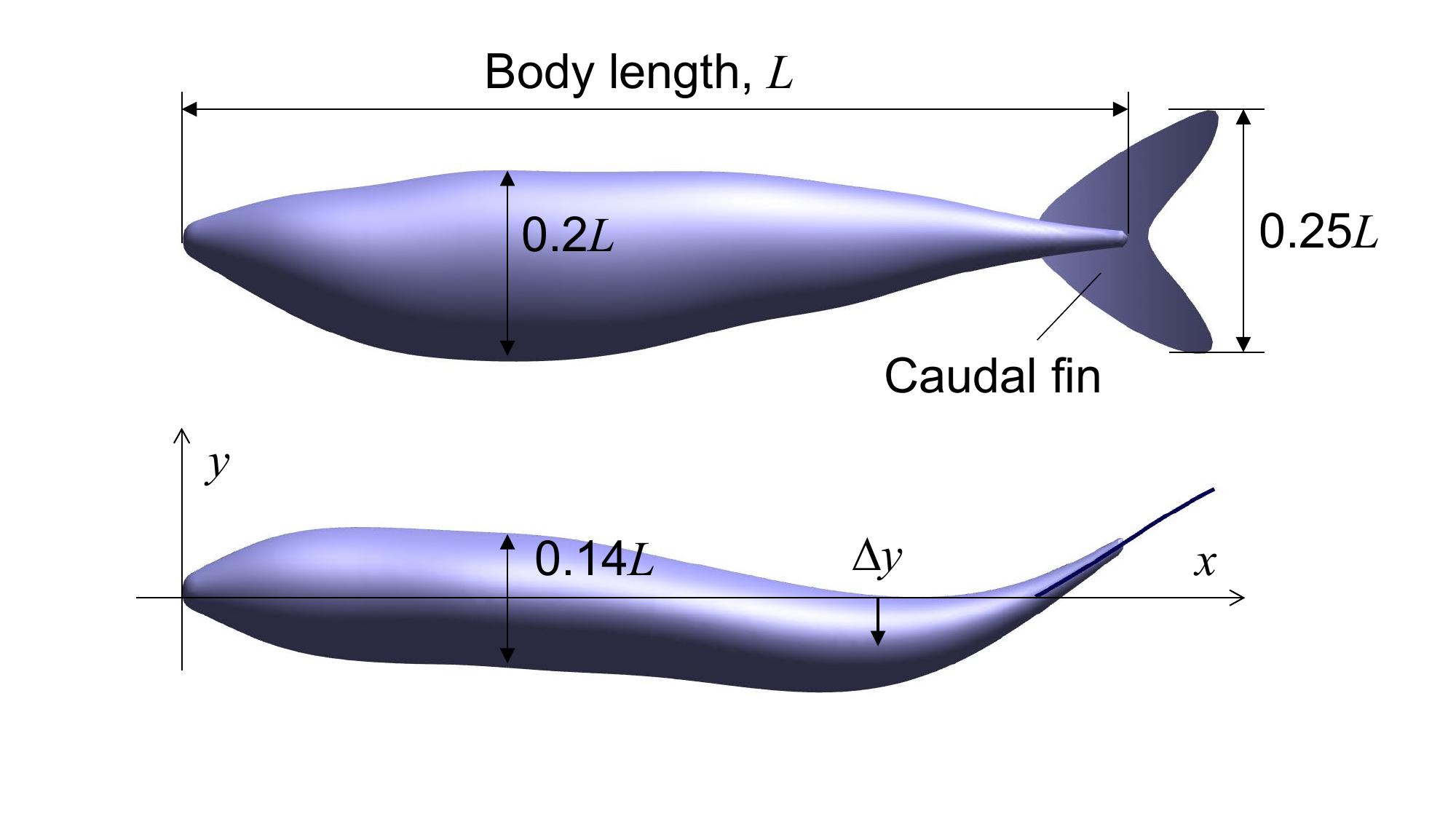}
    \end{center}
    \caption{3D fish model of a carangiform swimmer employed in the present study. The model is based on the common Mackerel (Scomber scombrus).}
    \label{fig:Fishmodel}
\end{figure}

In the present study, we have employed high-fidelity direct numerical simulations (DNSs) of a carangiform fish model for wide range of Reynolds numbers to firstly investigate the relationship between the swimming performance of carangiform swimmers and the Reynolds and Strouhal numbers. We subsequently focus on deriving scaling laws to estimate thrust, power, cost of transport, efficiency, and swimming velocity based on morphology, kinematics, and scale effects. These laws are validated using our DNSs and corroborated with prior experimental and computational studies. Throughout, we highlight the broader implications of the analysis, particularly the role of newly identified parameters, not just for understanding biological swimming but also for informing the design and optimization of bioinspired underwater vehicles.
\section{Methods}

\subsection{Kinematic model of a carangiform swimmer}
The 3D fish model used in the current study is exactly the same as in our previous studies\cite{seo2022improved,Zhou2025hydrodynamical} and is based on the common mackerel (\emph{Scomber scombrus}), which is a well-known example of a carangiform swimmer. The model consists of the body and the caudal fin, and the caudal fin is modeled as a zero-thickness membrane (see Figure \ref{fig:Fishmodel}). The caudal fins of fish are generally very thin (membrane-like) and flexible, and can display significant curvature. The shapes of the caudal fin can vary significantly, but a forked shape with two lobes is quite common. While the two lobes can be significantly unequal in some fish\cite{lauder2000function}, a homocercal tail with two equal lobes is the most common shape in modern teleost (bony) fish, and is adopted here. 

A carangiform swimming motion is prescribed by imposing the following lateral displacement of the centerline of the
fish body extending into the caudal fin: 
\begin{equation}
\Delta y (x,t) = A(x) \sin \left[2 \pi (x/\lambda - ft) \right],   
\label{eq:carangi}
\end{equation}
where $\Delta y$ is the lateral displacement, $x$ is the axial coordinate along the body starting from the nose, $f$ is the tail beat frequency, $\lambda$ is the undulatory wavelength, and $A(x)$ is the amplitude envelope function given by
\begin{equation}
A(x)/L = a_0 +a_1(x/L)+a_2(x/L)^2,      
\label{eq:A}
\end{equation}
where $L$ is the body length. 
The amplitude is set to increase quadratically from the nose to the tail, and the peak-to-peak amplitude at the tips of the caudal fin is designated as $A_F$. The parameters are set to the following values: $a_0$ = 0.02,  $a_1$ = -0.08, and  $a_2$ = 0.16 based on literature \cite{videler1984fast}. This results in a peak-to-peak tail-beat amplitude of $A_F/L = 0.2$, which is inline with the value found to be typical for carangiform swimmers \cite{bainbridge1958speed,videler1984fast}. For carangiform swimmers, the wavelength, $\lambda$ is close to the body length\cite{videler1984fast} and we set the wavelength equal to the body length in the present study; $\lambda=L$. The Reynolds numbers based on body length and tail beat frequency, $\textrm{Re}_L = L^2 f / \nu$ are set to 500, 1000, 2000, 5000, 10000, 25000 and 50000 which enable us to investigate the swimming performance over a wide range of Reynolds numbers. The fish is tethered to a fixed location in an incoming current in the simulations. 
In the experiments of Videler and Hess\cite{videler1984fast}, from where the above swimming kinematics were extracted, it was reported that the inline swimming velocity oscillation was less than 2\% of the swimming speed and the lateral whole body oscillation velocity was less than 4\% of the body length per tail-beat. This provides strong justification for the use of the ``tethered fish'' model.
Multiple simulations are performed, varying the speed of the incoming current, $U$, and through trial-and-error, the terminal speed at which the mean surge force on the fish is nearly zero is found for each Reynolds number. This terminal condition is used for all the analysis in the paper.

\subsection{Computational Methodology}
The flow simulations are performed by solving the incompressible Navier-Stokes equations:
\begin{equation}
\nabla  \cdot \vec u = 0,\,\,\,\,\,\,\frac{{\partial \vec u}}{{\partial t}} + (\vec u \cdot \nabla )\vec u + \frac{{\nabla p}}{\rho } = \nu {\nabla ^2}\vec u    
\end{equation}
by using a sharp-interface, immersed boundary flow solver, Vicar3D\cite{mittal2008versatile}. In the above equation, $\vec{u}$ is the flow velocity vector, $p$ is the pressure, and $\rho$ and $\nu$ are the density and kinematic viscosity of the water. The equations are discretized with a second-order finite difference method in time and space. The details for the numerical methods employed in the flow solver can be found in Ref.\cite{mittal2008versatile}. This flow solver resolves the complex flow around moving/deforming bodies on the non-body-conformal Cartesian grid by using a sharp-interface, immersed boundary method. The same solver was successfully used in our previous study to investigate the hydrodynamic interactions in fish schools\cite{seo2022improved}. The solver has also been extensively validated for a variety of laminar/turbulent flows\cite{mittal2008versatile} and applied to a wide range of studies in bio-locomotion flows\cite{seo2019mechanism,zhou2024effect,kumar2025batwings}.

As noted above, in the present study, the prescribed carangiform swimming motion (Eq. \ref{eq:carangi}) is imposed on the fish, which is tethered in an incoming steady flow with a velocity equal and opposite to the swimming velocity, $U$. The fish body and caudal fin are meshed with triangular surface elements and immersed into the Cartesian volume mesh, which covers the flow domain. The flow domain size is set to $8L\times 10L \times 10L$. In our previous study\cite{seo2022improved}, we have performed a grid convergence study for a swimming fish at $\textrm{Re}_L=5000$ and found that the grid with $640\times 320\times 240$ (about 49 million) grid points was sufficient to obtain converged results. In the present study, to go to higher Reynolds numbers, we have employed a refined grid with $1200 \times 540\times 360$ (about 233 million) grid points. The minimum grid spacing (cell size) is $0.002L$ and the body length is covered by 500 grid points. The time-step size used in the simulation is $\Delta t=0.0005/f$, which resolves one tail beat cycle with 2000 time-steps. The grid convergence test for this resolution is presented in Appendix \ref{sec:GC}. We used this high-resolution grid for the cases with the Reynolds number 5000, 10000, 25000, and 50000. The simulations of the low Reynolds number cases ($\textrm{Re}_L<5000$) are performed on the grid with $640 \times 320\times 240$ points in which the fish body is covered by 200 grid points. A no-slip boundary condition on the fish body and fin surfaces is applied by using the sharp-interface, immersed boundary method\cite{mittal2008versatile}. A zero-gradient boundary condition for the velocity and pressure is applied on the domain boundaries except the inflow. 
\subsection{Hydrodynamic Metrics}
Forces and mechanical power are calculated by the surface integrals:
\begin{equation}
\vec F = \int {\left( {p\vec n + \vec \tau } \right)dS} ,\,\,\,W = \int {\left( {p\vec n + \vec \tau } \right) \cdot \vec vdS} 
\label{eq:force}
\end{equation}
where $\vec{n}$ is the surface normal unit vector (pointing toward the body), $\vec{\tau}$ is the viscous stress, and $\vec{v}$ is the body velocity on the surface. Following our previous study \cite{seo2022improved}, the force on the fish is separated into four components for the detailed analysis: Pressure ($F_{p,\textrm{body}}$) and viscous ($F_{s,\textrm{body}}$) forces on the fish body, and pressure ($F_{p,\textrm{fin}}$) and viscous ($F_{s,\textrm{fin}}$) forces on the caudal fin. This is done by calculating the integral of the pressure ($p\vec{n}$) and viscous stress ($\vec\tau$) in Eq.(\ref{eq:force}) separately. The hydrodynamic power is also decomposed in the same way. In the previous study, we have found that the fish body mostly produces viscous drag, while the caudal fin generates pressure thrust. The Froude efficiency, $\eta$ is considered as a main efficiency metric and defined by
\begin{equation}
\eta  = \frac{{{{\bar F}_T}{U}}}{{\bar W}},
\label{eq:eta}
\end{equation}
where $F_T$ is the total thrust and $W$ is the total expended power, $U$ is the terminal swimming speed, and the overbar denotes time average over one tail-beat cycle.

\begin{table}[h]
    \begin{center}
    \begin{tabular}{|c|c|c|c|c|c|c|c|c|c|c|}
        \hline
        $\textrm{Re}_L$ & $U/(Lf)$ & $F^*_{p,\textrm{body}}$ & $F^*_{s,\textrm{body}}$ & $F^*_{p,\textrm{fin}}$ & $F^*_{s,\textrm{fin}}$ & $W^*_{p,\textrm{body}}$ & $W^*_{s,\textrm{body}}$ & $W^*_{p,\textrm{fin}}$ & $W^*_{s,\textrm{fin}}$ & $W^*_{\textrm{total}}$  \\ \hline
        500 & 0.22  & -1.7 & 6.0 & -6.0 & 1.8 & -1.6 & -0.8 & -4.7 & -0.4 & -7.5 \\ \hline
        1000 & 0.28 & -1.1 & 4.9 & -4.9 & 1.1 & -1.3 & -0.5 & -4.0 & -0.2  & -6.0\\ \hline
        2000 & 0.36 & -0.64 & 4.1 & -4.1 & 0.67 & -1.0 & -0.3 & -3.5 & -0.12 & -4.9 \\ \hline 
        5000  & 0.48 & -0.17 & 3.3 & -3.7 & 0.51 & -0.78 & -0.17 & -3.7 & -0.07 & -4.7 \\ \hline
        10000 & 0.58 &  0.12 & 2.7 & -3.4 & 0.42 & -0.64 & -0.10 & -3.7 & -0.05 & -4.5\\ \hline
        25000  & 0.68  & 0.38 & 2.0 & -2.7 & 0.31 & -0.47 & -0.06 & -3.5 & -0.03  & -4.1\\ \hline
        50000  & 0.72  & 0.44 & 1.5 & -2.2 & 0.23 & -0.39 & -0.04 & -3.2 & -0.02  &  -3.6 \\ \hline
    \end{tabular}
    \end{center}
    \caption{Forces and hydrodynamic powers on the free swimming fish at various Reynolds numbers. $Re_L=L^2f/\nu$, $F^*$ is the time averaged force normalized by $(1/2)\rho (Lf)^2 L^2$. $W^*$ is the time averaged power normalized by $(1/2)\rho (Lf)^3 L^2$. Negative force values denote thrust (force in the swimming direction), and the negative power is the rate of work done by the fish. All $F^*$ and $W^*$ values in the table are to be multiplied by $\times10^{-3}$.}
    \label{tab:allcases}
\end{table}

\begin{figure}
    \centering
    (a)\includegraphics[trim={5cm 0 0 5cm},clip,width=0.35\linewidth]{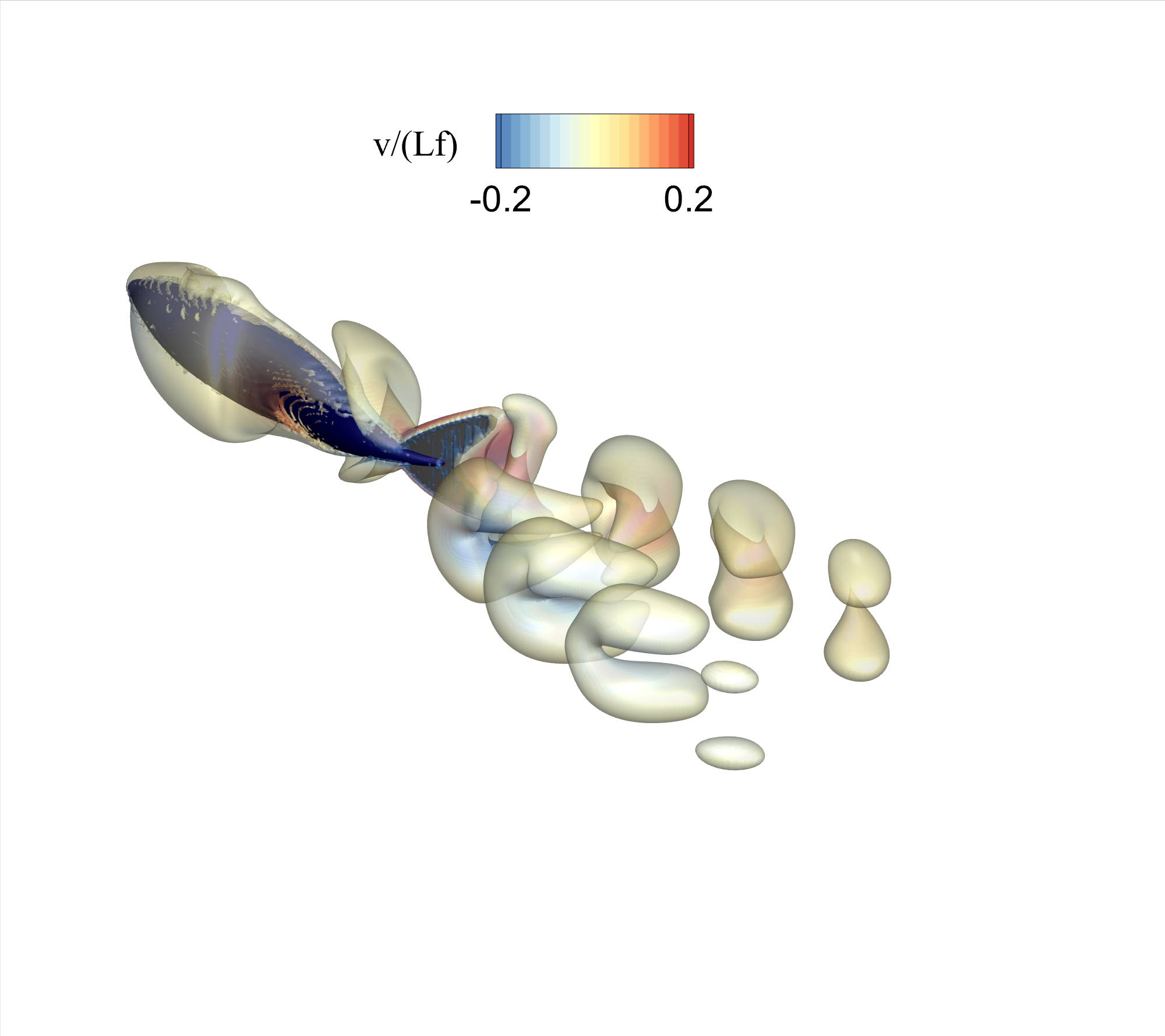} %1000
    (b)\includegraphics[trim={5cm 0 0 5cm},clip,width=0.35\linewidth]{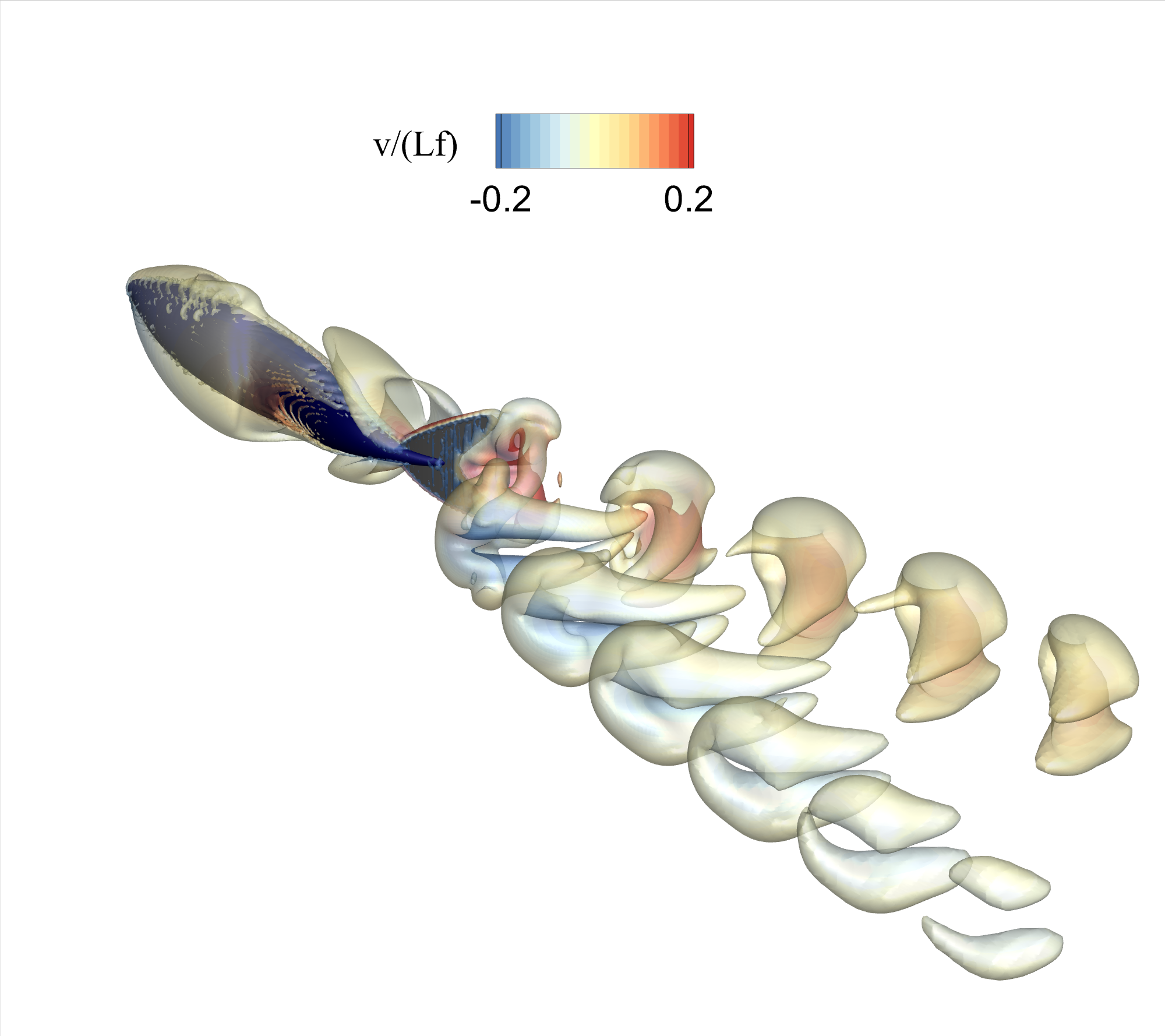} %2000
    (c)\includegraphics[trim={5cm 0 0 5cm},clip,width=0.35\linewidth]{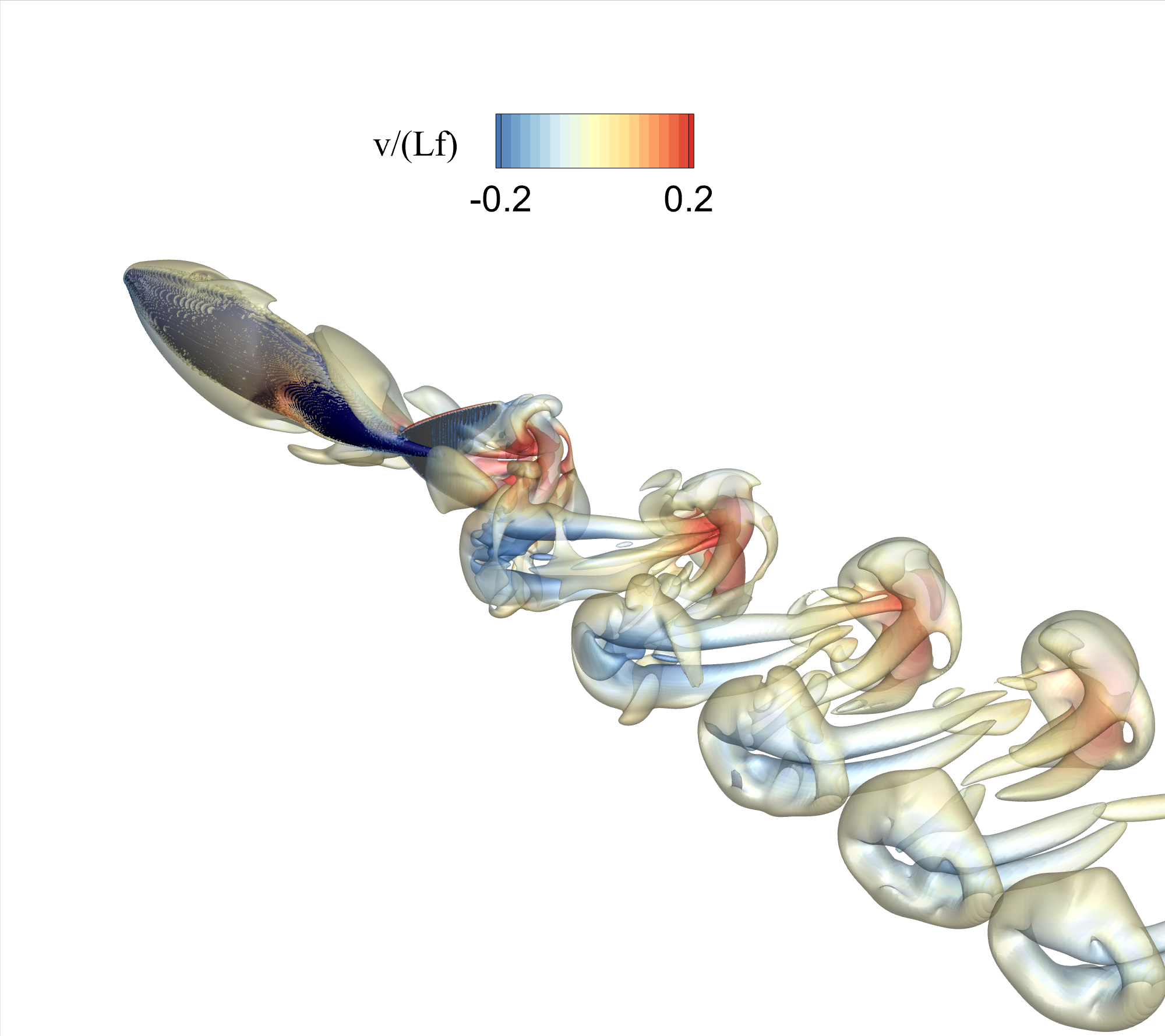} %5000
    (d)\includegraphics[trim={5cm 0 0 5cm},clip,width=0.35\linewidth]{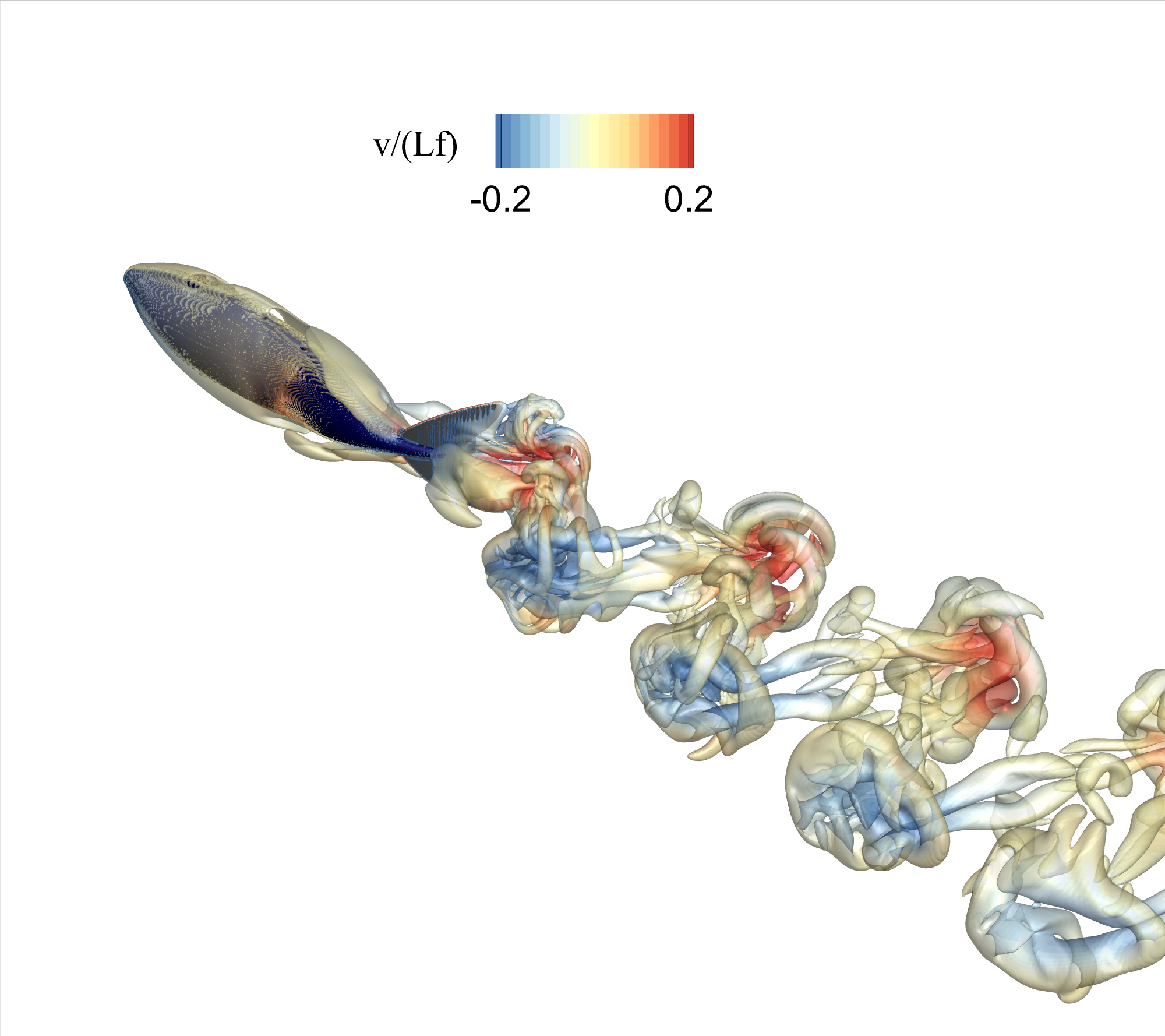} %10000  
    (e)\includegraphics[trim={5cm 0 0 5cm},clip,width=0.35\linewidth]{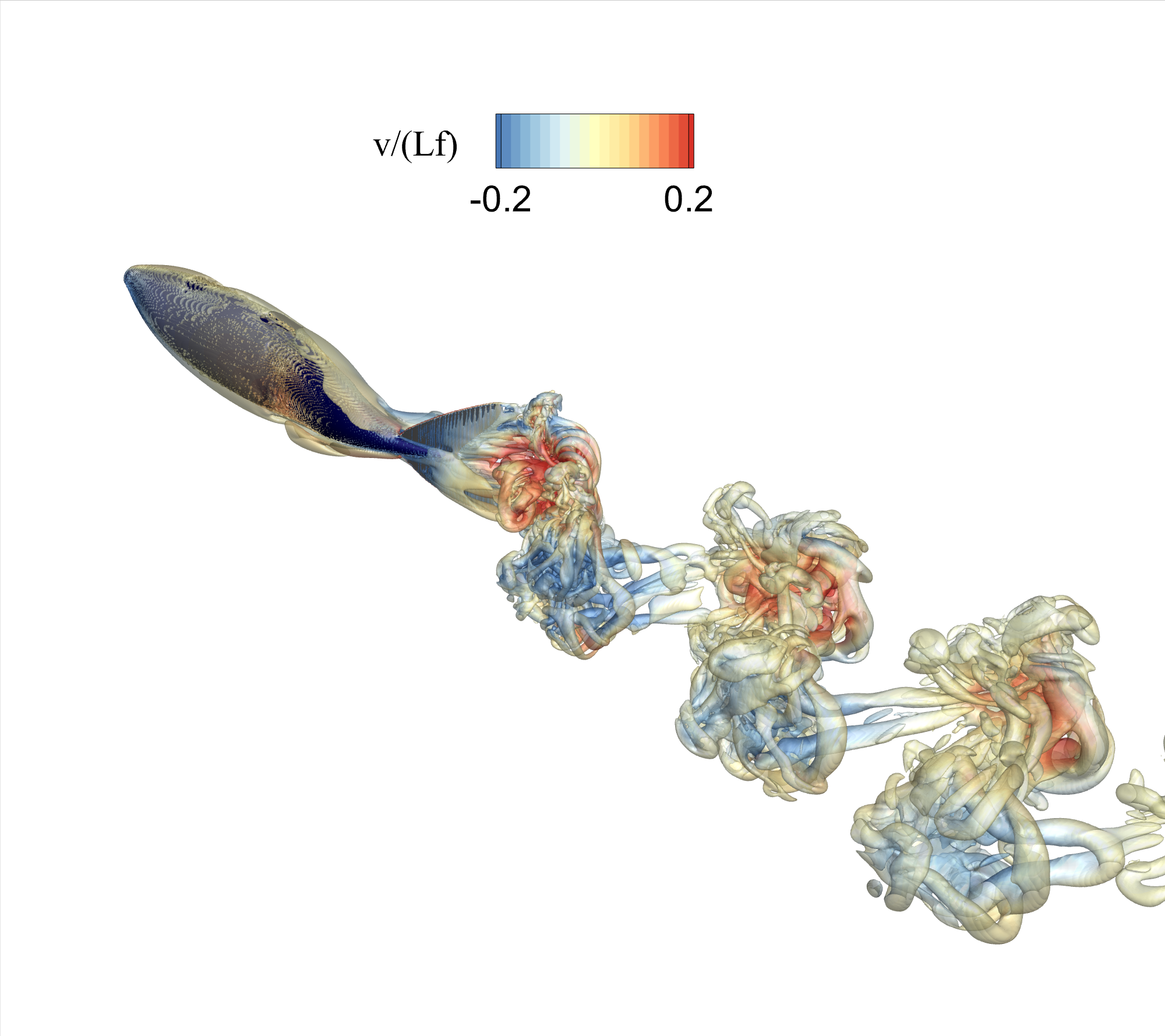} %25000
    (f)\includegraphics[trim={5cm 0 0 5cm},clip,width=0.35\linewidth]{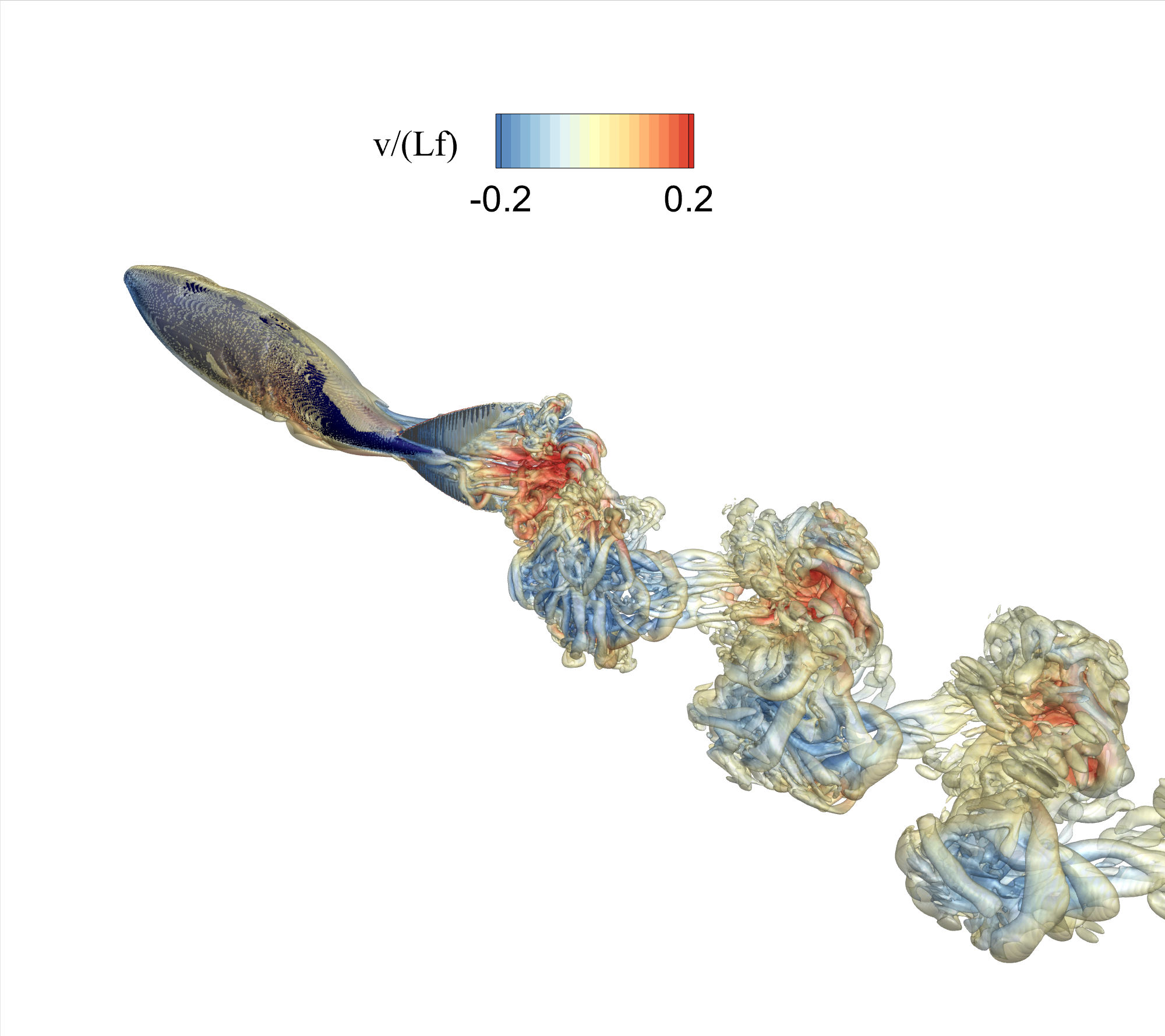} %50000
    \caption{Three-dimensional vortical structures around the swimming fish visualized by the iso-surface of the second invariant of velocity gradient, $Q=0.1f^2$, colored by the lateral velocity ($v$) at various Reynolds numbers. $\textrm{Re}_L=$ (a) 1000, (b) 2000, (c) 5000, (d) 10000, (e) 25000, (f) 50000.}
    \label{fig:vorQ}
\end{figure}
\section{Results}
\subsection{Terminal Swimming Speed and Forces}
The DNSs of the swimming fish model performed in the present study are summarized in Table \ref{tab:allcases}. In this table, the Reynolds numbers are based on body length and tail-beat frequency: $\textrm{Re}_L=L^2f/\nu$, where $L$ is the body length from head to tail. The time-averaged force components in the surge direction ($x$) as well as the mechanical power are also tabulated. Note that the negative force value represents thrust (force in the swimming direction), and the positive value represents drag (force in the opposite direction to the swimming). One can see that the caudal fin is mainly generating pressure thrust, while on the body, the viscous shear drag is dominant. The free swimming, terminal speeds that result in almost 0 net force in the surge direction, are found in the tabulated $U/(Lf)$, i.e., the advance ratio, which is equal to the body lengths traveled per tail-beat.

Three dimensional vortical structures around the swimming fish at various Reynolds numbers are visualized in Fig.\ref{fig:vorQ} by the second invariant of velocity gradient, $Q=\frac{1}{2} (|| {\bf \Omega}||^2- || {\bf S}||^2)$, where ${\bf S}$ and ${\bf \Omega}$ are symmetric and anti-symmetric components of velocity gradient tensor, respectively. At low Reynolds numbers, an alternating horseshoe-like vortex street is observed (Figs.\ref{fig:vorQ}(a-b)). At higher Reynolds numbers, the structure changes to alternating vortex rings connected by elongated vortices between them (Figs.\ref{fig:vorQ}(c-d)). The vortices in the wake break into smaller eddies and exhibit complex structures at further higher Reynolds numbers (Figs.\ref{fig:vorQ}(e-f)). The wake characteristics will be discussed further in the following section.

With the free-swimming speed ($U$) found by the simulations, the data in Table \ref{tab:allcases} are converted to the force coefficients defined in a traditional way:
\begin{equation}
{C_x} = \frac{{{{\bar F}_x}}}{{\frac{1}{2}\rho {U^2}{S_x}}},    
\end{equation}
where $S_x$ is the frontal area of the fish body, whose value is about $0.023L^2$ for the current model, and $F_x$ is the force in the surge direction tabulated in Table \ref{tab:allcases} for each component. 
The force coefficients are tabulated in Table \ref{tab:free}.
The Strouhal number based on the tail beat amplitude, $\textrm{St}_A=fA_F/U$, the Reynolds number, $\textrm{Re}_U=UL/\nu$, and the Froude efficiency for whole fish ($\eta_\textrm{fish}$) and caudal fin ($\eta_\textrm{fin}$) are also calculated and listed in the Table \ref{tab:free}. For the Froude efficiency of the caudal fin, the force and power due only to the pressure are considered, since the viscous force and power are very small compared to the pressure ones. This is also to investigate the effect of Strouhal number on the Froude efficiency, which will be discussed in the later section.  

\begin{table}[h]
    \centering
    \begin{tabular}{|c|c|c|c|c|c|c|c|}
        \hline
        $\textrm{Re}_U$ & $\textrm{St}_A$ & $C_{p,\textrm{body}}$ & $C_{s,\textrm{body}}$ & $C_{p,\textrm{fin}}$ & $C_{s,\textrm{fin}}$ & $\eta_\textrm{fish}$    &  $\eta_\textrm{fin}$    \\ \hline
        110    & 0.91     & -1.6     & 5.49     & -5.49    & 1.6    & 0.22  & 0.28  \\ \hline
        284    & 0.70     & -0.6     & 2.66     & -2.63    & 0.57   & 0.28  & 0.35  \\ \hline
        720    & 0.56     &  -0.21   & 1.37     & -1.37    & 0.26   & 0.34  & 0.41  \\ \hline
        2400   & 0.42     & -0.03    & 0.62     & -0.70    & 0.096  & 0.41  & 0.48   \\ \hline
        5800   & 0.34     & 0.016    & 0.35     & -0.43    & 0.054  & 0.44  & 0.53   \\ \hline
        17000  & 0.29     & 0.036    & 0.19     & -0.26    & 0.03   & 0.46  & 0.52   \\ \hline
        36000  & 0.28     & 0.037    & 0.12     & -0.18    & 0.019  & 0.43  & 0.49    \\ \hline    
    \end{tabular}
    \caption{Force coefficients and Froude efficiencies.}
    \label{tab:free}
\end{table}

\begin{figure}
    \begin{center}
    \includegraphics[trim={0 2cm 0 2cm},clip,width=\textwidth]{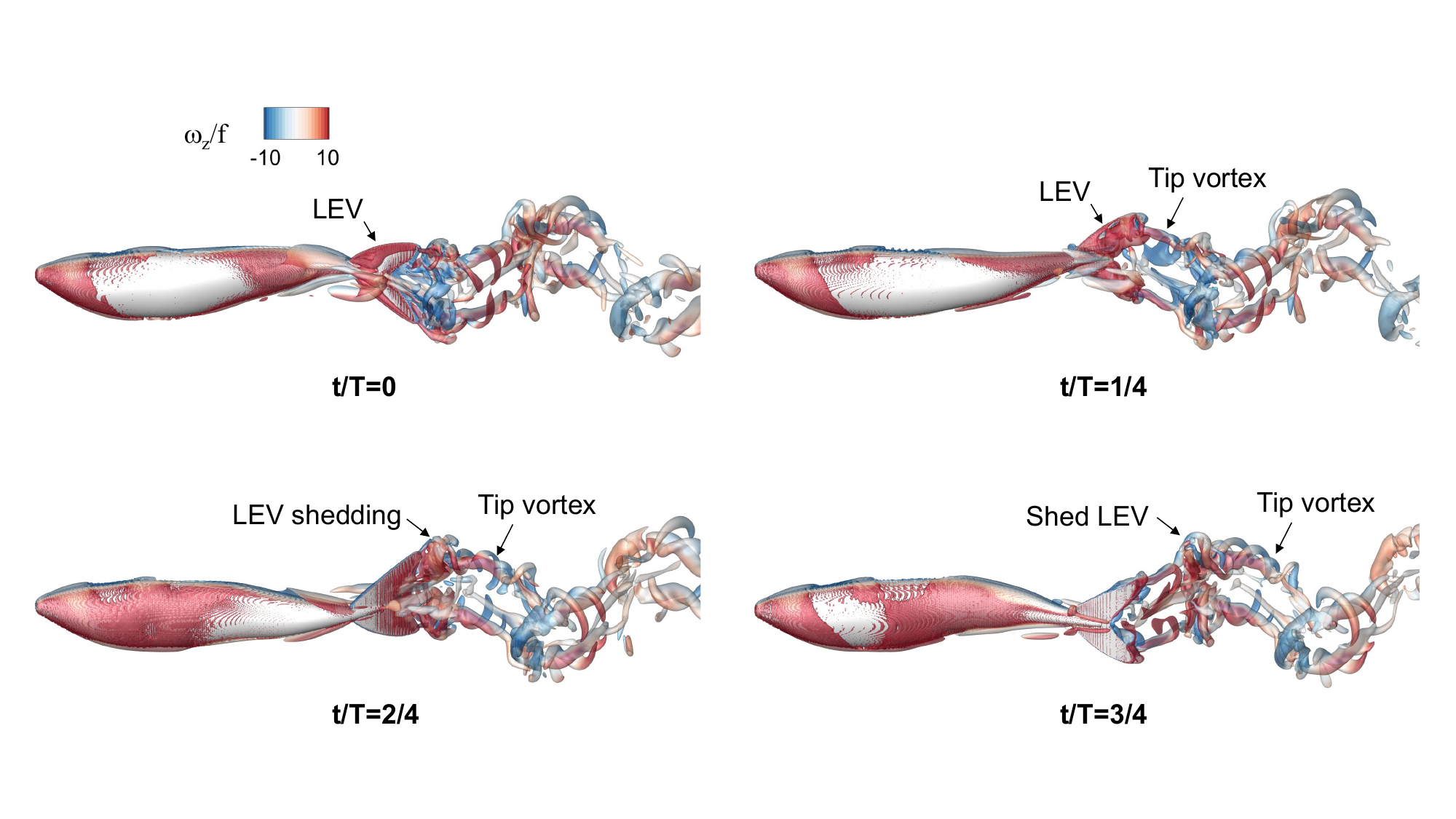}
    \end{center}
    \caption{Evolution of the vortical structure in the wake of a swimming fish at $\textrm{Re}_L=10000$. The vortical structure is visualized by the iso-surface of $Q=10f^2$ colored by the normalized depthwise vorticity, $\omega_z/f$. $T=1/f$ is the tail-beat period.}
    \label{fig:wake}
\end{figure}

\subsection{Wake characteristics}
The wake of swimming fish exhibits characteristic vortical structure as shown in Fig.\ref{fig:vorQ}. The evolution of this vortical structure is examined in Fig.\ref{fig:wake} for $\textrm{Re}_L=10000$ case. As discussed in our previous study\cite{seo2022improved}, the leading edge vortex (LEV) on the caudal fin is the key vortical structure associated with the thrust generation mechanism. The formation of the LEV on the caudal fin is clearly visible in the middle of upstroke (if viewed from the top) in Fig.\ref{fig:wake} at $t/T=0$ ($T=1/f$ is the tail-beat period). The LEV keeps growing, and it detaches from the fin at the end of the upstroke ($t/T=1/4$). During this process, the tip vortices are also being generated from the tips of the caudal fin and they are convected downstream. These make elongated vortical structures as denoted in Fig.\ref{fig:wake}. As the caudal fin moves in the other direction, the LEV is shed from the caudal fin ($t/T=2/4$) and also convected downstream. The ring (or horseshoe-like) vortical structures are therefore generated by the shed LEVs connected by the tip vortices. The particular shape of the vortex ring/chain may depend on the shape of the caudal fin as well. The process of the wake vortex evolution is generally the same for all Reynolds number cases, but the Strouhal number plays a role in the form of the wake structure. The Reynolds number also plays an additional role in the dissipation of the vortical structure as well as the additional instability resulting in smaller eddies, especially at high Reynolds numbers.
\begin{figure}
    \begin{center}
    \includegraphics[width=\textwidth]{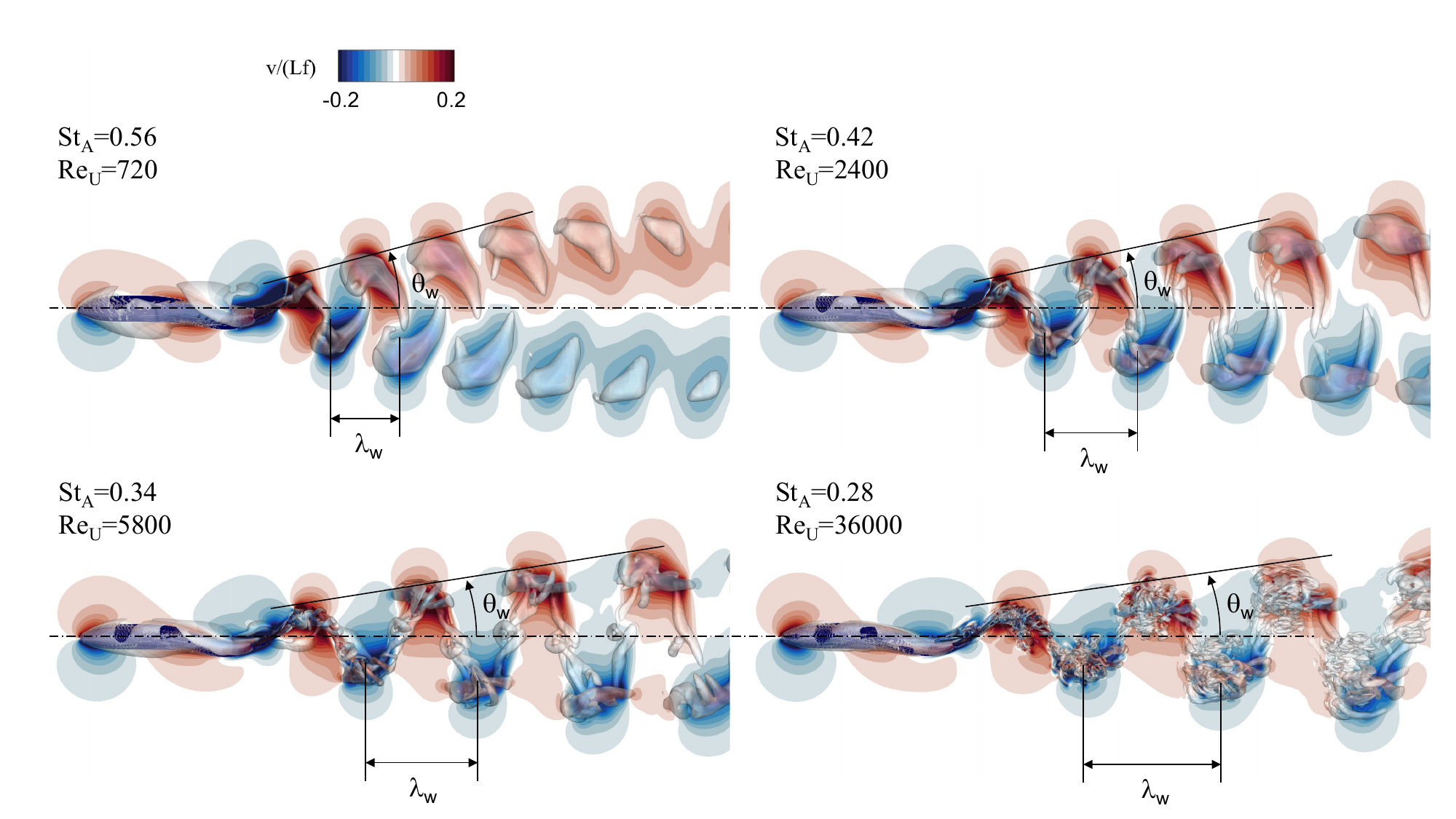}
    \end{center}
    \caption{Characterization of the wake structure. $\lambda_w$: wake wavelength. $\theta_w$: wake spreading angle. The vortical structure is visualized by the iso-surface of $Q$ along with the lateral velocity contours. $\lambda_w/A_F=1/\textrm{St}_A$, $\theta_w=\tan^{-1}(\textrm{St}_A/2)$.}
    \label{fig:wakeangle}
\end{figure}

\begin{figure}
    \begin{center}
    (a)\includegraphics[width=0.4\textwidth]{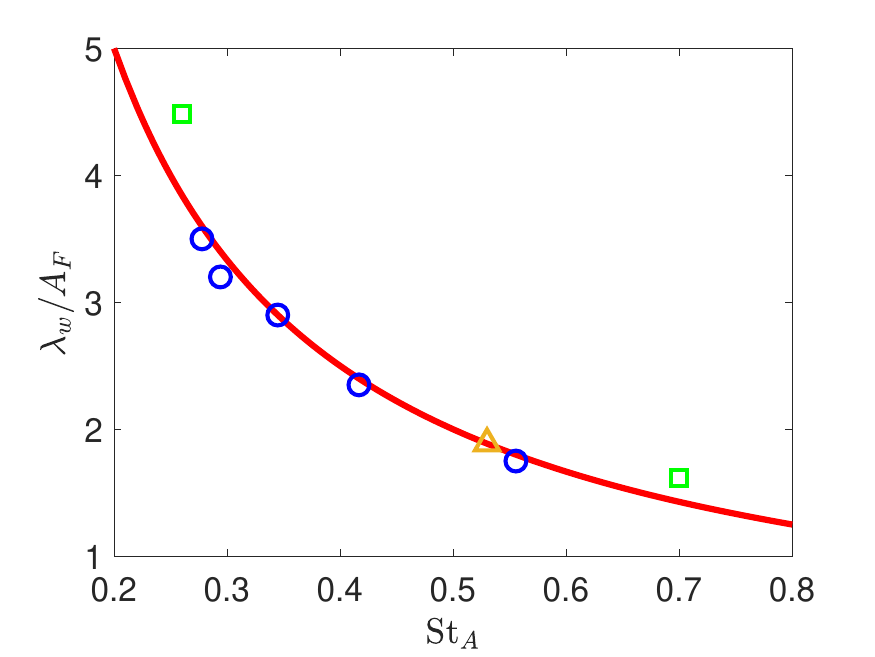} 
    (b)\includegraphics[width=0.4\textwidth]{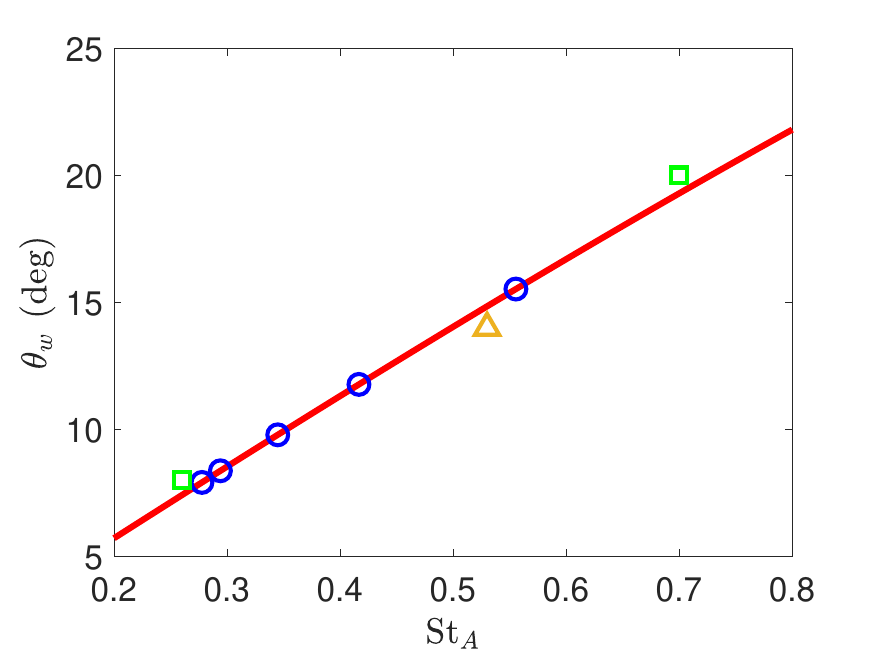} 
    \end{center}
    \caption{Wake characteristics as a function of Strouhal number. (a) Wake wavelength, $\lambda_w$. (b) Wake spreading angle, $\theta_w$. Sold line: Present scaling law, Circle: Present DNS data, Square: Data measured from the results of Borazjani and Sotiropoulos\cite{borazjani2008numerical} (Figs. 8B and 8C). Triangle: Measured from the result of Maertens et al.\cite{maertens2017optimal} (Fig. 19(c)).}
    \label{fig:wakeSt}
\end{figure}
The distance between the shed LEVs is determined by the swimming speed ($U$) and the tail-beat frequency ($f$). Thus, the wake wavelength, i.e. the distance between vortices, should be given by $\lambda_w=U/f$, and this depends on the Strouhal number. The normalized wake wavelength can be written as a function of Strouhal number: $\lambda_w/A_F=1/\textrm{St}_A$. 
The lateral motion of the caudal fin results in a strong lateral velocity component ($v$) in the wake, as shown by the color contours in Fig.\ref{fig:vorQ}, and this results in a lateral spread of the wake. It is observed that the wake vortices are convected in the lateral direction with a speed close to $fA_F/2$, especially in the near-downstream region. Since the wake vortices are also convected in the streamwise direction with the swimming speed, $U$, the wake spreading angle in the near wake can be estimated by $\theta_w=\tan^{-1}[fA_F/(2U)]=\tan^{-1}(\textrm{St}_A/2)$. Thus, two main parameters characterizing the wake structure, $\lambda_w$ and $\theta_w$ are both functions of the Strouhal number, $\textrm{St}_A$. 

The identification of the wake structure at various $\textrm{St}_A$ (and thus $\textrm{Re}_U$) is shown in Fig.\ref{fig:wakeangle}, where $\lambda_w$ and $\theta_w$ are measured from the DNS results. 
$\lambda_w$ is measured from the lateral velocity ($v$) contours by the distance between the local peaks of $v$, and $\theta_w$ is measured by following the outlines of the $Q$ iso-surfaces as depicted in Fig.\ref{fig:wakeangle}.
At higher Reynolds number, the free swimming Strouhal number gets smaller, and this makes the wake narrower with the longer wavelength. 
At $\textrm{Re}_U=36000$ and $\textrm{St}_A=0.28$, the wake spreading angle, $\theta_w$ is found to be only about $8^\circ$. As will be shown later, the minimum Strouhal number for this swimmer is estimated to be 0.23, and for this condition, the wake spreading angle will be about $6.6^\circ$ based on the present scaling law, and this small angle might be difficult to detect, especially in the near wake. 

The measured wake wavelength and spreading angle are plotted along with the present scaling laws in Fig. \ref{fig:wakeSt} for the present simulation data. 
For comparison, we measured these metrics from other carangiform swimmer simulation results\cite{borazjani2008numerical,maertens2017optimal}. The wake wavelength and spreading angle are measured from the voritcity contours presented in the papers (Figs. 8B \& 8C in Ref.\cite{borazjani2008numerical}, and Fig. 19(c) in Ref.\cite{maertens2017optimal}) and they are also plotted in the Fig.\ref{fig:wakeangle}.
Despite slight deviations mainly caused by the uncertainties in measuring wake characteristic metrics from the contour plots, the data follow the present scaling laws reasonably well. In particular, Borazjani and Sotiropoulos\citep{borazjani2008numerical} suggested that the wake at low Strouhal numbers is a ``single vortex row'' wake as opposed to the higher Strouhal number wake, which is a ``double vortex row'' wake. Our analysis of their data suggests that the difference between the two is primarily the \emph{magnitude} of the wake divergence angle, which is much smaller (but finite) for the low Strouhal number case. Indeed, the formation of a single row vortex wake would require that the lateral velocity imparted by the caudal fin be negligible compared to the swimming velocity, and this is not realizable in steady terminal swimming.

\begin{figure}
    \begin{center}
    \includegraphics[trim={0 5cm 0 6cm},clip,width=\textwidth]{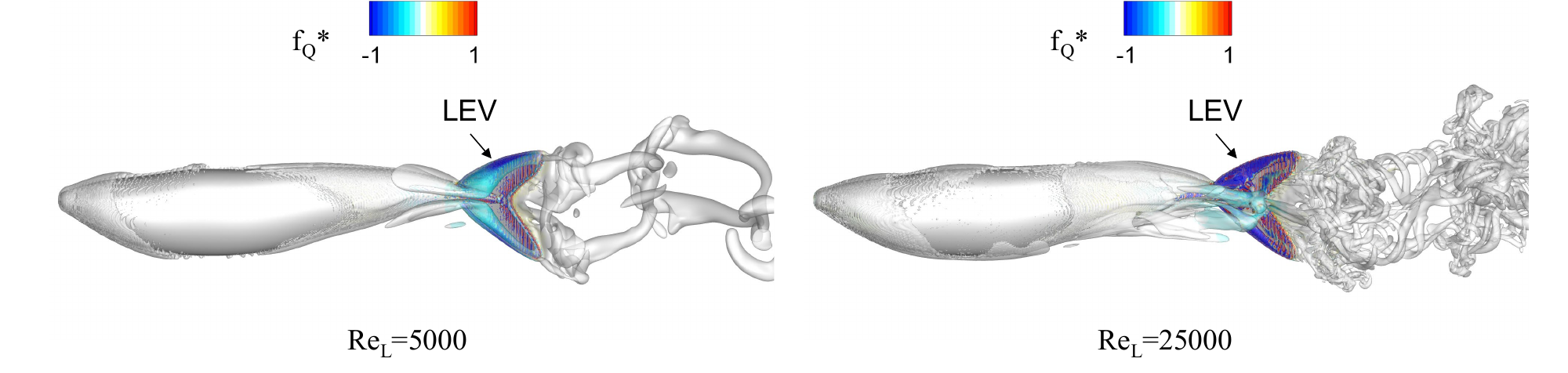}
    \end{center}
    \caption{Plot showing the importance of the LEV on the caudal fin for the generation of thrust. Plot shows iso-surface of $Q=10f^2$ colored by the normalized vortex-induced force density, $f_Q^*=f_Q/(\rho Lf^2)$, where $f_Q=-2\rho\psi Q$ and $\psi$ is the influence potential associated with the force in the surge direction on the caudal fin (this is based on the force-partitioning methods described briefly in Appendix \ref{FPM}). Negative value of force density corresponds to thrust.}
    \label{fig:fQ}
\end{figure}

\subsection{Thrust scaling}
The rest of the paper employs and/or introduces a number of dimensional as well as dimensionless parameters, and for ease of reading, we have included a table of key parameters along with their definitions and brief explanations in Appendix \ref{keyparam}. 

For sub-carangiform, carangiform, and thunniform swimmers, thrust is mainly generated by the caudal fin.
According to our findings from the force partitioning method (FPM) analysis \citep{seo2022improved}, the thrust generated by the caudal fin is primarily associated with the leading edge vortex (LEV). 
The FPM analysis\cite{menon2022contribution}, which is briefly described in Appendix \ref{FPM}, provides the vortex-induced force density field, which shows the contribution of local vortical structure on the force generation. 
The vortex-induced force density is defined by $f_Q=-2\rho \psi Q$, where $Q$ is the second invariant of velocity gradient and $\psi$ is the influence potential associated with the body of interest. 
Figure \ref{fig:fQ} shows the vortical structures colored by $f_Q$ for the force in the surge direction generated by the caudal fin at two different Reynolds numbers. 
One can clearly see that the vortex-induced force density is concentrated on the LEV of the caudal fin. 
At higher Reynolds number, the size of the LEV gets smaller, while the force density increases, thereby maintaining the dominant role of the LEV in thrust generation. 
More details about the application of the FPM to a caudal fin swimmer can be found in Ref.\cite{seo2022improved}.

\begin{figure}
    \begin{center}
    \includegraphics[width=0.6\textwidth]{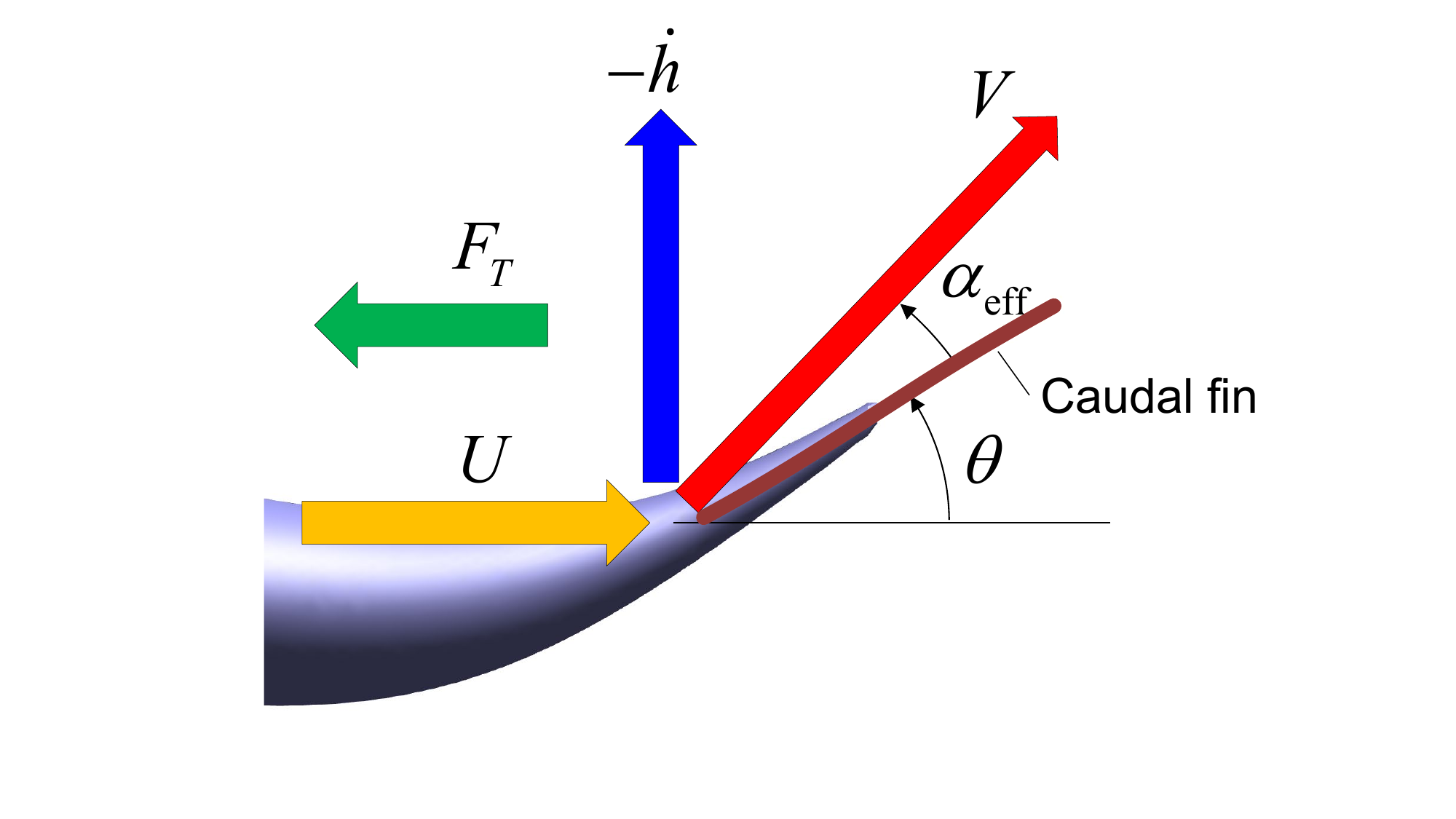}
    \end{center}
    \caption{Effective angle of attack, $\alpha_{\textrm{eff}}$, on the caudal fin}
    \label{fig:alpha_eff}
\end{figure}

Our previous study on flapping foils showed that the force generated by the pitching and heaving foil is also mainly associated with the LEV\citep{raut2024hydrodynamic}, and we have developed a LEV-based model to predict the thrust of flapping foils. The caudal fin of the fish can also be considered as a pitching and heaving foil (see Fig.\ref{fig:alpha_eff}) where $h$ is the heaving displacement, $\dot{h}$ is the heaving velocity, and $\theta$ is the pitching angle. It follows that the LEV-based model can be extended to the generation of thrust by a caudal fin in a BCF swimmer, and this model (described below) forms the basis of our scaling analysis. 

Following our previous work \citep{raut2024hydrodynamic}, the strength of the LEV should be proportional to the component of the net relative flow velocity, $V=\sqrt{U^2+ \dot{h}^2}$, normal to the chord of the foil. The magnitude of this velocity component is related to the instantaneous effective angle of attack ($\alpha_\textrm{eff}$) on the caudal fin (see Fig.\ref{fig:alpha_eff}). The circulation $\Gamma$ for the foil is then proportional to this velocity component by
\begin{equation}
    \Gamma \propto c V \sin \alpha_\textrm{eff},
    \label{eq:gamma}
\end{equation}
where $c$ is the chord length of the fin. 
The instantaneous effective angle of attack, $\alpha_\textrm{eff}$, is given by
\begin{equation}
    \alpha_\textrm{eff}(t) = -\tan^{-1}\left[{\dot{h}(t)}/{U}\right] - \theta(t)
    \label{alphaEffDef}
\end{equation}
By applying the Kutta–Joukowski theorem, Eq. (\ref{eq:gamma}) provides the scaling of the force generated by the flapping foil:
\begin{equation}
\frac{{{F_N}}}{{(1/2)\rho {V^2}{S_f}}} \propto \sin {\alpha _\textrm{eff}},
\label{eq:FN}
\end{equation}
where $F_N$ is the force normal to the foil surface and $S_f$ is the area of the foil. The scaling of the thrust component, $F_T=F_N \sin \theta$, is therefore given by
\begin{equation}
\frac{{{F_T}}}{{(1/2)\rho {V^2}{S_f}}} \propto \sin {\alpha _\textrm{eff}}\sin \theta, 
\label{eq:FT1}
\end{equation}
Based on the above relation, a mean thrust factor, $\Lambda_T$ can be defined as
\begin{equation}
{\Lambda _T} = \overline {\sin {\alpha _\textrm{eff}}\sin \theta },
\label{eq:sasq}
\end{equation}
where bar denotes average over the flapping cycle, and the mean thrust coefficient, $C_T$ should be proportional to this factor, i.e., $C_T\propto \Lambda_T$. This is obtained under the potential flow framework by applying the Kutta-Joukowski theorem and assuming zero drag on the fin.
This linear relationship has been verified extensively for pitching and heaving foils in the previous study of Raut et al. \citep{raut2024hydrodynamic} by conducting 462 distinct simulations of flapping foils. The data from the simulations fits linearly to this model with an $R^2$ value of 0.91, indicating a high level of accuracy in the model. 

The above model can be applied to derive a scaling for the thrust force generated by the caudal fin. Based on carangiform swimming kinematics (Eq.(\ref{eq:carangi})), the heaving ($h(t)$) and pitching motion ($\theta(t)$) of the caudal fin, which is located at $x=L$ can be given by:
\begin{equation}
\begin{aligned}
h(t) &= \Delta y(L,t) = ({A_F}/2)\sin ( t^*),\\
\dot h (t) &=  - \pi f{A_F}\cos ( t^*),\\
\theta(t)  &= {\tan ^{ - 1}}\left[ {{{\left( {{\partial (\Delta y)}}/{{\partial x}} \right|}_{x = L}}} \right],
\end{aligned}
\label{eq:h_dot}
\end{equation}
where
\begin{equation}
 {{{\left. {{\partial (\Delta y)}}/{{\partial x}} \right|}_{x = L}}} = (\pi {A_F}/\lambda)\cos ( t^*) + {(dA/dx)_{x = L}}\sin ( t^*). 
\label{eq:dydx}
\end{equation}
and $t^*=2\pi(L/\lambda -ft)$.
The above equation shows that the amplitude growth rate, $dA/dx$, at the tail affects the caudal fin pitching angle. 
The second term on the right hand side of Eq.(\ref{eq:dydx}) modulates the pitching amplitude and phase, which can be quantified via the following parameter:
\begin{equation}
A'^*=\frac{1}{2\pi} \frac{\lambda}{A(L)} \left({\frac{dA}{dx}}\right)_{x=L} = \frac{1}{2\pi} \frac{\lambda}{L}\left(\frac{{{a_1} + 2{a_2}}}{{{a_0} + {a_1} + {a_2}}}\right),
\label{eq:Aps}
\end{equation}
where $a_0, a_1,$ and $a_2$ are the second-order polynomial coefficients in the quadratic amplitude envelope function, Eq. (\ref{eq:A}).
$A^{\prime *}$ includes a measure of the growth rate of the amplitude envelope at the tail, and the normalized wavelength of the body wave, $\lambda/L$. With this parameter, pitching amplitude and phase modulations are given by ${R_\theta } = \sqrt {1 + {A^{\prime *}}^ 2}$ and ${\phi _\theta } = {\tan^{ - 1} A^{\prime *}}$, respectively, and the modulated amplitude is defined by $A_F^*={A_F}R_\theta/\lambda$. The pitching angle of the caudal fin is then given by
\begin{equation}
\theta(t)  = {\tan ^{ - 1}}\left[ {\pi A_F^* \cos \left( {t^* - {\phi _\theta }} \right)} \right].
\label{eq:theta}
\end{equation}
The above expressions for the heaving and pitching of the caudal fin can be derived for any undulatory motion kinematics ($\Delta y(x,t)$) given by the amplitude envelope function and the traveling wave equation in the form of Eq.(\ref{eq:carangi}). Alternatively, they can also be derived directly from the heaving and pitching motion of the fin.

\begin{figure}
    \centering
    (a)\includegraphics[trim={0 2.5cm 0 3cm},clip,width=0.4\textwidth]{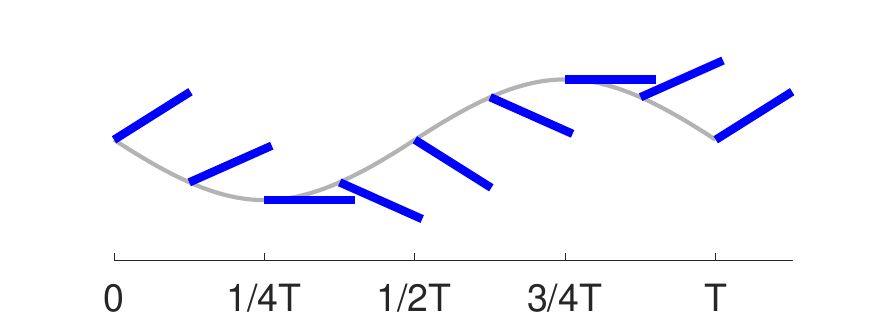}\\
    (b)\includegraphics[trim={0 2.5cm 0 3cm},clip,width=0.4\textwidth]{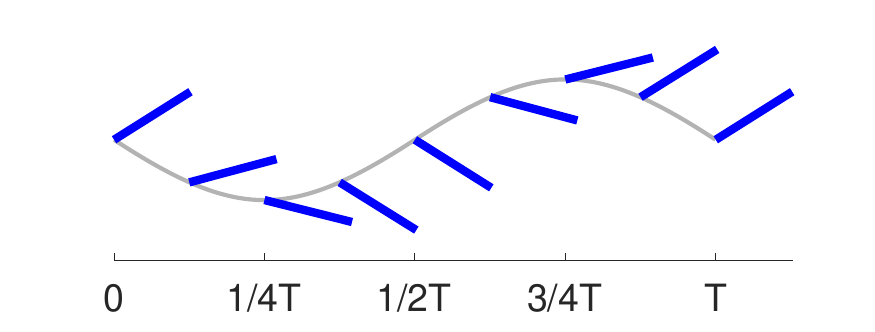}\\
    (c)\includegraphics[trim={0 2.5cm 0 3cm},clip,width=0.4\textwidth]{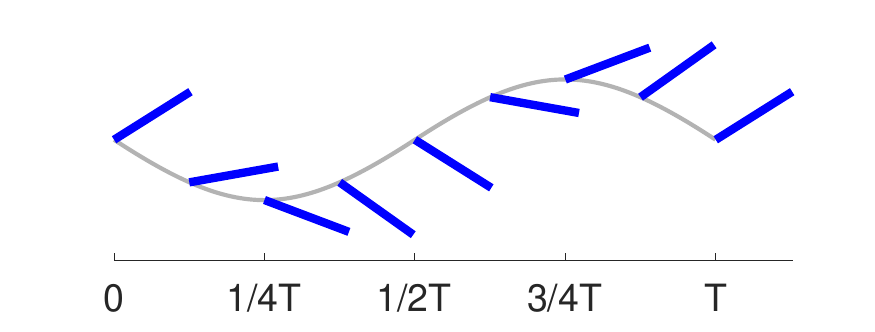} 
    \caption{Heaving and pitching motion of the caudal fin for various $A'^*$ values. The caudal fin represented by the blue straight line is plotted with the temporal interval of $T/8$. The gray line shows the heaving profile. (a) $A'^*=0$ ($R_\theta =1, \phi_\theta=0, A_F^*=0.2$), (b) $A'^*=0.4$ ($R_\theta=1.08, \phi_\theta=21.8^\circ ,A_F^*=0.215$), (c) $A'^*=0.6$ ($R_\theta=1.17, \phi_\theta=31^\circ, A_F^*=0.233$). The tail-beat amplitude, wavelength, and caudal fin length are set to $0.2L$, $L$, and $0.15L$, respectively.}
    \label{fig:finAps}
\end{figure}

\begin{figure}
    \centering
    \includegraphics[width=0.99\linewidth]{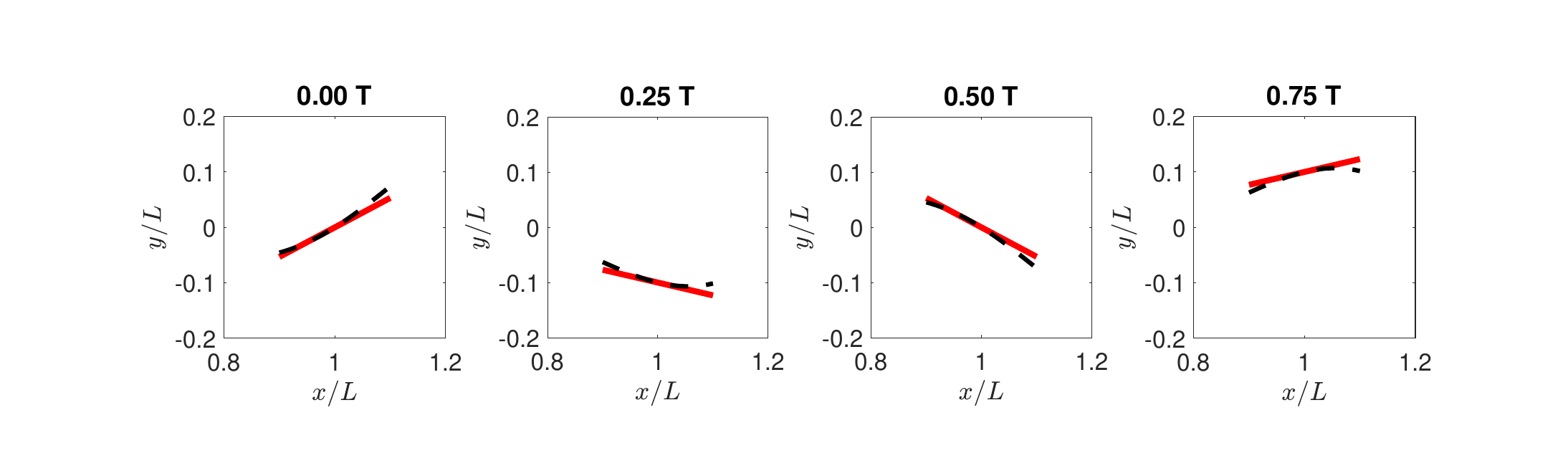}
    \caption{Caudal fin motion comparison. Solid line: pitching and heaving motion given by Eqs.(\ref{eq:h_dot})-(\ref{eq:theta}). Dashed line: undulatory wave motion given by Eq.(\ref{eq:carangi}).}
    \label{fig:tailcomp}
\end{figure}

Thus, $A^{\prime *}$ emerges as an independent non-dimensional parameter in the pitch variation of the caudal fin for BCF swimming. 
The pitching and heaving motions of the caudal fin given by Eqs.(\ref{eq:h_dot})-(\ref{eq:theta}) are plotted in Fig.\ref{fig:finAps} for three different $A'^*$ values: 0, 0.4, and 0.6.  
While $A'^*$ also affects $A_F^*$, which determines the maximum pitch angle of the fin, $A'^*$ may be best viewed as a measure of the phase angle mismatch between pitch and rate of heave introduced by the kinematics, with direct impact on the effective angle-of-attack. 
As one can see in Fig.\ref{fig:finAps}, the most noticeable change in the caudal fin kinematics due to $A'^*$ is the pitch angle at the maximum heave displacement (or at the zero rate of heave, $1/4T$ and $3/4T$). As will be shown, this affects the effective angle-of-attack, and thus the thrust and power as well. All of the parameters introduced here are derived from the BCF kinematics prescribed in Eqs. (\ref{eq:carangi}) and (\ref{eq:A}). For the current kinematics, $A_F^*=0.214$, $A^{\prime *}=0.38$, $R_\theta=1.071$, and $\phi_\theta=21^\circ$. 
In figure \ref{fig:tailcomp}, the caudal fin motion modeled by pitching and heaving (Eqs.(\ref{eq:h_dot})-(\ref{eq:theta})) is compared with the one prescribed by the undulatory motion equation (Eq.(\ref{eq:carangi})). 
It shows that the present pitching and heaving formulations represent the caudal fin kinematics quite well, although there are small differences due to the additional deformation in the undulatory wave motion. The maximum difference between the two is found to be about 2\% of the body length.

The LEV-based model suggests that the thrust scales with the product of $\sin{\alpha_{\textrm{eff}}}$ and $\sin\theta$ (Eq. \ref{eq:FT1}). For the caudal fin kinematics given by Eqs.(\ref{eq:h_dot})-(\ref{eq:theta}), one can get:
\begin{equation}
\begin{aligned}
\sin {\alpha _{{\rm{eff}}}} = & \sin \left[ {{{\tan }^{ - 1}}\left( {\pi {\rm{S}}{{\rm{t}}_A}\cos (t^*)} \right) - {{\tan }^{ - 1}}\left( {{\pi A_F^*}\cos (t^* - {\phi _\theta })} \right)} \right] \\
= &\frac{{\pi {\rm{S}}{{\rm{t}}_A}\cos (t^*) - {\pi A_F^*} \cos (t^* - {\phi _\theta })}}{{\sqrt {1 + {{(\pi {\rm{S}}{{\rm{t}}_A})}^2}{{\cos }^2}(t^*)} \sqrt {1 + {{{\left(\pi A_F^*\right)}}^2}{{\cos }^2}(t^* - {\phi _\theta })} }},
\end{aligned}
\end{equation}
and
\begin{equation}
\sin \theta  = \sin \left[ {{{\tan }^{ - 1}}\left( {{\pi A_F^*}\cos (t^* - {\phi _\theta })} \right)} \right] = \frac{{{\pi A_F^*}\cos (t^* - {\phi _\theta })}}{{\sqrt {1 + {{{\left(\pi A_F^*\right)}}^2}{{\cos }^2}(t^* - {\phi _\theta })} }},
\end{equation}
where
$\textrm{St}_A=fA_F/U$.
The thrust factor is then written in terms of the swimming kinematics parameters:
\begin{equation}
{\Lambda _T} = \overline { \left\{ \frac{{{\pi^2 A_F^*}\left[ {{\rm{S}}{{\rm{t}}_A}\cos (t^*)\cos (t^* - {\phi _\theta }) - { A_F^*}{{\cos }^2}(t^* - {\phi _\theta })} \right]}}{{\sqrt {1 + {{(\pi {\rm{S}}{{\rm{t}}_A})}^2}{{\cos }^2}(t^*)} \left[ {1 + {{{\left(\pi A_F^*\right)}}^2}{{\cos }^2}(t^* - {\phi _\theta })} \right]}} \right\} }.
\label{eq:Lambda_T}
\end{equation}
The integral to perform averaging in Eq.(\ref{eq:Lambda_T}) may be challenging, and an approximate expression is proposed by simplifying the denominator as the following:
\begin{equation}
{\Lambda _T} \approx \frac{{{\pi^2 A_F^*}\left( { {\rm{S}}{{\rm{t}}_A}\cos {\phi _\theta } - { A_F^*}} \right)}}{{2\sqrt {1 + \sigma {{(\pi {\rm{S}}{{\rm{t}}_A})}^2}} \left[ {1 + \sigma {{{\left(\pi A_F^*\right)}}^2}} \right]}},
\label{eq:Lambda_T_app}
\end{equation}
where $\sigma$ is an empirical parameter determined to have a value of about 0.63 by using a non-linear curve fitting with a RMS error of 4\% for ${\rm{0 < S}}{{\rm{t}}_A} \le 1$, $0.1 \le {A_F^*} \le 0.5$, which covers a large range of possible values of these kinematic parameters for BCF swimming. The details can be found in Appendix \ref{sec:appx}.  Alternatively, the integral can be performed numerically if all the kinematic parameters ($\textrm{St}_A$, $A_F^*$, and $A^{\prime *}$) are given. 

Based on Eqs.(\ref{eq:FT1}) and (\ref{eq:sasq}), the mean thrust coefficient, $C_T$, is then given by
\begin{equation}
\begin{aligned}
{C_T} = & \frac{{{{\bar F}_T}}}{{\frac{1}{2}\rho {U^2}{S_x}}}= \frac{{{\beta _T}\frac{1}{2}\rho \overline {{V^2}{S_f}\sin {\alpha _{{\rm{eff}}}}\sin \theta } }}{{\frac{1}{2}\rho {U^2}{S_x}}} \approx {\beta _T}\frac{{{S_f}}}{{{S_x}}}\left[ {1 + \sigma {{(\pi {\rm{S}}{{\rm{t}}_A})}^2}} \right]{\Lambda _T} \\ \approx & \beta_T   \frac{\pi^2}{{2}}  \frac{S_f}{S_x} 
\frac{A_F}{\lambda}
\left( {{\rm{S}}{{\rm{t}}_A} - {{ A_F^* R_\theta}}} \right)\frac{{\sqrt {1 + \sigma {{(\pi {\rm{S}}{{\rm{t}}_A})}^2}} }}{{1 + \sigma {{{\left(\pi A_F^*\right)}}^2}}} 
\end{aligned}
\label{eq:CT}
\end{equation}
where $S_f$ is the area of the caudal fin, $S_x$ is the fish frontal area in the surge direction, and $\beta_T$ is a constant of proportionality that we expect is mostly related to the shape of the fin. Note that ${R_\theta }\cos {\phi _\theta } = 1$ by definition.  
The above equation shows that thrust coefficient is a function of body morphology (the parameter $S_f/S_x$, which is equal to 1.17 for the current model) and fin morphology ($\beta_T$), BCF kinematics ($A_F^*$ and $A^{'*}$, and the swimming velocity, which is embedded in $\text{St}_A$).
Since Eq.(\ref{eq:FN}) is based on the Kutta-Joukowski theorem, in theory one may derive the value of $\beta_T$ by applying a potential flow model. However, in reality, $\beta_T$ may also depend on the fin flexibility, because deformation can result in camber along the chord (see Fig.\ref{fig:tailcomp}), and also interactions with flow structures from the body and any upstream fins.

\begin{figure}
    \begin{center}
    (a)\includegraphics[width=0.4\textwidth]{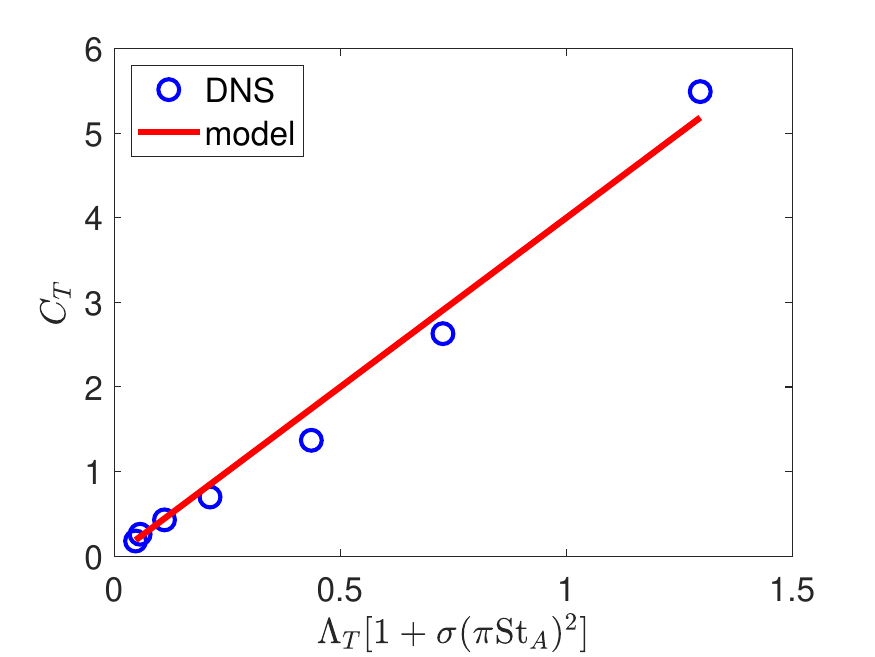} 
    (b)\includegraphics[width=0.4\textwidth]{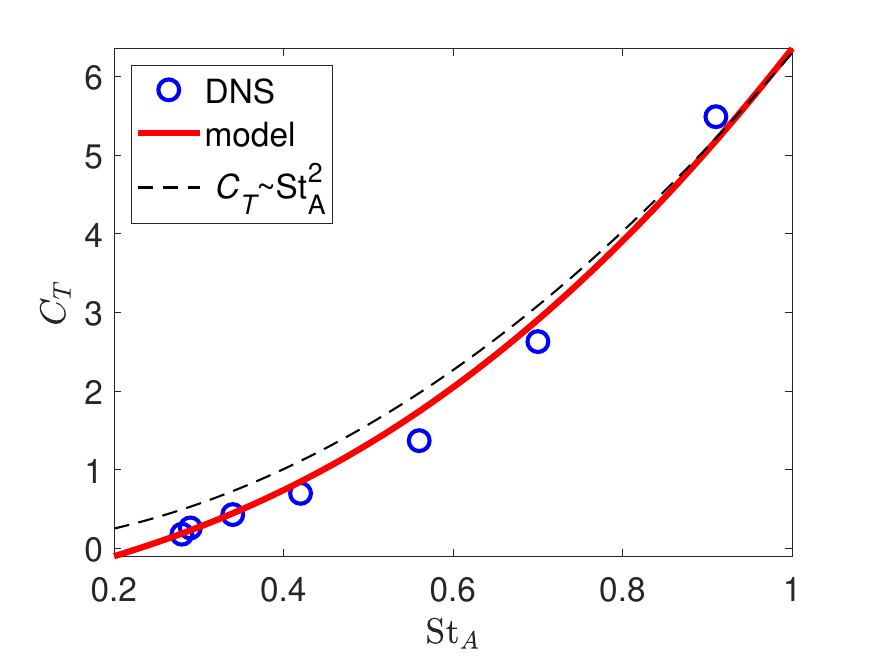} 
    \end{center}
    \caption{Thrust scaling of carangiform swimmers. (a) Correlation between the mean thrust coefficient and thrust factor. (b) Thrust coefficient as a function of Strouhal number. Dashed line: Asymptotic scaling, $C_T\sim\textrm{St}_A^2$.}
    \label{fig:CT}
\end{figure}

The thrust scaling derived above is applied to the present DNS results. As noted earlier, the thrust on the swimming fish is mainly due to the pressure force on the caudal fin. Thus, it is assumed that $C_T \approx -C_{p,\textrm{fin}}$. The data from the DNSs is fitted onto Eq.(\ref{eq:CT}) in Fig.\ref{fig:CT}(a), and it shows excellent linear correlation with $R^2=0.98$. The regression estimates the constant $\beta_T$ to be 3.43. In Fig.\ref{fig:CT}(b), the thrust coefficient is plotted as a function of Strouhal number ($\textrm{St}_A$) by using Eq.(\ref{eq:CT}) along with the data from the DNS, and it shows that the thrust coefficient of carangiform swimmers can be predicted by the present scaling with reasonable accuracy. 

In some previous studies \cite{gazzola2014scaling,ventejou2025universal}, the thrust was simply scaled by  $\sim\rho(\pi fA_F)^2=\rho V_{\max}^2$, where $V_{\max}=\pi fA_F$ is the maximum lateral velocity at the tail. This yields the scaling for thrust coefficient: $C_T \sim\textrm{St}_A^2$. 
Floryan\citep{floryan2018efficient} et al. also proposed the thrust scaling as $C_T \sim \textrm{St}_A^2-C_{D,0}$, where $C_{D,0}$ is a coefficient for an ``offset drag", which is a drag on the foil at $\textrm{St}_A \rightarrow 0$.
Equation (\ref{eq:CT}) shows that, if the Strouhal number is very high ($\textrm{St}_A>>A_F^* R_\theta$ and $\textrm{St}_A>>1/(\pi \sqrt{\sigma})\approx0.4$), $C_T$ indeed scales with $\sim\textrm{St}_A^2$. This asymptotic scaling is also plotted in the Fig.\ref{fig:CT}(b). At low Strouhal numbers, however, the present result suggests that $C_T$ may only increase linearly with $\textrm{St}_A$. 
The present formulation also shows that there is an ``offset drag" on the caudal fin at $\textrm{St}_A=0$ but this is associated with the fin kinematics.

The thrust scaling, Eq.(\ref{eq:CT}), indicates that to generate positive thrust (i.e. to overcome the offset drag) in carangiform type BCF swimming, the Strouhal number has to be greater than a certain minimum value, $\textrm{St}_\textrm{min}$, which is given by
\begin{equation}
{\textrm{St}_{\min }} = A_F^* R_\theta  = ({A_F}/{\lambda})\left( {1 + {A^{\prime *}}^2} \right).
\label{eq:Stmin}    
\end{equation}
The $\textrm{St}_{\min}$ can be considered as the minimum Strouhal number for free swimming at zero body drag or at $\textrm{Re}_U \rightarrow \infty$. 
If $\textrm{St}_A<\textrm{St}_\textrm{min}$, the fin will generate drag instead of thrust, and the fish will decelerate. For the present fish model, the minimum Strouhal number is estimated to be 0.23. 

We also note that, for the given kinematics, the minimum Strouhal number corresponds to the maximum swimming speed such as $\textrm{St}_\textrm{min} \equiv A_F f/U_\textrm{max}$, where $U_\textrm{max}$ is the maximum swimming speed with zero body drag, the parameter $\textrm{St}_\textrm{min}$ provides the following scaling for the non-dimensional maximum free swimming speed, $U^*_{\max}$ :
\begin{equation}
  U^*_{\max} = {{{U_{\max }}}}/{({\lambda f})} = {{{U_{\max }}}}/{{{U_c}}} = {\left( {1 + {A^{\prime *}}^2} \right)^{ - 1}},
    \label{eq:Umax}
\end{equation}
where $U_c=\lambda f$ is the wave speed of the undulatory motion. This has several important implications. First, this indicates that the swimming speed $U$ cannot exceed $U_c$ during steady (terminal) swimming, a result known at least as far back as the work of Lighthill\cite{lighthill1960note}.  Second, for the present swimming kinematics, $U_\textrm{max}/U_c$ is about 0.87, but we note that this maximum speed would only be achieved for the caudal fin swimmer if the drag on the body to which the fin is attached, were zero. In reality, the body of the fish will generate a non-zero drag, thereby reducing the velocity below the maximum achievable velocity. Indeed, the average swimming speed of mackerel reported is about 0.81$U_c$ \citep{videler1984fast}, which is slightly lower than the maximum possible velocity. However, during deceleration, $U/U_c$ could exceed the maximum value given by Eq.(\ref{eq:Umax}). $U/U_c>U^*_{\max}$ corresponds to $\textrm{St}_A<\textrm{St}_\textrm{min}$, and the fin generates drag instead of thrust as discussed above.

\begin{figure}
    \centering
    (a)\includegraphics[width=0.4\linewidth]{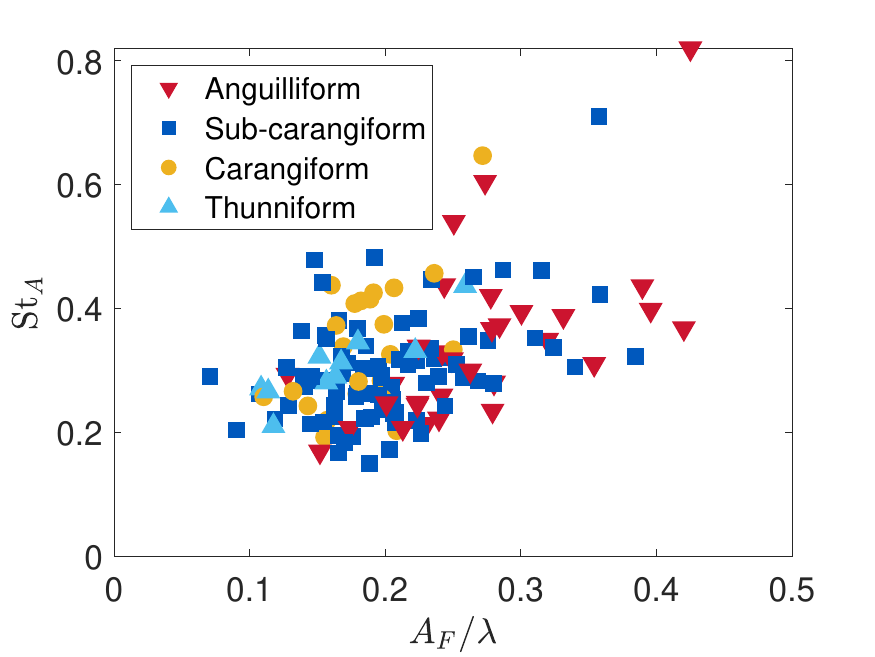}
    (b)\includegraphics[width=0.4\linewidth]{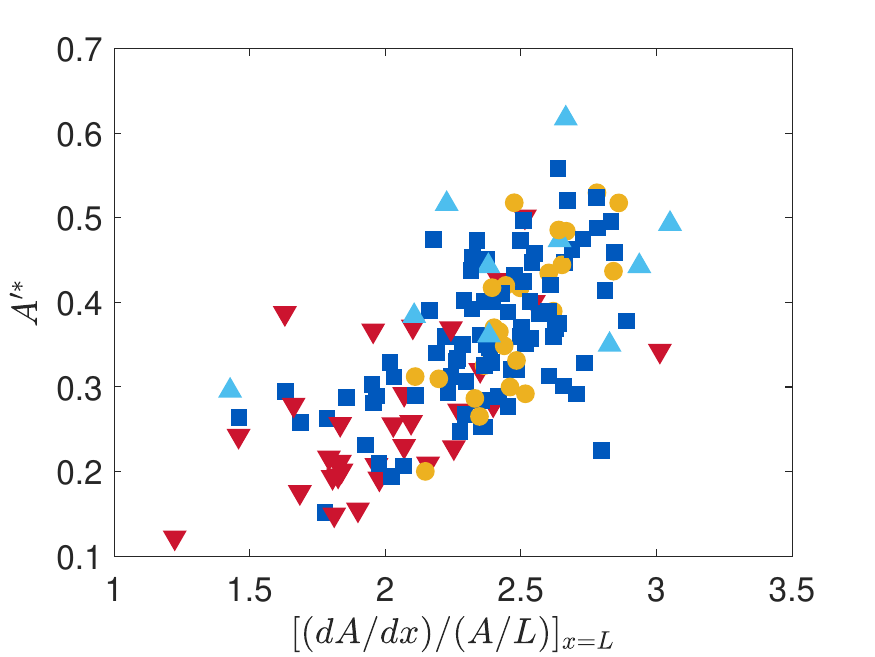}
    \caption{Kinematic parameters for BCF swimmers extracted from the data by Di Santo et al.\cite{di2021convergence} for various swimming modes.(a) Strouhal number, $\textrm{St}_A$ and normalized tail-beat amplitude, $A_F/\lambda$. 
    (b) Normalized amplitude envelope slope at the tail, $[(dA/dx)/(A/L)]_{x=L}$, and the kinematic parameter, $A'^*$.
    %Delta: Anguilliform, Square: Sub-carangiform, Circle: Carangiform, Triangle: Thunniform.
    }
    \label{fig:ApsQs}
\end{figure}

Third, the maximum velocity $U^*_{\max}$ is a function of $A'^*$ only, which depends on the BCF kinematics. Specifically, in addition to the wavelength $\lambda/L$, the parameter $A'^*$ depends on the slope of the body amplitude envelope, $dA/dx$, at the fin location. While several studies have shown that the shape of the body envelope\cite{di2021convergence} can vary significantly for different BCF swimmers and researchers have also simulated BCF swimmers with different amplitude envelopes\cite{li2021fishes, borazjani2010role}, the criticality of this feature in determining the maximum swimming speed of caudal fin swimmers has never been emphasized before. To understand the variability in this parameter for fish, we have extracted the kinematic parameters of swimming fish from Di Santo et al.\cite{di2021convergence}, who studied the swimming kinematics of a large number of BCF swimmers. Their study included a total of 151 cases with various swimming modes, including anguilliform, sub-carangiform, carangiform, and thunniform -- and quantified the amplitude envelope functions for these swimmers. 
The Strouhal number and normalized tail-beat amplitude are plotted in Fig.\ref{fig:ApsQs}(a). Most of data points are in the Strouhal number range from 0.2 to 0.4, and the normalized tail-beat amplitude between 0.1 and 0.3.
The kinematic parameter, $A'^*$, and the normalized amplitude envelope slope at the tail, $[(dA/dx)/(A/L)]_{x=L}$ are plotted in Fig.\ref{fig:ApsQs}. 
Note that the current model does not apply to anguilliform propulsion since these swimmers do not have a prominent caudal fin, but this data is included for the sake of completeness.
The figure suggests a large variability of $A'^*$ from 0.12 to 0.62 for these BCF swimmers. This is partially due to the fact that $\lambda/L$ varies within and among species from 0.5 to 1.5\cite{di2021convergence}, but part of this is also due to the amplitude envelope slope at the tail. We also note that while anguilliform and thunniform swimmers mostly occupy the two ends of the distribution, sub-carangiform and carangiform swimmers span almost the entire range of these two variables.

Fourth, it is noteworthy that the scaling of the maximum swimming speed reveals no dependence on fin morphology. This suggests that the maximum achievable speed is governed for the most part by the BCF kinematics. This insight offers practical guidance for the design of bio-robotic underwater vehicles (BUVs), highlighting how speed can be maximized by optimizing kinematic parameters.

\subsection{Scaling of Hydrodynamic Power}
The LEV-based model introduced in the previous section can also be applied to derive a scaling law for the mechanical power expended by the caudal fin. As shown in Table \ref{tab:allcases}, the mechanical power expended by the caudal fin is mostly due to the work done against the pressure load, and the contribution of viscous force to this is negligible. Thus, the mechanical power can be calculated by
\begin{equation}
{W_\textrm{fin}}\approx  - \int_{{\rm{fin}}} {\Delta p \,{n_y} \, \Delta \dot y} \,dS \approx  - {F_L}\dot h,
\label{eq:W}
\end{equation}
where $\Delta p$ is the pressure difference across the caudal fin, $n_y$ is the lateral component of the surface normal unit vector, $F_L$ is the lateral pressure force on the caudal fin, and the negative sign is introduced to represent power input by the fish. 
The mechanical power associated with pitching can be expressed as $r_\theta F_N \dot{\theta}$, where $r_\theta$ is the distance from the rotation center to the pressure center of the caudal fin. This term can, in principle, be included in the total power budget. However, our analysis indicates that the hydrodynamic power due to pitching is significantly smaller than the heaving power. This is primarily because $r_\theta$ is very small (approximately $0.005L$ in the present case) and because the phase difference between $F_N$ and $\dot{\theta}$ is close to $\pi/2$, which minimizes the effective contribution of pitching to the total power. Based on these factors, our estimates show that pitching power is less than 6\% of the heaving power. Consequently, only the heaving power associated with the lateral force is considered in the scaling analysis.

By employing the LEV-based model described in the previous section, the lateral force can be expressed by
\begin{equation}
\frac{{{F_L}}}{{(1/2)\rho {V^2}{S_f}}} \propto \sin {\alpha _\textrm{eff}}\cos \theta.    
\end{equation}
Based on this, a mean power factor can be defined as 
\begin{equation}
{\Lambda _W} =  - \overline {(\dot h/U)\sin {\alpha _\textrm{eff}}\cos \theta },
\end{equation}
and the normalized mean power is expected to be proportional to this factor.
By using the swimming kinematics equations, the power factor is written as
\begin{equation}
{\Lambda _W} = \overline { \left\{ \frac{{\pi^2 {\rm{S}}{{\rm{t}}_A}\left[ { {\rm{S}}{{\rm{t}}_A}{{\cos }^2}(t^*) - { A_F^*}\cos (t^*)\cos (t^* - {\phi _\theta })} \right]}}{{\sqrt {1 + {{(\pi {\rm{S}}{{\rm{t}}_A})}^2}{{\cos }^2}(t^*)} \left[ {1 + {{{\left(\pi A_F^*\right)}}^2}\cos^2(t^* - {\phi _\theta })} \right]}} \right\} }.  
\end{equation}
Like the thrust factor (see Appendix \ref{sec:appx}), an approximate expression is proposed as
\begin{equation}
{\Lambda _W} \approx \frac{{\pi^2 {\rm{S}}{{\rm{t}}_A}\left( { {\rm{S}}{{\rm{t}}_A} - { A_F^*}\cos {\phi _\theta }} \right)}}{{2\sqrt {1 + \sigma {{(\pi {\rm{S}}{{\rm{t}}_A})}^2}} \left[ {1 + \sigma {{{\left(\pi A_F^*\right)}}^2}} \right]}}.
\label{eq:Lambda_W}
\end{equation}
The mean power coefficient is then given by
\begin{equation}
\begin{aligned}
C_W= & \frac{{{{\bar W}_{{\rm{fin}}}}}}{{\frac{1}{2}\rho {U^3}{S_x}}} \approx {\beta _W} \frac{S_f}{S_x}\left[ {1 + \sigma {{(\pi {\rm{S}}{{\rm{t}}_A})}^2}} \right]{\Lambda _W} \\ \approx & {\beta _W}\frac{S_f}{S_x}\frac{{{\pi ^2}}}{2}{\rm{S}}{{\rm{t}}_A}\left( {{\rm{S}}{{\rm
{t}}_A} - A_F/\lambda} \right)\frac{{\sqrt {1 + \sigma {{(\pi {\rm{S}}{{\rm{t}}_A})}^2}} }}{{1 + \sigma {{{\left(\pi A_F^*\right) }}^2}}}.
\end{aligned}
\label{eq:CW}
\end{equation}

\begin{figure}
    \begin{center}
    (a)\includegraphics[width=0.4\textwidth]{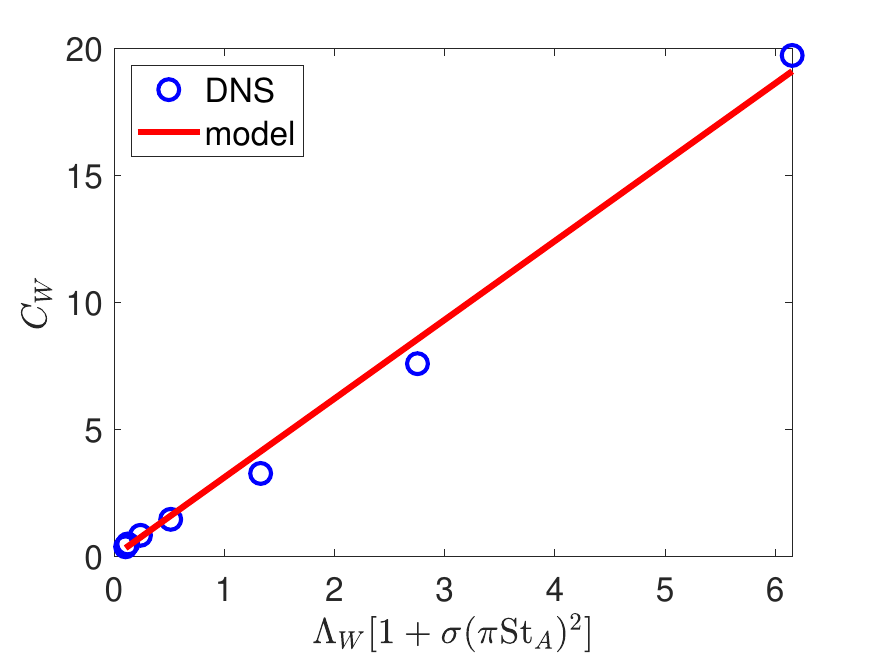} 
    (b)\includegraphics[width=0.4\textwidth]{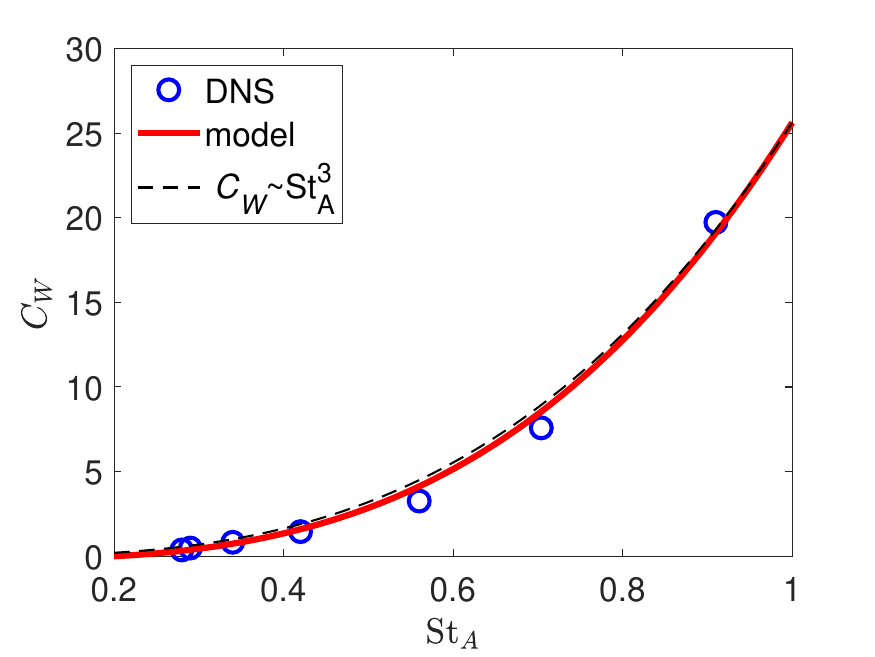} 
    \end{center}
    \caption{Power scaling for the caudal fin of carangiform swimmers. (a) Correlation between the power coefficient and power factor. (b) Power coefficient as a function of Strouhal number. Dashed line: Asymptotic scaling, $C_W\sim\textrm{St}_A^3$.}
    \label{fig:CW}
\end{figure}

The data from the DNSs are fitted to the suggested scaling, Eq.(\ref{eq:CW}) in Fig.\ref{fig:CW}(a), and an excellent linear correlation ($R^2=0.99$) with $\beta_W=2.55$ is observed. In Fig.\ref{fig:CW}(b), the power coefficient is plotted as a function of the Strouhal number ($\textrm{St}_A$) by using Eq.(\ref{eq:CW}) along with the data from the DNS, which shows that the present scaling predicts the power coefficient very well. 
In many previous studies, the mechanical power was scaled by $C_W\sim\textrm{St}_A^3$\cite{floryan2018efficient,das2022contrasting,quinn2014unsteady} and this scaling is also plotted in Fig.\ref{fig:CW}(b). We note that the present data fits well to the $\textrm{St}_A^3$ scaling as well. However the scaling in Eq. (\ref{eq:CW}) provides additional useful information. For instance
it shows that $C_W>0$ for $\textrm{St}_A>A_F/\lambda$, a condition that is identical to $U_c>U$. Thus, the rate of work for the fish is positive if the undulatory wave speed is faster than the swimming speed, a result that agrees with Lighthill's slender swimmer theory\cite{lighthill1960note}.

\subsection{Froude Efficiency}
\begin{figure}
    \centering
    \includegraphics[width=0.5\linewidth]{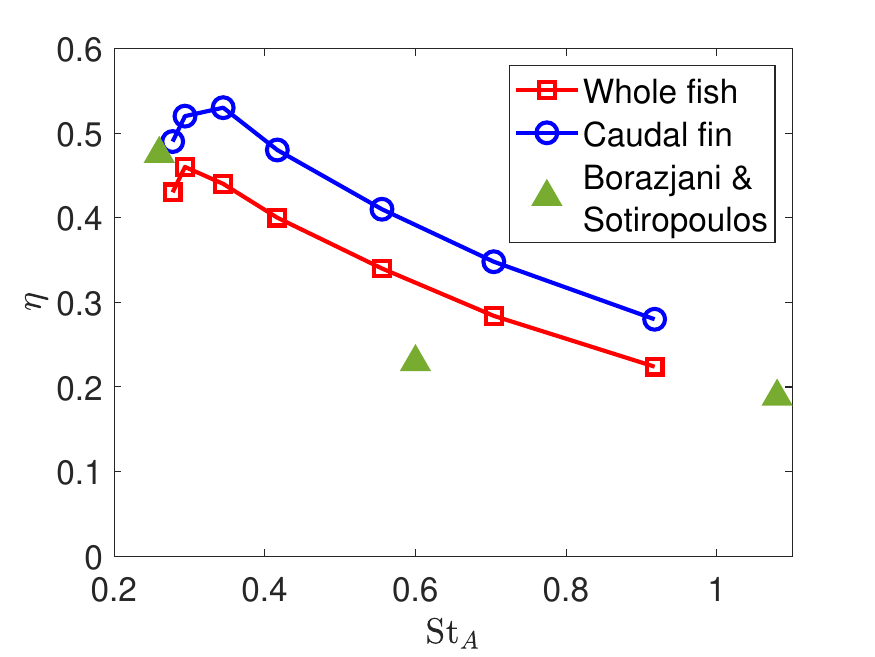}
    \caption{Froude efficiency of the whole fish (square symbols) and the caudal fin (circle) for the present simulation cases. Triangle: Data from the previous computational study\cite{borazjani2008numerical}, which employed swimming kinematics similar to the current study.}
    \label{fig:etafish}
\end{figure}

The Froude efficiency is an often-used metric for evaluating swimming performance, and it has been shown that the Froude efficiency of flapping foils/fins primarily depends on the Strouhal number\cite{triantafyllou2000hydrodynamics,floryan2018efficient,yoshizawa2024waveform}.
The Froude efficiency for the caudal fin is defined by
\begin{equation}
{\eta _{{\rm{fin}}}} ={{{{\bar F}_{T,\textrm{fin}}}U}}/{{{{\bar W}_{{\rm{fin}}}}}}.
\end{equation}
The efficiencies computed from the DNS data are plotted in Fig. \ref{fig:etafish}, and we note that the Froude efficiency increases as Strouhal number decreases, but there is a peak around $St_\textrm{A}\approx 0.3$ and the maximum efficiency is about 0.54.
The Froude efficiencies for the whole fish ($\eta_\textrm{fish}$) are also computed by Eq.(\ref{eq:eta}) and plotted in Fig. \ref{fig:etafish}. For the whole fish, the total thrust is evaluated by the sum of all negative force components in Table \ref{tab:allcases}, and the total power is the sum of all power components, including the viscous power. The efficiency is lower for the whole fish as compared to the caudal fin because of the additional work done by the undulatory motion of the body of the fish. We have also included data from the previous computational study\cite{borazjani2008numerical} which employed similar carangiform swimming kinematics in Fig. \ref{fig:etafish}, and these are in general agreement with our predictions, although the efficiency at the intermediate value of Strouhal number is a bit lower than our DNS. 

As noted earlier, only the pressure load is considered for the thrust and power on the caudal fin, and we can obtain a scaling for the fin efficiency by using the thrust and power scalings derived in the previous sections. 
Since the caudal fin Froude efficiency is also given by the ratio of the non-dimensional thrust to the power coefficients, i.e.  $\eta_\text{fin}= C_T/C_W$, 
from Eqs.(\ref{eq:CT}) and (\ref{eq:CW}), an efficiency factor can be defined as:
\begin{equation}
{\Lambda _\eta } = \frac{{{\Lambda _T}}}{{{\Lambda _W}}} \approx \frac{{\left({A_F/\lambda}\right) \left( {{\textrm{St}_A} - A_F^*{R_\theta }} \right)}}{{{\rm{S}}{{\rm{t}}_A}\left( {{\rm{S}}{{\rm{t}}_A} - A_F/\lambda} \right)}},
\end{equation}
and the Froude efficiency is expected to be linearly proportional to this factor, i.e. 
\begin{equation}
{\eta _{{\rm{fin}}}} = {\beta _\eta }\frac{{ \left({A_F/\lambda}\right) \left( {{\rm{S}}{{\rm{t}}_A} - A_F^*{R_\theta }} \right)}}{{{\rm{S}}{{\rm{t}}_A}\left( {{\rm{S}}{{\rm{t}}_A} - A_F/\lambda} \right)}}
\label{eq:etaSt}
\end{equation}
where $\beta_\eta$ is again a constant related to the fin shape which should ideally be close in value to $\beta_T/\beta_W$ of which value is $1.34$ for the present fish model. Note that $C_W>0$ for $\textrm{St}_A>A_F/\lambda$, thus the Froude efficiency is defined for $\textrm{St}_A>A_F/\lambda$. However, as shown in Eq.(\ref{eq:Stmin}), $C_T>0$ for $\textrm{St}_A>A_F^* {R_\theta}$ and therefore $\eta_\textrm{fin}>0$ for $\textrm{St}_A>A_F^* {R_\theta}$, i.e. $\textrm{St}_A>\textrm{St}_\textrm{min}$.

\begin{figure}
    \centering
    \includegraphics[width=0.5\linewidth]{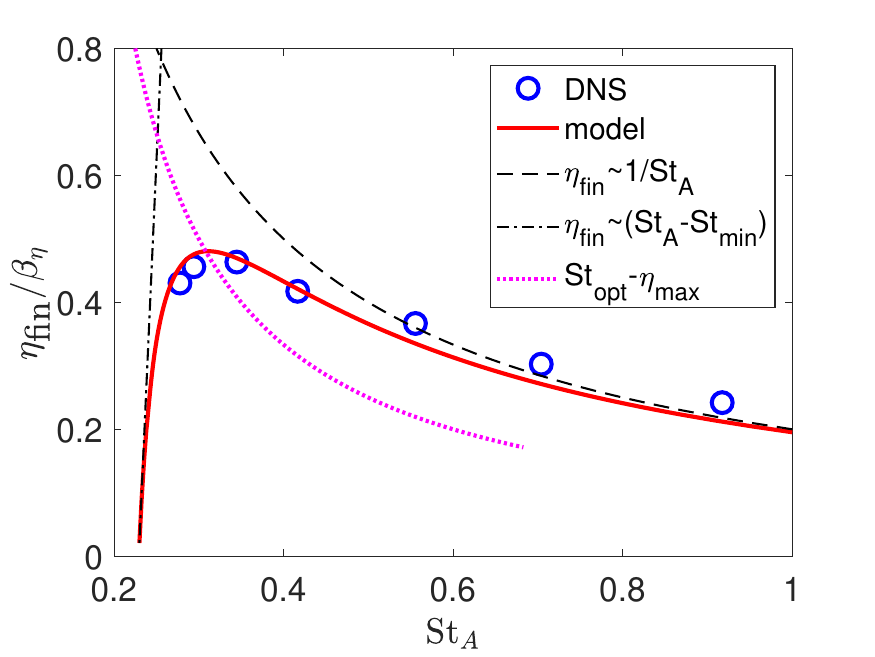}
    \caption{
    Caudal fin Froude efficiency as a function of Strouhal number. 
    Solid line: Eq.(\ref{eq:etaSt}). 
    Symbols: DNS data from Table \ref{tab:free}. 
    Dashed line: Asymptotic line for high Strouhal number, ${A_F}/(\lambda \textrm{St}_A)$. 
    Dash-dot line: Asymptotic line for low Strouhal number, $(\textrm{St}_A - \textrm{St}_{\min})/(\textrm{St}_\textrm{min}{A'^*}^2)$. 
    Dotted line: $\textrm{St}_{\textrm{opt}} - \eta_{\max} $ curve (Eqs. (\ref{eq:Stopt}) and (\ref{eq:etamax})).
    }
    \label{fig:etaSt}
\end{figure}

The efficiency scaling, Eq.(\ref{eq:etaSt}), is plotted in Fig.\ref{fig:etaSt} along with the present DNS data. 
To remove the effect of fin morphology and show the scaling with the kinematic parameters more clearly, we have plotted $\eta_{\textrm{fin}}/\beta_\eta$ instead of the actual Froude efficiency.
The constant $\beta_\eta$ is found to be 1.15 for the best fit, which is slightly different from $\beta_T/\beta_W=1.34$. 
This is because $\beta_T$ and $\beta_W$ were obtained as best-fits to their respective data and they contain individual statistical 
uncertainties. Operations such as division between quantities with uncertainties can increase the error. Thus, we choose to obtain $\beta_\eta$ directly from the least-square fitting to the DNS data on $\eta$ since this should result in a more accurate fit.
We find that the proposed efficiency scaling curve with this fit captures the trend of the DNS results very well.

At high Strouhal numbers, ${\eta _{{\rm{fin}}}}/{\beta _\eta }\sim A_F/(\lambda\textrm{St}_A)$, and the efficiency slowly decreases with increasing $\textrm{St}_A$. 
On the other hand, at low Strouhal numbers, a Taylor series expansion suggests that ${\eta _{{\rm{fin}}}}/{\beta _\eta }\sim ({\rm{S}}{{\rm{t}}_A} - {\rm{S}}{{\rm{t}}_{\min }})/(\textrm{St}_\textrm{min}A{'^*}^2)$, where $\textrm{St}_\textrm{min}=A_F^*{R_\theta}$. The value of $1/(\textrm{St}_\textrm{min}A{'^*}^2)$ is about 30 for the present swimming kinematics, which is typical for carangiform swimmers, and this means that the efficiency increases very rapidly with increasing $\textrm{St}_A$ from $\textrm{St}_\textrm{min}$. Two asymptotic lines are also plotted in Fig. \ref{fig:etaSt}. The combination of these two asymptotic behaviors results in a peak efficiency at a certain optimal value of Strouhal number. 
From Eq. (\ref{eq:etaSt}), the optimal Strouhal number that maximizes the efficiency can be found to be
\begin{equation}
{\rm{S}}{{\rm{t}}_{{\rm{opt}}}} = \frac{A_F}{\lambda}\left( {1 + A{'^*}^2 + A{'^*}\sqrt {1 + A{'^*}^2} } \right) = {\rm{S}}{{\rm{t}}_{\min }}\left( {1 + \frac{{A{'^*}}}{{\sqrt {1 + A{'^*}^2} }}} \right),    
\label{eq:Stopt}
\end{equation}
and the maximum efficiency at $\textrm{St}_A=\textrm{St}_\textrm{opt}$ is given by
\begin{equation}
{\eta _{\max }}/{\beta _\eta } = 1 - 2A{'^*}\left( {\sqrt {1 + A{'^*}^2}  - A{'^*}} \right).
\label{eq:etamax}
\end{equation}
A curve for the optimal Strouhal number ($\textrm{St}_\textrm{opt}$) versus the maximum efficiency ($\eta_{\max}/\beta_\eta$) with varying $A'^*$ is plotted in Fig. \ref{fig:etaSt} as well. For the smaller $A'^*$, $\textrm{St}_\textrm{opt}$ gets smaller and $\eta_{\max}/\beta_\eta$ becomes larger.
For the present fish model with $A'^*=0.38$, $\textrm{St}_\textrm{opt}=0.31$, and this value is in the middle of the well-known optimal Strouhal number range, $0.2<\textrm{St}_A<0.4$ for swimming and flying animals\cite{taylor2003flying}. The maximum Froude efficiency for the present fish model is found from Eq.(\ref{eq:etamax}) to be ${\eta _{\max }}=0.474{\beta _\eta }$ ($\beta_\eta=1.15$). 

It is interesting to note that $\eta_\textrm{max}/\beta_\eta$, which removes the effect of fin morphology from efficiency, is a function of $A'^*$ only, and it decreases monotonically with increasing $A'^*$. This parameter also determines the optimal Strouhal number as shown in Eq.(\ref{eq:Stopt}), and $\textrm{St}_\textrm{opt}/\textrm{St}_{\min}$ is also a function of $A'^*$ only. 
In fact, the rapid drop in efficiency at low Strouhal numbers is also due to this parameter. If $A'^*=0$ (thus $R_\theta=1$), the efficiency would increase monotonically with decreasing $\textrm{St}_A$. As explained earlier, $A'^*$ affects the pitching amplitude and phase. More importantly, it results in a non-zero pitching angle when the heaving velocity is 0 ($\dot{h}=0$) (see Fig.\ref{fig:finAps} as well), which makes the fin produce drag at that instance. Because of this drag, the minimum Strouhal number to produce positive mean thrust ($\textrm{St}_A>(A_F/\lambda)(1+{A'^*}^2)$) is slightly higher than the one for positive mean power input ($\textrm{St}_A>A_F/\lambda$), and this is the reason for the rapid drop of the efficiency at low Strouhal numbers. The drag due to $A'^*$ is similar to the ``offset drag" (the drag at zero angle of attack) for a pitching and heaving hydrofoil\cite{dong2006wake,floryan2018efficient}, and this is the reason why the efficiency curve for the pitching and heaving foil looks quite similar to the one for swimming fish\cite{quinn2015maximizing,triantafyllou2000hydrodynamics}. 

We note that in many previous studies\cite{triantafyllou2000hydrodynamics,floryan2018efficient,yoshizawa2024waveform} the swimming efficiency has been presented as a function of $\textrm{St}_A$ - the Strouhal number based on the tail-beat amplitude. However, our analysis of caudal fin swimmers indicates that the Froude efficiency given by Eq.(\ref{eq:etaSt}) is in fact a function of $\textrm{St}_A/(A_F/\lambda)=\lambda f/U= \textrm{St}_\lambda$ which is different from $\textrm{St}_A$ since it replaces $A_F$ with $\lambda$. Thus, $A_F$ does not directly affect efficiency because efficiency involves a ratio of thrust and power, and both of these quantities are affected similarly by the tail-beat amplitude. The effects of tail-beat amplitude and frequency on the thrust and power scalings are further examined in Appendix \ref{amp_and_freq}.

\begin{figure}
    \centering
    (a)\includegraphics[width=0.4\linewidth]{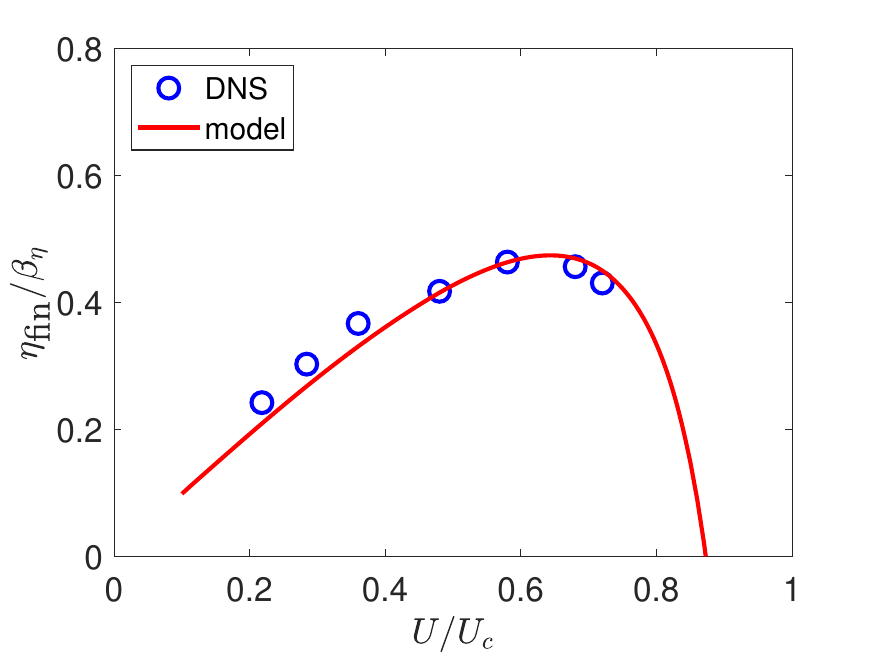}
    (b)\includegraphics[width=0.4\linewidth]{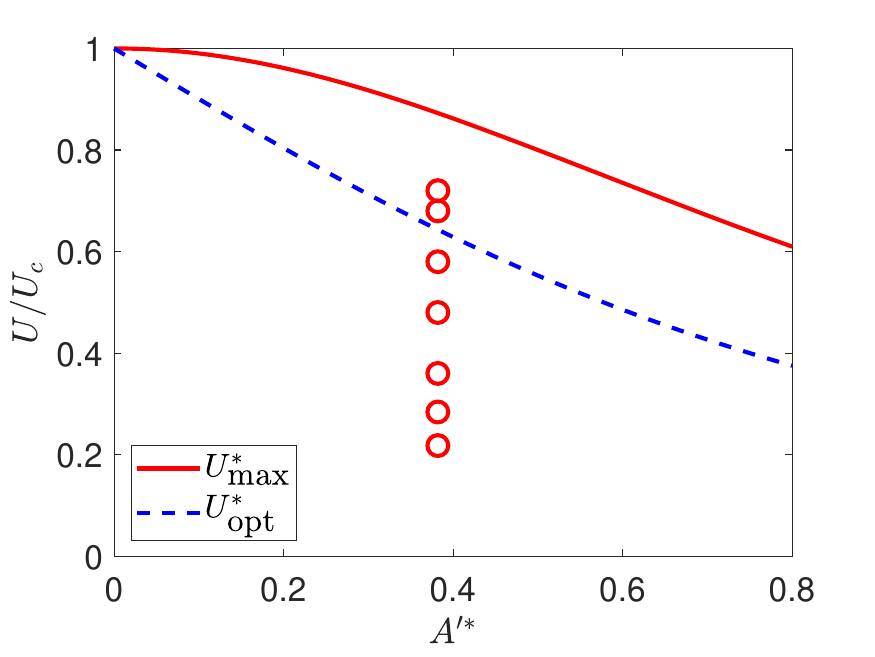}
    \caption{(a): Froude efficiency as a function of $U/U_c=1/\textrm{St}_\lambda$, Lines: Eq.(\ref{eq:eta_UUc}), Symbols: Present DNS data. 
    (b): The maximum and optimal swimming speeds. Symbols: Present DNS data at various Reynolds numbers.}
    \label{fig:UUc}
\end{figure}

It is noted that $\textrm{St}_\lambda=\lambda f/U$ is a Strouhal number based on the undulatory wavelength, and this is equal to the ratio of the wave speed to the swimming speed, $U_c/U$. 
The ratio $U/U_c$ is often called ``slip" or ``slip ratio" and has been considered as an important parameter in many previous studies.
We note that the slip ratio is equal to the inverse of the Strouhal number based on the undulatory wavelength, i.e. $U/U_c=1/\textrm{St}_\lambda$.
The Froude efficiency, therefore, can simply be written as a function of $U/U_c$:
\begin{equation}
    {\eta _{{\rm{fin}}}}/{\beta _{^\eta }} = {\Lambda _\eta } = \frac{U}{{{U_c}}} - A{'^*}^2\frac{{{{(U/{U_c})}^2}}}{{1 - U/{U_c}}},
    \label{eq:eta_UUc}
\end{equation}
and this is verified with the present DNS data in Fig. \ref{fig:UUc}(a). The above expression agrees with Lighthill's slender swimmer theory\cite{lighthill1960note}, in which the Froude efficiency was given as a function of $U/U_c$ and $dA/dx$ was also considered as an important parameter. 
Equation (\ref{eq:eta_UUc}) suggests the optimal swimming speed (or the optimal slip ratio) that maximizes the efficiency, which is given by
\begin{equation}
U_\textrm{opt}^*=\frac{{{U_{{\rm{opt}}}}}}{{{U_c}}} = 1 - \frac{{A{'^*}}}{{\sqrt {1 + A{'^*}^2} }}.
\label{eq:Uopt}
\end{equation}
We draw the reader's attention to the asymmetric nature of the efficiency curve about $U_\textrm{opt}/U_c$ in Fig. \ref{fig:UUc}(a). In particular, this implies that there is a significant power penalty for swimming at speeds higher than the optimal speed, and therefore, swimmers might not be able to sustain such speeds for long, since it would drain the energy reserves rapidly. 
\begin{figure}
    \centering
    \includegraphics[width=0.5\linewidth]{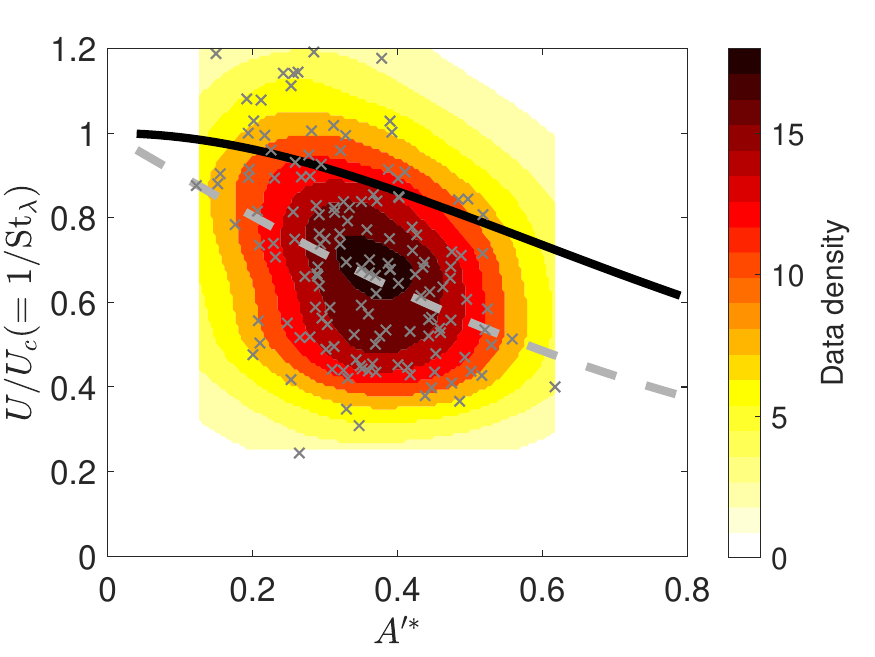}
    \caption{Relation between the slip ratio ($U/U_c=1/\textrm{St}_\lambda$) and the kinematic parameter, $A'^*$. Symbols: Data from Di Santo et al.\cite{di2021convergence} for various BCF swimmers. Contour: Data density map generated by the inverse distance kernel. Solid line: maximum slip ratio (Eq. \ref{eq:Umax}), Dashed line: optimal slip ratio (Eq.\ref{eq:Uopt}).}
    \label{fig:UUData}
\end{figure}
As with the maximum slip ratio (Eqs.(\ref{eq:Umax})), the optimal slip ratio is also determined by the kinematic parameter, $A'^*$ only, and this dependency is plotted in Fig. \ref{fig:UUc}(b).
This further emphasizes the exceptional importance of this parameter, $A'^*$ in determining swimming performance. 
The present DNS data for the mackerel model are also plotted in the figure \ref{fig:UUc}(b). Since the present simulation cases employed the same kinematics, $A'^*=0.38$ is also the same for all cases. The slip ratio, $U/U_c$, however, increases monotonically for the higher Reynolds number. The lowest value, $U/U_c=0.22$ corresponds to $\textrm{Re}_U=110$, and the highest value, 0.72 is for $\textrm{Re}_U=36,000$. This suggests that, although the optimal swimming condition is determined by the kinematics, the actual swimming status depends on the Reynolds number as well. In fact, the kinematics employed in the current study are based on mackerel swimming at an average Reynolds number of about 600,000 \cite{videler1984fast}, so it is not surprising that when these same kinematics are used for a swimmer at Re of O(1000), the swimmer's velocity is significantly below the optimal or maximum value. However, even for Reynolds number slightly greater than 10,000, the swimmer already achieves very close to the optimal speed. 

In Fig.\ref{fig:UUData}, slip ratio data for various BCF swimmers from Di Santo et al.\cite{di2021convergence} are plotted along with the optimal and maximum slip ratios given by Eq.(\ref{eq:Uopt}) and Eq.(\ref{eq:Umax}), respectively. The data points are quite scattered, mainly because of the uncertainties in the data acquisition in live-animal experiments. 
The data density map, however, clearly shows that most of the data points are clustered around the optimal slip ratio (dashed line), and the overall distribution of data density generally follows the trend (increasing $U/U_c$ with decreasing $A'^*$) suggested in the present study. 
We note that some of the data in plot suggests that fish are swimming faster than the maximum slip ratio and even exceeding unity. This is likely because while the scaling laws derived here are for terminal swimming, where thrust balances drag, in experiments, the fish could be accelerating or decelerating. For a decelerating fish, the slip ratio could exceed unity.
This analysis with this large data set also provides strong evidence for the notion that these BCF swimmers swim at a speed that maximizes efficiency. Finally, the data also suggest that $A'^* \sim 0.4$ is the most common value among these swimmers.

\subsection{Cost-of-Transport (COT)}
Swimming performance has been quantified by the Froude efficiency in many previous studies \cite{lighthill1960note,floryan2018efficient,yoshizawa2024waveform}, but the Froude efficiency could be ambiguous because it is often difficult to separate out the drag and thrust\cite{bale2014energy} for a swimmer at terminal speed. The COT is another metric that can be used to estimate the performance of animal locomotion. The COT can simply be defined by $\textrm{COT}=\bar{W}/U$; this represents the energy expended to travel a fixed distance, but this is a dimensional quantity and therefore difficult to interpret. Bale et al.\cite{bale2014energy} proposed a non-dimensional COT by normalizing the power and swimming speed by the tail-beat velocity and wave speed, respectively. This non-dimensional COT is given by
\begin{equation}
    C_\text{COT} =\frac{\bar{W}}{U}\frac{{\lambda f}}{{\frac{1}{2}\rho {{(\pi A_Ff)}^3}{S_f}}}=  \frac{1}{\pi^3} \frac{{{S_x}}}{{{S_f}}}\frac{{{C_W}}}{{ {\textrm{St}_A}^2A_F/\lambda}}.
\end{equation}
By using the power coefficient scaling (Eq.(\ref{eq:CW})), this can be expressed as
\begin{equation}
{C_\text{COT}} = {\frac{{{\beta _W}}}{{2\pi }} } \left( {\frac{{{\rm{S}}{{\rm{t}}_A} - A_F/\lambda}}{{{\rm{S}}{{\rm{t}}_A}A_F/\lambda}}} \right)\left( {\frac{{\sqrt {1 + \sigma {{(\pi {\rm{S}}{{\rm{t}}_A})}^2}} }}{{1 + \sigma {({{\pi A_F^*}})^2}}}} \right).
\label{eq:CE}
\end{equation}
\begin{figure}
    \centering
    (a)\includegraphics[width=0.4\linewidth]{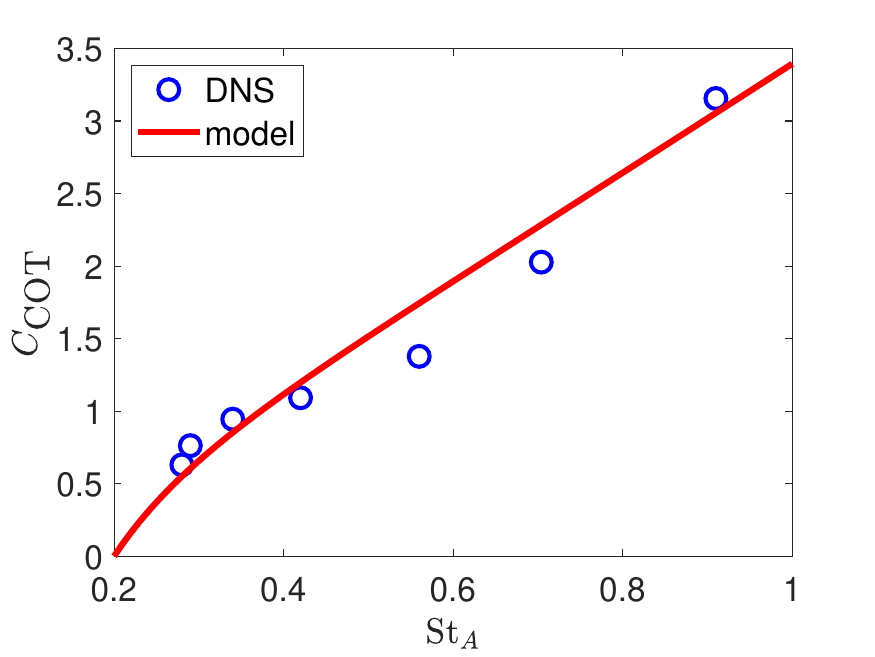}
    (b)\includegraphics[width=0.4\linewidth]{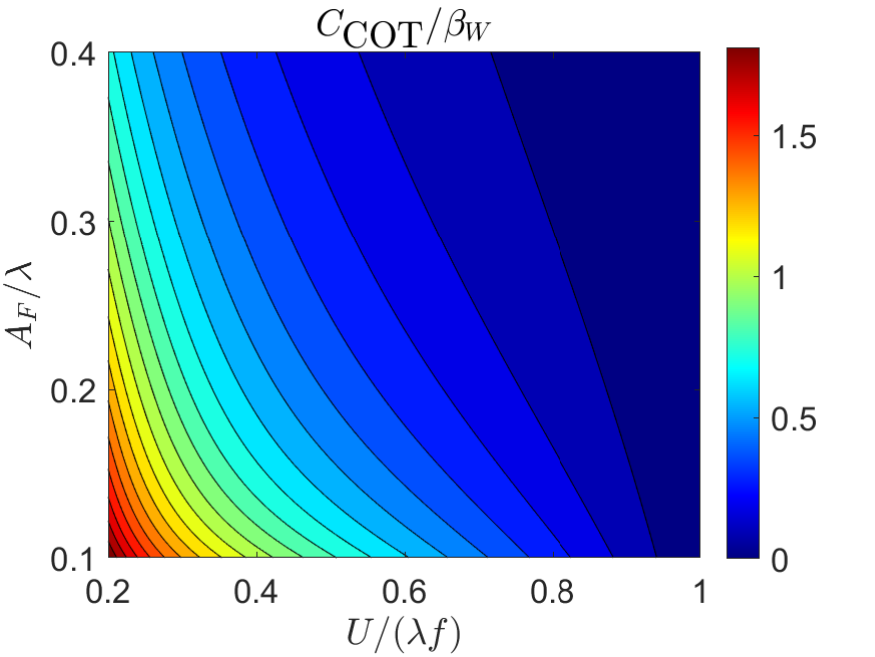}
    \caption{(a) Non-dimensional cost of transport as a function of Strouhal number, Solid line: Eq.(\ref{eq:CE}), Symbols: DNS data. (b) Effect of tail-beat amplitude and frequency on the non-dimensional COT.}
    \label{fig:CE}
\end{figure}
The above formula is plotted in Fig.\ref{fig:CE}(a) along with the DNS data, and the agreement is reasonably good. We note that, unlike the Froude efficiency, the non-dimensional COT decreases monotonically for the lower Strouhal number. 
Based on data on swimming energy expenditure, Videler and Nolet\cite{videler1990costs} reported that dimensionless COT decreased with increasing scale (size) of the swimmer.
As shown in previous studies\cite{gazzola2014scaling,ventejou2025universal}, and will also be shown in Sec.\ref{sec:StRe}, a higher Reynolds number corresponds to a lower Strouhal number for free swimming and it also corresponds to an overall increase in the size of the swimmer. The trend in Fig.\ref{fig:CE}(a) is therefore, inline with the result of Videler and Nolet.
Note, however, that $\textrm{St}_A=(A_F/\lambda)(f\lambda/U)=(A_F/\lambda)/(U/U_c)$, and thus the COT may depend both on the slip ratio and the amplitude separately. In Fig.\ref{fig:CE}(b), we examine the effect of tail-beat amplitude and slip ratio on the non-dimensional COT. It is interesting to note that, while the higher tail-beat amplitude may correspond to the higher Strouhal number, increasing tail-beat amplitude decreases the COT, especially when $U/(\lambda f)$ is low. 
In contrast, at higher values of $U/(\lambda f)$, the dependency on the tail-beat amplitude is much weaker, and this trend is observed for thrust and power as well (see Appendix \ref{amp_and_freq}).

\subsection{Scaling of Drag on the Body}
The scaling laws for thrust, power, and efficiency derived in the previous section suggest optimal values of Strouhal number and swimming speed. However, these have all been derived based on the consideration of the forces on the caudal fin only, but the actual cruising speed is determined by the balance between the thrust produced by the fin and the drag generated by the body of the swimmer. Thus, in order to fully understand the swimming performance, the drag on the fish body has to be considered, and this is the objective of the following sections. In this section, we begin by extracting a scaling for the drag on the fish body from our DNS data.  

\begin{figure}
    \centering
    (a)\includegraphics[trim={2cm 0 5cm 20cm},clip,width=0.45\linewidth]{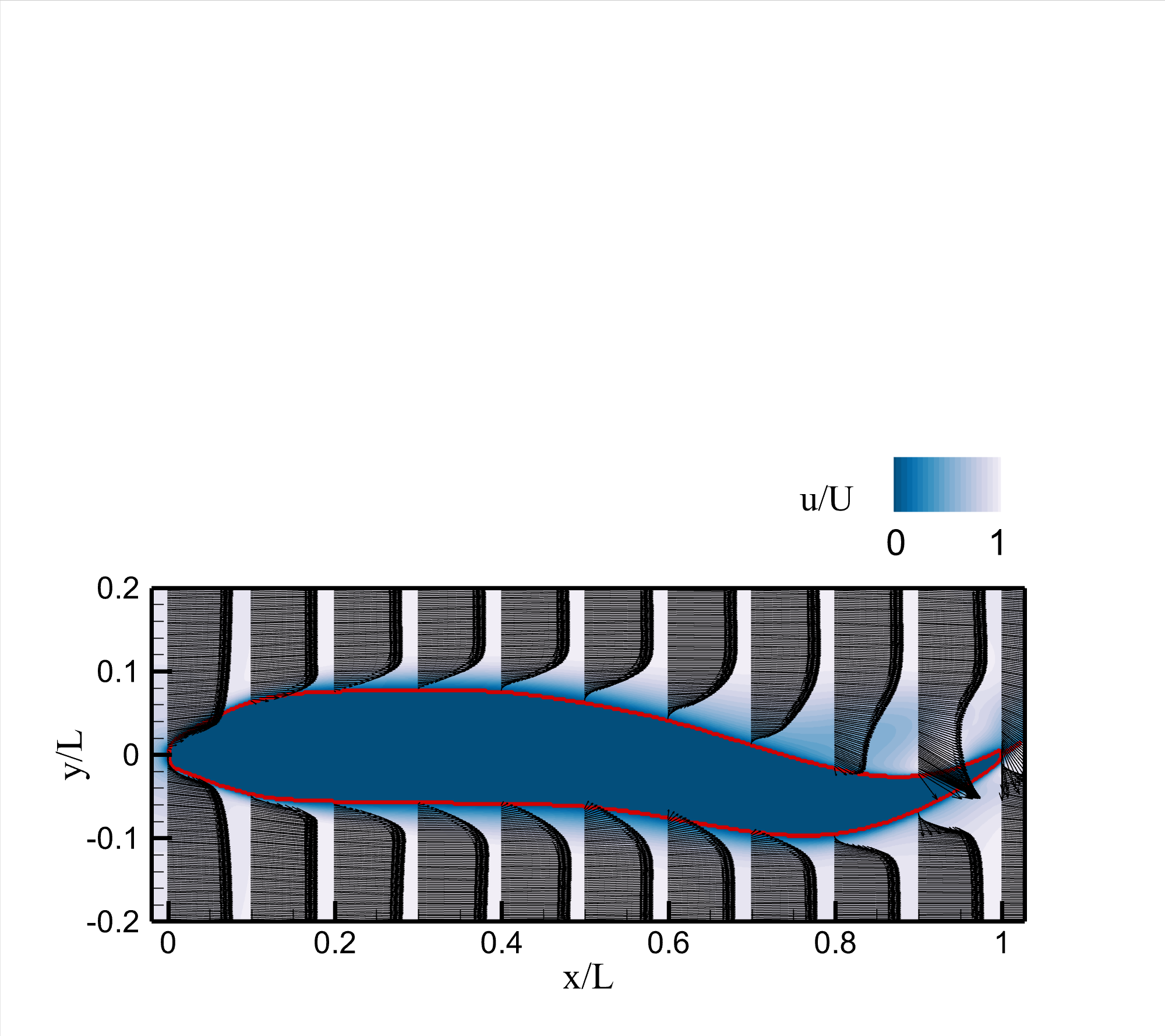}
    (b)\includegraphics[trim={2cm 0 5cm 20cm},clip,width=0.45\linewidth]{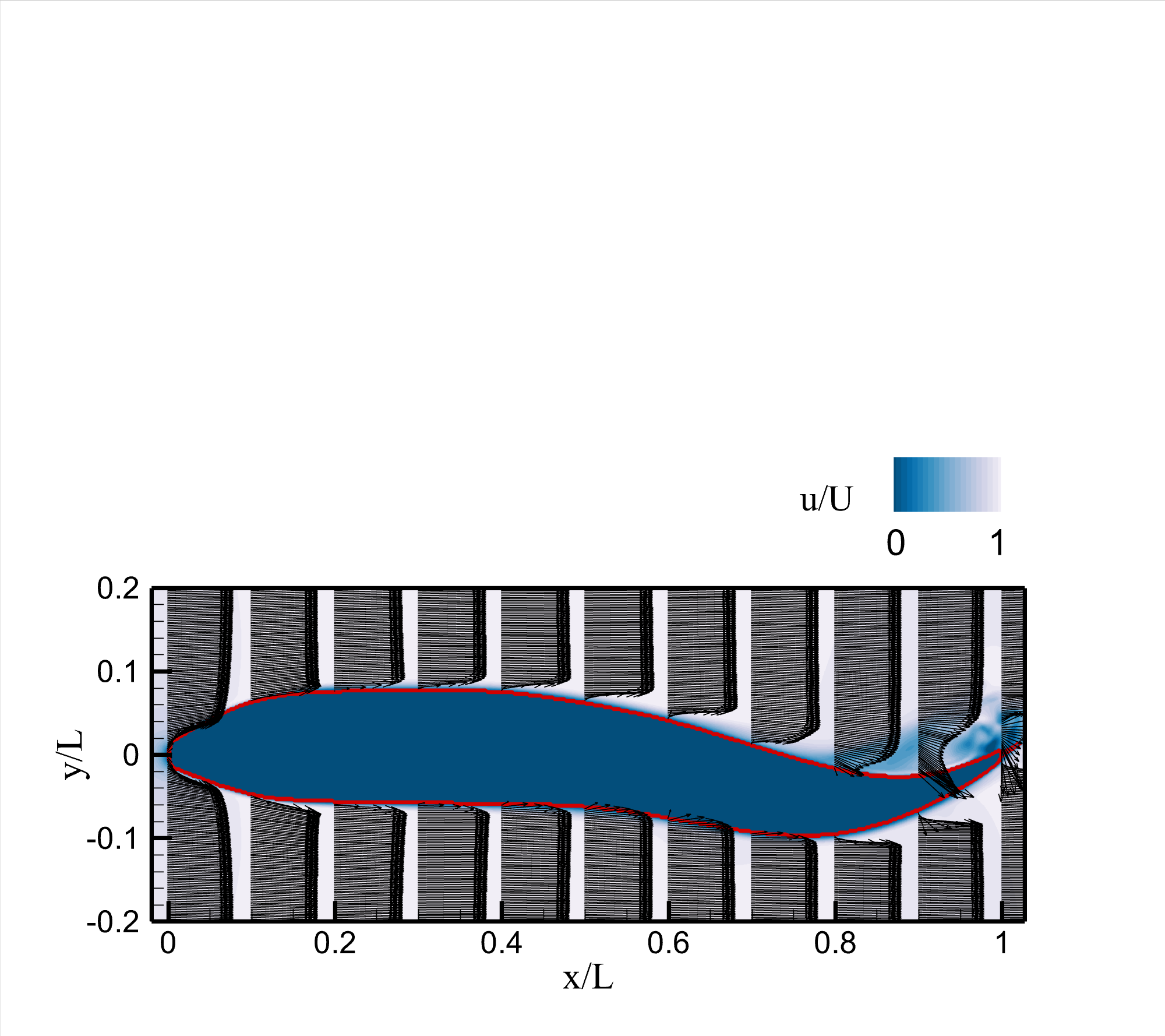}
    \caption{Instantaneous boundary layer velocity profiles around the fish body. (a): $\textrm{Re}_U=2400$. (b): $\textrm{Re}_U=36000$}
    \label{fig:URe}
\end{figure}

The drag coefficient is defined by
\begin{equation}
{C_D} = \frac{{{{\bar F}_D}}}{{\frac{1}{2}\rho {U^2}{S_x}}},    
\end{equation}
where $F_D$ is the drag on the body. 
As evident from Table \ref{tab:allcases}, the drag on the fish is mainly due to the shear force on the fish body, i.e., skin friction. This force scales as $F_s \propto  \mu U S_b/\delta$ where $S_b$ is the surface area of the body and $\delta$ is the boundary layer thickness. The boundary layer thickness scales as $L/\text{Re}^\kappa_U$, where $\kappa$ mostly depends on the state of the boundary layer. This gives $F_s = C_f  \mu U S_b\textrm{Re}^\kappa_U/L$ where $C_f$ is the coefficient of skin friction drag force. The drag coefficient can then be written as 
\begin{equation}
{C_D} = \frac{2 C_f \left( S_b/S_{x} \right)  }{{{{{\mathop{\rm Re}\nolimits} }_U}^{1-\kappa}}},
\label{eq:CD}
\end{equation}
For the present fish model $S_b/S_{x} = 16.5$ ($S_b=0.38L^2$), and to complete the scaling law for the drag on the fish, we need to estimate $C_f$, which is mostly related to the shape of the body, and $\kappa$. Based on the boundary layer theory, the drag coefficient due to the skin friction scales with $1/(\textrm{Re}_U)^{1/2}$ ($\kappa=1/2$) for a laminar boundary layer over a stationary surface. The surface of the fish body, however, moves in a direction perpendicular to the outer flow ($U$), and this affects the boundary layer thickness on either side of the body of the fish as shown in Fig.\ref{fig:URe} for example. 

\begin{figure}
    \centering
    \includegraphics[width=0.4\linewidth]{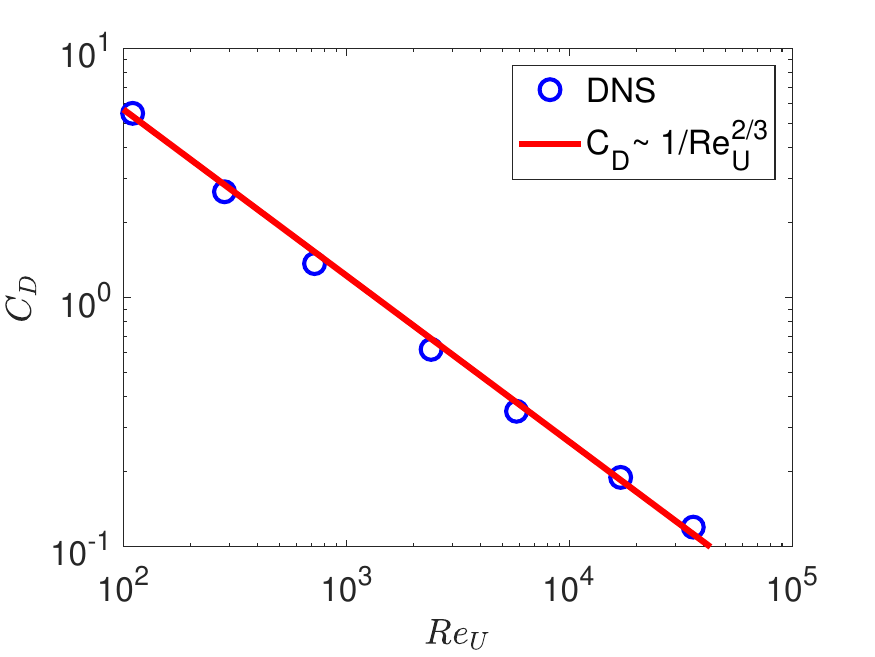}
    \caption{Scaling of the coefficient of drag on the body with Reynolds number.}
    \label{fig:CD}
\end{figure}

Ehrenstein et al.\cite{ehrenstein2014skin} examined the scaling of the drag coefficient for a oscillating flat plate as a canonical model of a flapping propulsor and proposed the scaling of the drag coefficient as a function of the Reynolds number as well as the mean wall normal velocity magnitude. Since the normal velocity is related to the Strouhal number, in general, it may be possible to derive the drag scaling as a function of both Reynolds and Strouhal numbers\cite{das2022contrasting}. However, during terminal swimming, Reynolds and Strouhal numbers are not independent as shown here as well as in the previous studies\cite{gazzola2014scaling,ventejou2025universal,das2022contrasting}, and thus the drag for terminal swimming can be scaled with the Reynolds number only. This could be the reason why, in many previous studies, the body drag was scaled by the Reynolds number only\cite{ventejou2025universal,gazzola2014scaling,li2021fishes,eloy2012optimal}, and we have followed the same approach here. In this regards, the boundary layer thinning effect due to the wall normal velocity should be embedded in the exponent, $\kappa$.
A previous computational study by Li et al.\cite{li2021fishes} showed that $\kappa$ is between 0 and $1/2$ for an undulating fish.
We therefore employ regression on our DNS data to determine the two parameters, and the best fit (with an $R^2=0.99$; data fit shown in Fig.\ref{fig:CD}) is found for $\kappa=1/3$ and $C_f=3.7$. We employ these values in the rest of the scaling development.
We note that, however, the drag coefficient scaling given by Eq.(\ref{eq:CD}) is strictly valid for the terminal swimming because of the aforementioned reason.

\subsection{Scaling between Strouhal and Reynolds numbers}
\label{sec:StRe}
For free swimming at a constant, terminal speed, the thrust balances the drag. The present scaling laws for the thrust and drag show that the thrust primarily depends on the Strouhal number, while the drag depends on the Reynolds number. Thus, the balance between the thrust and drag provides the scaling between the two important non-dimensional numbers: the Strouhal and Reynolds numbers.

\begin{figure}
    \centering
    (a)\includegraphics[width=0.4\linewidth]{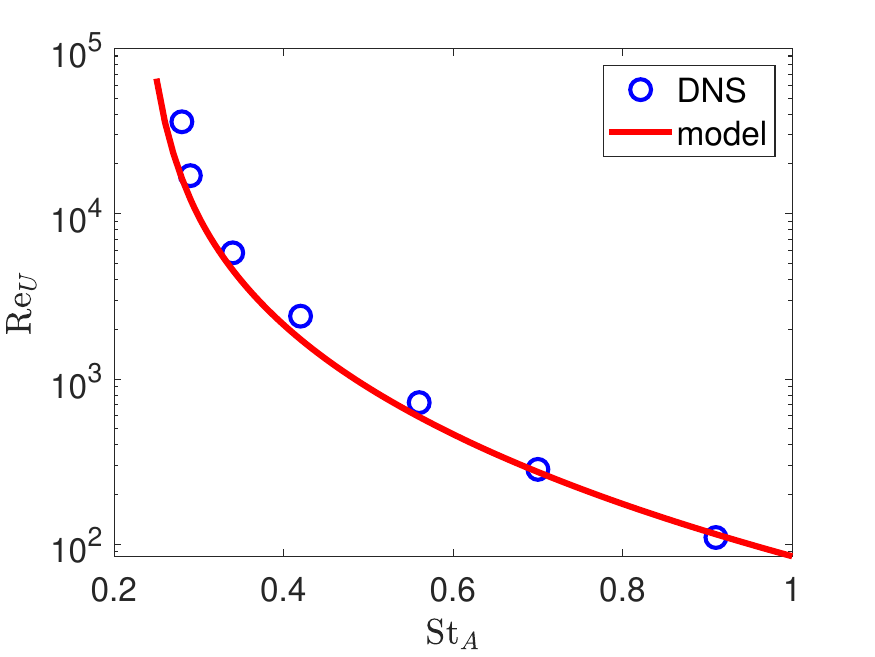}
    (b)\includegraphics[width=0.4\linewidth]{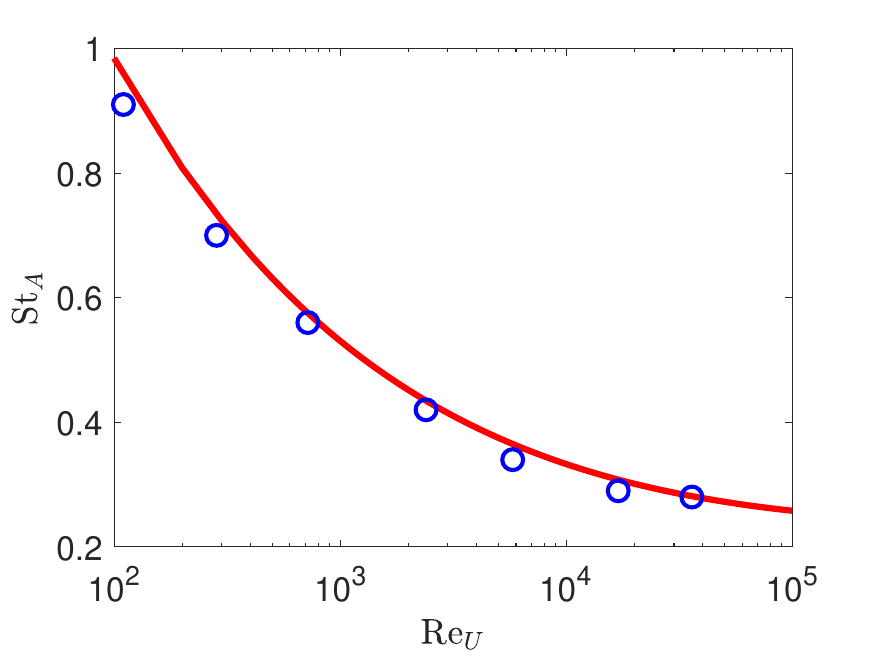}
    \caption{Relation between the Reynolds and Strouhal numbers for free swimming fish. (a) $\textrm{Re}_U$ as a function of $\textrm{St}_A$, Solid line: Eq.(\ref{eq:StRe}), Symbols: DNS data from Table \ref{tab:free}. (b) $\textrm{St}_A$ as a function of $\textrm{Re}_U$, Solid line: Eq.(\ref{eq:ReSt2}), Symbols: DNS data from Table \ref{tab:free}.}
    \label{fig:StRe}
\end{figure}

From Eqs.(\ref{eq:CD}) and (\ref{eq:CT}), for $C_D=C_T$, we get:
\begin{equation}
\frac{2}{{{\pi ^2}}}\frac{K_\text{morph}}{{{{\mathop{\rm Re}\nolimits} }_U^{2/3}}}\, =({A_F}/\lambda)\left({\textrm{St}_A - \textrm{St}_{\min}} \right)\frac{{\sqrt {1 + \sigma {{(\pi\textrm{St}_A)}^2}} }}{{1 + \sigma {{({\pi A_F^*})}^2}}},    
\end{equation}
where $\textrm{St}_{\min}=A_F^*R_\theta$ and $K_\text{morph} = 2({C_f}/{\beta_T})({S_b}/{S_f})$ is a morphological parameter related to the shape and relative sizes of the body and caudal fin. By definition, $K_\text{morph}$ is proportional to the ratio of the body surface area to the fin area ($S_b/S_f$), $C_f$ depends on the body shape and surface condition (for instance, surface roughness), and $\beta_T$ is associated with the fin shape. For the present fish model, $K_\text{morph}$ is equal to  30.7. The above relation gives an expression for $\textrm{Re}_U$ as a function of $\textrm{St}_A$:
\begin{equation}
{{\mathop{\rm Re}\nolimits} _U} = \frac{{2\sqrt 2 {K_\text{morph}^{3/2}}}}{{{\pi ^3}}}{\left[ {\frac{{1 + \sigma {{({\pi A_F^*})}^2}}}{({A_F/\lambda)\left( {\textrm{St}_A-\textrm{St}_{\min}} \right)\sqrt {1 + \sigma {{(\pi {\rm{S}}{{\rm{t}}_A})}^2}} }}} \right]^{3/2}}.
\label{eq:StRe}    
\end{equation}
Equation (\ref{eq:StRe}) is plotted in Fig.\ref{fig:StRe}(a) along with the DNS data in Table \ref{tab:free}, and it shows that the data follows the proposed relation very well.

The above expression can be written for $\textrm{St}_A$ as a function of $\textrm{Re}_U$, but the exact expression may look complicated. Alternatively, for the following conditions: ($\textrm{St}_A>1/(\sqrt{\sigma}\pi)$), $\sqrt {{\rm{1 + }}\sigma {{{\rm{(}}\pi {\rm{S}}{{\rm{t}}_A})}^2}}  \approx \sqrt \sigma  \pi {\rm{S}}{{\rm{t}}_A}$, the following approximate relationship can be derived for 
\begin{equation}
{\rm{S}}{{\rm{t}}_A} = \frac{\textrm{St}_{\min}}{2} + \sqrt {{{\left( {\frac{\textrm{St}_{\min}}{2}} \right)}^2} + \left[ {\frac{{1 + \sigma {{({\pi A_F^*})}^2}}}{{\sqrt \sigma  A_F/\lambda}}} \right]\frac{{2K_\text{morph}}}{{{\pi ^3}{{\mathop{\rm Re}\nolimits} }_U^{2/3}}}}.
\label{eq:ReSt2}    
\end{equation}
This expression shows that, if ${{\mathop{\rm Re}\nolimits} _U} \to \infty$, $\textrm{St}_A$ approaches the minimum value, $\textrm{St}_\textrm{min}=A_F^*R_\theta$. 
Equation (\ref{eq:ReSt2}) is plotted along with the DNS data in Fig.\ref{fig:StRe}(b).
As evident from the figure, the above relation is actually quite accurate over a wide range of Reynolds numbers, including the low Reynolds number regime ($\textrm{Re}_U<1000$) and this expression can be rewritten in the following simpler form:
\begin{equation}
\frac{{{\textrm{St}_A}}}{{{\textrm{St}_{\min}}}} = \frac{1}{2} + \frac{1}{2}\sqrt {1 + {{\left( {\frac{{{{{\mathop{\rm Re}\nolimits} }_\text{cr}}}}{{{{{\mathop{\rm Re}\nolimits} }_U}}}} \right)}^{2/3}}},  
\label{eq:StAStmin}
\end{equation}
where $\textrm{Re}_\text{cr}$ is a critical Reynolds number defined by
\begin{equation}
{{\mathop{\rm Re}\nolimits} _\text{cr}}^{2/3} = \left( {\frac{{8K_\text{morph}}}{{\sqrt \sigma  \pi {\rm{S}}{{\rm{t}}_{\min }}}}} \right)\left( {\sigma  + \frac{1}{{{\pi ^2}{\rm{S}}{{\rm{t}}_{\min }A_F/\lambda }}}} \right).    
\label{eq:Recr}
\end{equation}

\begin{figure}
    \centering
    \includegraphics[width=0.5\linewidth]{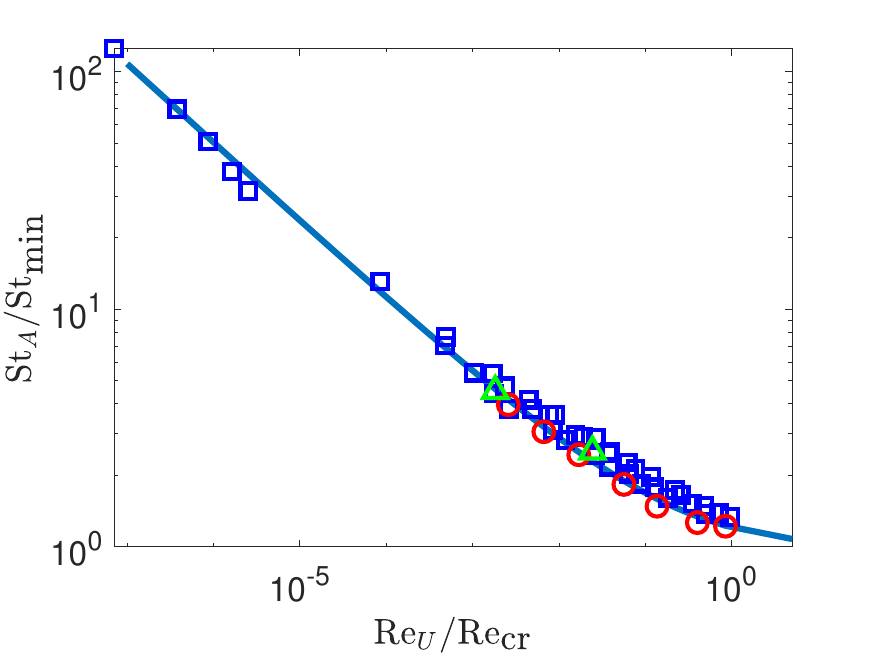}
    \caption{Relation between the Strouhal and Reynolds numbers. Solid line: Eq.(\ref{eq:StAStmin}). Symbols: Results from fish swimming simulations. Circle: Present DNS data. Square: Data from Li et al.\cite{li2021fishes}, $A'^*=0.35$, $\textrm{Re}_U=7.17\sim6070$. Triangle: Data from Borazjani and Sotiropoulos\cite{borazjani2008numerical}, $A'^*=0.36$, $\textrm{Re}_U=300, 4000$.}
    \label{fig:StReDNSothers}
\end{figure}
The critical Reynolds number, $\textrm{Re}_\text{cr}$, depends on kinematic as well as morphological parameters
and for the present fish model, $\textrm{Re}_\text{cr}$ is about 42,600. The significance of this parameter can be ascertained from Eq.(\ref{eq:StAStmin}) which indicates that if $\textrm{Re}_U$ is sufficiently higher than $\textrm{Re}_\text{cr}$, $\text{St}_A \sim \text{St}_\text{min}$ thereby resulting in a linear relationship between the swimming speed and the tail-beat frequency.
The average value of $\textrm{St}_{\min}$ (Eq.(\ref{eq:Stmin}) computed for the BCF swimmer data in Di Santo et al.\cite{di2021convergence} is found to be about 0.24, which coincides with the Strouhal number given by the Bainbridge equation for $U \gg 1 [L/s]$.
Gazzola et al.\cite{gazzola2014scaling} also suggested that, for turbulent flows at high Reynolds numbers, the Strouhal number would be a constant with little or no influence of the Reynolds number. This is in line with the present scaling as well.
On the other hand, if $\textrm{Re}_U$ is much smaller than $\textrm{Re}_\text{cr}$, the above relation gives $\textrm{St}_A \sim \textrm{Re}_U^{-1/3}$. 
This is also in line with the scaling law proposed by Gazzola et al.\cite{gazzola2014scaling} and Vent'ejou et al.\cite{ventejou2025universal}. They obtained the scaling $\textrm{St}\sim\textrm{Re}^{-1/4}$ in the laminar flow regime by assuming $C_D\sim\textrm{Re}^{-1/2}$, but if we apply the current drag scaling, $C_D\sim\textrm{Re}^{-2/3}$, to their theory, it yields $\textrm{St}\sim\textrm{Re}^{-1/3}$. 
This scaling is also quite similar to $\textrm{St}\sim\textrm{Re}^{-0.375}$ proposed by Das et al.\cite{das2022contrasting} for a flapping foil at $\textrm{Re}\le 1000$.
Thus, $\textrm{Re}_\text{cr}$ is the Reynolds number below which the swimming speed is dependent on viscous effects, but above that, the swimming becomes increasingly independent of these effects.
Equation (\ref{eq:StAStmin}) is plotted in Fig.\ref{fig:StReDNSothers} along with the present DNS data as well as the data from other computational studies\cite{li2021fishes,borazjani2008numerical}. Here, $\textrm{St}_\textrm{min}$ and $\textrm{Re}_\text{cr}$ are computed by using the kinematics data used in each study, while the morphological parameter, $K_\text{morph}$, is obtained by regression. $K_\text{morph}$ is found to be 40 for the fish model used in the study of Li et al.\cite{li2021fishes}, and $K_\text{morph}=84$ for the carangiform model used by Borazjani and Sotiropoulos\cite{borazjani2008numerical}. The figure shows that the scaling law given by Eq.(\ref{eq:StAStmin}) provides a very good prediction of the relationship between the Strouhal and Reynolds numbers in carangiform swimming for a wide range of Reynolds numbers ($\textrm{Re}_U = 7.17 \sim 36000$, for the data set shown in the figure). 

\subsection{Significance of the Morphological Parameter $K_\text{morph}$}
The proposed relation between the Strouhal and Reynolds numbers depends not only on the kinematic parameters: $A_F/\lambda$, $A_F^*$, and $A'^*$, but also the morphological parameter $K_\text{morph} = 2({C_f}/{\beta_T})({S_b}/{S_f})$. In fact, the morphological parameter, $K_\text{morph}$ plays an important role in the balance between the drag and thrust, and therefore, the above relationship could be species- or even individual-specific. To understand the significance of this parameter, we have plotted Strouhal number against Reynolds number for three values of $K_\text{morph}$ in Fig. \ref{fig:StoptReDD}. 
In this figure, the Strouhal number is normalized by the optimal Strouhal number ($\textrm{St}_\textrm{opt}$), $\lambda/L$ is set to 1, and the suggested average values of $A_F/L=0.2$ and $A'^*=0.34$ for the BCF swimmers\cite{di2021convergence} are used to plot the curves. 
The experimental data by Di Santo et al.\cite{di2021convergence} collected for O(100) BCF swimmers are also plotted in the figure as a data density map. The optimal Strouhal number for each specimen is calculated via Eq. (\ref{eq:etaSt}) by using the kinematic data reported in the paper.  

Several interesting observations can be made from this plot:
\begin{enumerate}
\item The Reynolds number where the fish achieves optimal swimming (i.e. $\textrm{St}_A/\textrm{St}_\textrm{opt}=1$) is a strong function of $K_\text{morph}$. In particular, optimal swimming at high Reynolds number is associated with a larger value of $K_\text{morph}$, and vice-versa. 
\begin{figure}
    \centering
    \includegraphics[width=0.5\linewidth]{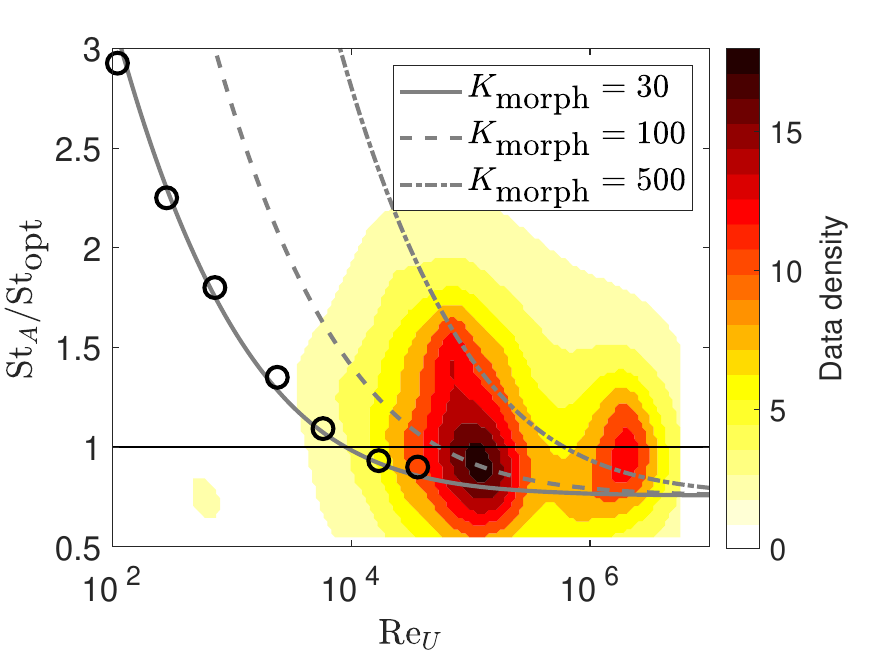}
    \caption{Effect of the morphological parameter, $K_\textrm{morph}$ on the relation between the Strouhal and Reynolds numbers (Eq.(\ref{eq:StRe})). The Strouhal number is normalized by the optimal Strouhal number (Eq.(\ref{eq:Stopt})) that maximizes the Froude efficiency. Contours: Data density map for the experimental data by Di Santo et al.\cite{di2021convergence}. Circular Symbols: Present DNS data ($K_\text{morph}=30.7$).}
    \label{fig:StoptReDD}
\end{figure}
\item Reynolds number can be considered a surrogate for the size and/or velocity. Thus, the curves in the figure implies that larger and/or faster fish in nature would tend to have a larger $K_\text{morph}$. The density map of data from Di Santo et al.\cite{di2021convergence} tends to confirm this notion. It shows high data density around the optimal Strouhal number, but the density map naturally clusters into two groups - one at a Reynolds number about $10^5$ and the other at about $2\times10^6$. The $\textrm{St}_A-\textrm{Re}_U$ curves intersect with the first group at $\textrm{Re}_U\sim10^5$ are for $K_\text{morph}=100$, and the other one at $\textrm{Re}_U\sim 2\times 10^6$ for $K_\text{morph}=500$. We note that the first group includes typical carangiform swimmers like mackerel and bluefish with body lengths of about 25 cm, and the second group includes large fish like American shad and yellowfin tuna with body lengths ranging from 0.5-1 m.
\item For our DNS data for the mackerel model, we note that while at high Reynolds numbers ($>$10,000), the swimming is close to the optimal condition, at lower Reynolds number, these same kinematics move the swimmer into a non-optimal range. This implies that smaller individuals of a given species would have to change their kinematics and/or reduce their $K_\textrm{morph}$ value to keep swimming in the optimal range. Indeed there is evidence that the caudal fin size in trout, which is a carangiform swimmer, grows slower than linear in proportion to the body length \cite{ellis2009further}, which would imply that the $K_\textrm{morph}$ increases with age (and therefore size) for this fish. In this regard, it would be interesting to examine the ontogenetic variation of $K_\textrm{morph}$ for other caudal fin swimmers.
\item The above observation has important implications for the design of BUVs, suggesting that simply scaling vehicle size while maintaining identical shapes and kinematics may result in suboptimal performance. The current analysis extends this idea further by offering a rationale for how body and fin shape should be coordinated with kinematics to enable optimal swimming across different scales.
\end{enumerate}

We note that $K_\text{morph}$ bears some similarity to the Lighthill number ($\textrm{Li}=S_b C_d/h^2$)\cite{eloy2012optimal} although a minor difference is that while $K_\textrm{morph}$ contains $\beta_T$ which explicitly accounts for fin shape, $\textrm{Li}$ contains no such dependence since it is based on slender body theory and does not account for caudal fin propulsion. Notwithstanding this, these parameters that emerge from two very different analyses confirm the important role of body morphology on the performance of BCF swimmers.

\section{Summary}
\subsection{Wake Topology}
Our simulations of terminal swimming show that the wake topology is closely related to the Strouhal number $\text{St}_A= fA_F/U$ which is a measure of the normalized lateral velocity imparted by the fin. Of particular importance is the wake spreading angle, which has been the subject of several studies and is shown to be proportional to the Strouhal number.  Since a high Strouhal number in terminal swimming corresponds to a low Reynolds number, fish at low Reynolds number generate wakes with a larger spreading angle (and vice-versa). This is the reason why the wake structures observed in previous computational studies performed for Reynolds numbers of $O(10^3)$\cite{borazjani2008numerical,li2019energetics,seo2022improved} showed a distinct double row of obliquely directed vortex streets. In contrast, at the generally higher Reynolds numbers in experimental studies with swimming fish \cite{blickhan1992generation,nauen2002hydrodynamics}, the free swimming Strouhal number and therefore the wake spreading angle decreases to low single digits where the wake (especially the near-wake) could appear as a single vortex street of linked vortex structures. However, the formation of a strictly single row vortex wake requires that the lateral velocity imparted by the caudal fin be much smaller than the swimming velocity, which is not realizable in steady terminal swimming. 

\subsection{Scaling Laws for Caudal Fin Swimmers}
\begin{table}[h]
    \centering
    \begin{tabular}{|c|c|c|c|}
    \hline
           Performance Parameter  & Morphology & Kinematics & Velocity \\ \hline
    $C_T$   & $\beta_T, S_f/S_x$  &  $A_F^*, A^{'*}, \text{St}_A$ & $\text{St}_A$ \\ \hline
    $C_W$   & $\beta_W, S_f/S_x$  &  $A_F^*, A^{'*}, \text{St}_A$ & $\text{St}_A$   \\ \hline
    $C_\text{COT}$   & $\beta_W$  &  $A_F^*, A^{'*}, \text{St}_A$ & $\text{St}_A$   \\ \hline
    $\eta_\textrm{fin}$   & $\beta_\eta$  &  $A_F^*, A^{'*}, \text{St}_A$ & $\text{St}_A$        \\ \hline
    $\eta_\text{max}$   & $\beta_\eta$  &  $A^{'*}$ & -   \\ \hline
    $\textrm{St}_\textrm{opt}$  &- & $A_F^*, A^{'*}$ & -   \\ \hline
    $\textrm{St}_{\min}$ & -& $A_F^*, A^{'*}$ & - \\ \hline
    $U_\textrm{opt}/U_c$ &- & $ A^{'*}$ & -  \\ \hline
    $U_{\max}/U_c $  & - & $ A^{'*}$ & -    \\ \hline \hline
    $C_D$  &  $C_f, \kappa, S_b/S_x$   &   $C_f, \kappa$ &  -     \\ \hline
    $K_\text{morph}$  &  $ C_f , \beta_T, S_b/S_f$  & $ C_f$  & -      \\ \hline
        $\text{Re}_\text{cr}$  &  $ K_\text{morph}$  &  $ K_\text{morph}, \text{St}_\text{min}, A_F^*$ & -      \\ \hline
    $\text{Re}_U$  &  $ \text{Re}_\text{cr}$    & $\text{St}_\text{min}, \text{Re}_\text{cr}, \text{St}_A$ &  $\text{St}_A$  \\ \hline
    \end{tabular}
    \caption{Dependencies of the swimming performance metrics on the key non-dimensional parameters. Definitions of the parameters and metrics are provided in the nomenclature section. Parameters are categorized into those associated with morphology, kinematics, and velocity. Parameters that span multiple categories are cross-listed in multiple columns. More details of these and other parameters can be found in Appendix \ref{keyparam}.}
    \label{tab:placeholder_label2}
\end{table}
Based on the leading-edge vortex (LEV) based model and the results from our simulation, we have derived the scaling laws for thrust, power, COT, and Froude efficiency. The scaling laws are given as functions of morphometric parameters as well as parameters associated with the midline kinematics of the swimmer. The present scaling laws therefore provide a path to predicting the forces and energetics of caudal fin swimmers from experimental measurements of midline kinematics and morphology. This would be very helpful for the experimental studies with real fish, because force and power are difficult to measure for live fish in the experiments. 
Among other dependencies we have found in the present study, it is interesting to note that the swimming efficiency depends mainly on the slip ratio, $U/U_c$. This is not a new finding, because the slip ratio has been considered as an important parameter since Lighthill's classical work\cite{lighthill1960note}, which employed slender body theory.

The scaling laws presented here should apply equally well to biorobotic autonomous underwater vehicles that are designed to mimic carangiform-like swimming kinematics.
In the present analysis, the effects of the center-of-mass (CoM) oscillation for a self-propelled body are not included. This is because such oscillations are expected to be very small for natural swimmers, as observed in the experimental study\cite{videler1984fast} and the effects on swimming performance would be negligible \cite{lighthill1960note}.  
Previous studies \cite{das2022contrasting,smits2019role} also show that the CoM oscillation does not change the scalings, although they can result in magnitude offsets.
Thus, we believe that, although the CoM oscillations may affect some of the coefficients such as $\beta_T$ and $\beta_W$, the scaling laws proposed here can still be generally applied to BCF-type swimming.

\subsection{The Importance of $A'^*$}
The parameter $A'^*$  appears in every single scaling law for the propulsive performance of the caudal fin. This parameter depends on the slope of the amplitude envelope function at the tail as well as the undulatory wavelength, and can be viewed primarily as a measure of the phase mismatch that is created between the heave velocity and the pitch of the caudal fin due to the BCF kinematics. 
Indeed, three metrics that define the operational bounds of the swimmer: $\eta_\text{max}/\beta_\eta$, $U_\textrm{opt}/U_c$ and $U_\textrm{max}/U_c$, are exclusively determined by $A'^*$. Experimental data from Di Santo et al. \cite{di2021convergence} that includes 151 individuals ranging from anguilliform, to sub-carangiform, carangiform and thunniform swimmers, shows clear convergence onto the $U_\textrm{opt}/U_c$ predicted from our scaling law that depends on $A'^*$, and therefore provides strong verification of the scaling analysis. 
\begin{figure}
    \centering
(a)\includegraphics[width=0.53\linewidth]{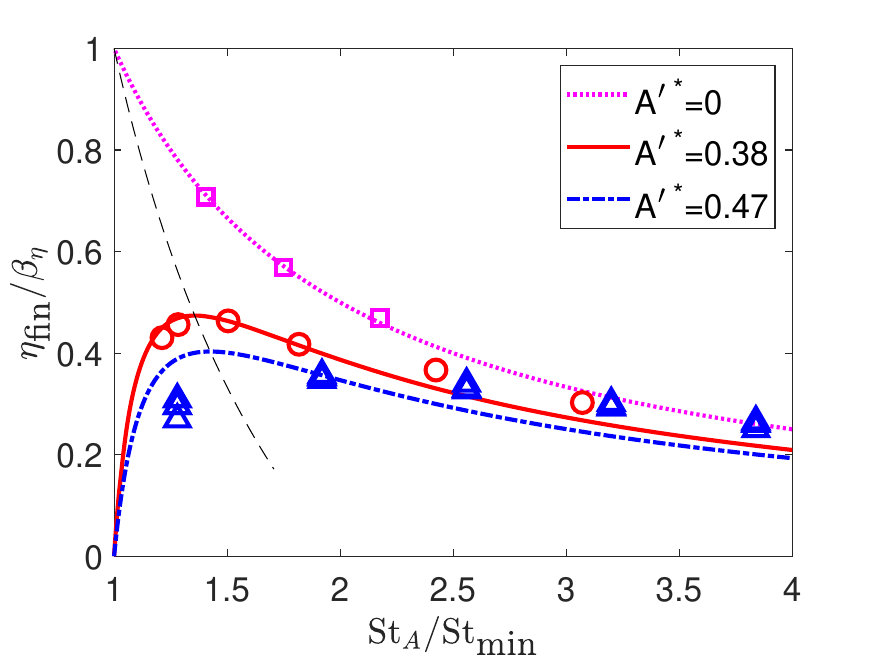}
(b)\includegraphics[width=0.35\linewidth]{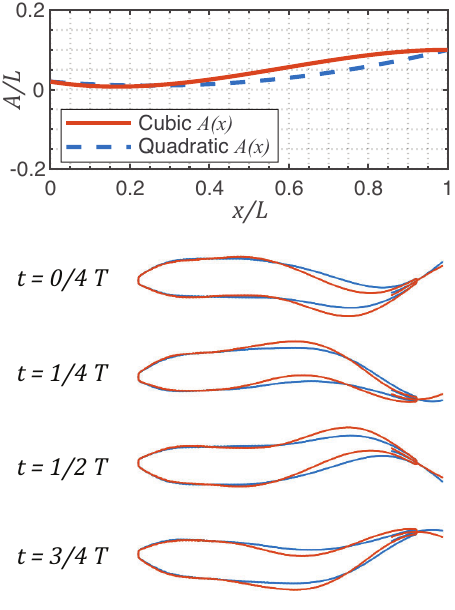}
    \caption{(a) Effect of $A'^*$ on the Froude efficiency of the fin. Dashed line: $\textrm{St}_\textrm{opt}-\eta_\textrm{max}$ curve. Circle: Present DNS data for the mackerel model ($A'^*=0.38$). Square: DNSs for the fish model with a cubic polynomial amplitude envelope ($A'^*=0$), see Appendix \ref{cubicfish}. 
    Triangle: Data from Huang et al.\cite{huang2023hydrodynamics} for thunniform swimmer ($A'^*=0.47$). 
    (b) Top: Amplitude envelope functions $A(x)$ along the fish body length: the original quadratic form (dashed line) and the modified cubic polynomial (solid line), which satisfies the condition $A'^*=0$. 
    Bottom: Time snapshots of the fish body undulation over one tailbeat cycle $T$ at $t = 0$, $t = 1/4T$, $t = 1/2T$, and $t = 3/4T$, showing the lateral deformation $\Delta y$ along the body of the cubic swimmer superposed on the corresponding shape of the original carangiform swimmer (blue).
    }
    \label{fig:etaSt_Aps}
\end{figure}

The present scaling analysis suggests that the efficiency of the caudal fin propulsion is inversely related to the value of  $A'^*$. In Fig. \ref{fig:etaSt_Aps}(a) the dashed lines shows the locus of values for the optimal condition for a range of $A'^*$ and we note that as $A'^*$ decreases, the maximum efficiency goes up while the optimal Strouhal number decreases. 
At the limit of $A'^*=0$, the efficiency scaling simply becomes $\eta_{\rm{fin}}/\beta_\eta=(A_F/\lambda)/\textrm{St}_A$ and $\textrm{St}_\textrm{opt}=\textrm{St}_\textrm{min}=A_F/\lambda$.
The maximum efficiency factor becomes 1, but thrust becomes zero at $\textrm{St}_A=\textrm{St}_\textrm{opt}$. 

We now verify this behavior predicted with respect to $A'^*$ in two ways: by comparing against data for a thunniform swimmer model, and by synthesizing and testing a fish model with swimming kinematics that corresponds to $A'^*=0$. For the thunniform swimmer model we have considered the simulation of a thunniform swimmer by Huang et al.\cite{huang2023hydrodynamics}. 
For the midline kinematics applied for the thunniform swimmer model, $A'^*=0.47$, and a best fit through the caudal fin Froude efficiency data from Huang et al.\cite{huang2023hydrodynamics}, we find that $\beta_\eta$ is 0.82. The data plotted in Fig. \ref{fig:etaSt_Aps}(a) show that the optimal Strouhal number for $A'^*=0.47$ is greater than the one for the mackerel model with $A'^*=0.38$, and the peak efficiency is lower. 

A stronger verification of the importance of $A'^*$ is for a swimmer with kinematics synthesized so as to satisfy $A'^*=0$. This is accomplished using a cubic polynomial amplitude envelope function which allows us to match all other kinematic parameters to the original carangiform swimmer model while also enforcing this new constraint. The new amplitude envelope and swimming kinematics are shown in Fig.\ref{fig:etaSt_Aps}(b) in comparison with the original ones. Additional details of the new fish model and the simulation results are summarized in Appendix \ref{cubicfish}. DNSs are performed with this new model for $\textrm{Re}_L=5,000, 10,000,$ and $25,000$. The present DNS results for $A'^*=0.38$ (mackerel model) and $A'^*=0$ (cubic amplitude envelope) are also plotted in Fig.\ref{fig:etaSt_Aps}(a) for comparison.
For the DNS results with $A'^*=0$, $\beta_\eta$ is found to be 1.0, and the caudal fin Froude efficiency matches very well with the proposed scaling law. Thus, as suggested by the scaling law, the efficiency factor increases significantly at low Strouhal numbers by reducing $A'^*$.

The above analysis not only verifies the predictions of the scaling law for a different class of swimming kinematics, it demonstrates that even with the same tail amplitude and Strouhal number, changes in the $A'^*$ value can have a significant (and predictable) effect on the propulsive performance. One implication of this finding is that experimental studies of midline kinematics in swimming fish (e.g., \cite{di2021convergence,li2021fishes}) should place greater emphasis on accurately capturing this parameter. Moreover, the widespread practice of modeling amplitude envelopes as quadratic functions—common in this field but unable to match the observed $A'^*$ values—warrants reconsideration.
This example also demonstrates how the scaling laws derived from basic principles of flow physics can be used to ``design'' kinematics to achieve desirable swimming performance. Thus, while the implication of this analysis for BCF swimming in fish is interesting (as shown above), this finding provides an important kinematic parameter for consideration in the design and control of BCF swimmer inspired underwater vehicles.   

Finally, the above analysis also raises an important question with respect to biological swimmers: if both peak efficiency and peak $U/U_c$ increase as $A'^*$ decreases, with the highest values of these swimming performance metrics occurring at $A'^* = 0$, why are BCF fishes swimming with $A'^*$ centered around $\sim 0.4$? Note again that $A'^* \neq 0$ may be interpreted as a phase mismatch between the pitch and and heave velocity, and $A'^* = 0.4$ corresponds to phase mismatch of about $20^\circ$. A plausible explanation lies in the nature of BCF locomotion, which combines active muscle-driven actuation in the anterior portion of the body with a (mostly) passive, traveling-wave response in the posterior region. The relatively free (elasticity-driven) motion of the tail allows the fish to achieve large tail amplitudes without additional muscular effort or control. However, this passive movement also limits precise control of the tail’s midline kinematics, naturally leading to a non-zero $A'^*$. If as an analogy, we consider a simple linear-elastic deformation of an elastic plate (or ribbon) that is oscillated transversely at a clamped end (which corresponds to the anterior portion of the fish body in this analogy) and free to oscillate at the other end (which would correspond to the tail), the free end would satisfy conditions of zero bending moment and zero shear, which correspond to $A^{''}(L)=0$ and $A^{'''}(L)=0$. In contrast, $A^{'*}=0$ which is associated with maximizing swimming performance requires that the tail be somehow ``clamped'' at the two extremes of the stroke, which is difficult to achieve for a biological swimmer.  Thus, BCF kinematics of biological swimmers are driven not just by hydrodynamic considerations (which is what the current model addresses), but also by constraints emanating from anatomy, muscle physiology and neuromuscular control, and this results in value of $A'^*$, which is noticeably larger than the optimal value.  Later in the paper and in Appendix \ref{cubicfish} we show that imposing $A'^* = 0$  does improve the swimming performance but this comes at the cost of a more complex midline kinematic profile.

\subsection{Effect of Scale, Kinematics and Morphology on Swimming Speed}
Predicting swimming speed given scale, kinematics and morphology is an important goal of such scaling analysis. Some of the early work  provided correlations that had no underpinnings in the flow physics of BCF swimming \cite{bainbridge1958speed,hunter1971swimming}. Other scaling laws have been based on dimensional analysis or heuristic relationships based on ideal flow assumptions \cite{eloy2012optimal,gazzola2014scaling,ventejou2025universal}. The current scaling analysis is based on the quantitative analysis of DNS data on thrust and drag of caudal fin swimmers \cite{seo2022improved}, which establishes the primacy of the caudal fin LEV in thrust generation. The final relationship provided here is between the free swimming Strouhal and Reynolds numbers, which is an implicit scaling for swimming velocity that incorporates scale, kinematics, and morphology.

This scaling analysis naturally introduces the parameter $K_\text{morph}$ that is connected to the fin and body morphology. This parameter is key in establishing the power-optimal Strouhal number for a given Reynolds number, the latter being a surrogate for scale. This has important implications not just for biological swimmers but for the design of BUVs since it provides a guiding principle for selecting the design of the body and fin shape for given kinematics in order to achieve optimal swimming at any given scale. In particular this scaling suggests that the caudal fin propulsor should grow slower than linear relative to the body size of the BUV implying that small(large) underwater vehicles should employ relatively larger(smaller) flapping foils for optimal operation.

\section*{Declaration of Interest}
The authors report no conflict of interest.

\begin{acknowledgments}
This work benefited from ONR Grants N00014-
22-1-2655 and N00014-22-1-2770 monitored by
Dr. Robert Brizzolara. This work used the computational resources at the Advanced Research Computing at Hopkins (ARCH) core facility (rockfish.jhu.edu), which is supported by the AFOSR DURIP Grant FA9550-21-1-0303. We acknowledge several fruitful discussion with Dr. Matthew McHenry and Ashley Peterson from University of California at Irvine. 
\end{acknowledgments}

\appendix
\section*{Appendix}
\subsection{Grid Convergence}
\label{sec:GC}
\begin{figure}[ht!]
\centering
(a)\includegraphics[width=0.45\textwidth]{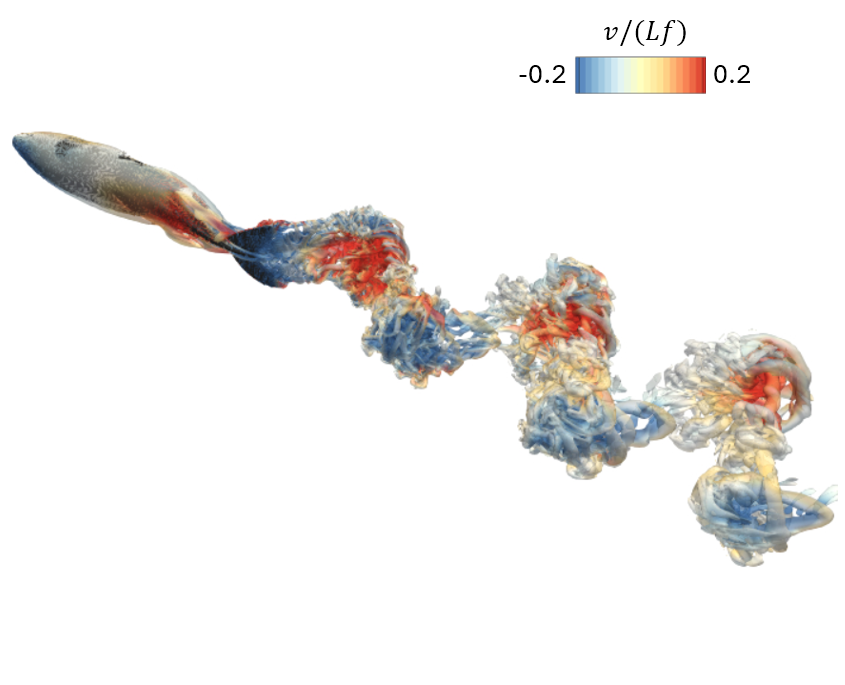}
(b)\includegraphics[width=0.35\textwidth]{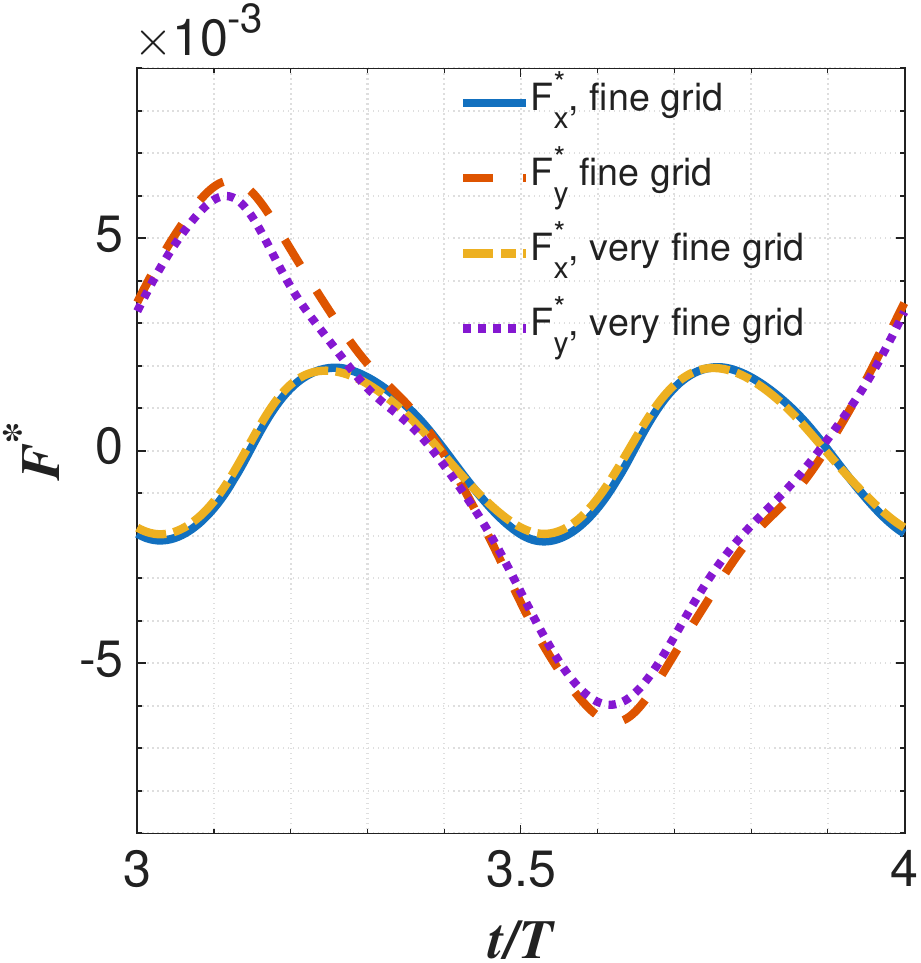}
\caption{Solitary fish swimming at $Re_L=50,000$ using the very fine grid. (a) Vortical structure of the very fine grid, visualized by the iso-surface of $Q=0.1f^2$ colored by the normalized lateral vorticity. (b) Time profiles of the total hydrodynamic force in the streamwise ($F^*_x$) and lateral ($F^*_y$)
directions, normalized by $(1/2)\rho (Lf)^2 L^2$. Fine grid: $1200\times 540\times 360$. Very fine grid: $1610 \times 1100 \times 600$.}
\label{gridConvergence}
\end{figure}

To assess grid convergence at high Reynolds numbers, we performed an additional simulation for a swimming fish at the highest Reynolds number, $Re_L=50,000$ using a very fine grid with $1610\times 1100\times 600$ (about 1 billion) points, in which the fish body was resolved by 667 grid points along its length. A snapshot of vortical structures around the fish obtained from this very fine grid is presented in Fig.\ref{gridConvergence}(a). The result is compared with the one on the fine grid ($1200\times 540\times 360$, about 233 million) employed in the present study. The time histories of the streamwise ($x$) and lateral ($y$) hydrodynamic forces plotted in Fig.\ref{gridConvergence}(b) showed a good agreement between the results on the fine and very fine grids. The root-mean-square errors are found to be about 3\% in $F^*_x$ and 1\% in $F^*_y$, where $F^*_x$ and $F^*_y$ are normalized total hydrodynamic forces by $(1/2)\rho (Lf)^2 L^2$. This confirms that the fine grid resolution ($1200\times 540\times 360$) is sufficient for the simulations at high Reynolds numbers, $5000 \le \textrm{Re}_L \le 50000$.

\subsection{Glossaryof Key Parameters} \label{keyparam}
Since the current analysis introduces and employs many parameters, we have provided this glossary to aid the reader. 
\renewcommand{\arraystretch}{0.8} % Keep your current row spacing
\begin{longtable}{|>{\raggedright\arraybackslash}p{1.3cm}|>{\raggedright\arraybackslash\baselineskip=0.6\normalbaselineskip}p{4cm}|>{\raggedright\arraybackslash\baselineskip=0.6\normalbaselineskip}p{12cm}|}
\caption{Nomenclature table defining and explaining the key parameters in this study.}
\label{tab:nomenclature} \\
\hline
\textbf{Symbol} & \textbf{Definition} & \textbf{Explanation} \\
\hline
\endfirsthead
\hline
\textbf{Symbol} & \textbf{Definition} & \textbf{Explanation} \\
\hline
\endhead
\hline \multicolumn{3}{r}{{Continued on next page}} \\
\endfoot
\hline
\endlastfoot
$\alpha_{\mathrm{eff}}$ & Effective angle-of-attack & The angle of the flow relative to the leading-edge of the fin that is induced by the undulatory movement of the body. \\
\hline
$\beta_T$ & Caudal fin shape parameter & A multiplicative parameter in the thrust relationship that is associated with the morphology of caudal fin. \\
\hline
$\lambda$ & Undulatory wavelength of BCF motion& Dimensions of length. \\
\hline
$\phi_\theta$ & Pitching phase modulation, $\phi_\theta = \tan^{-1} A'^*$ & The phase difference between the angle at the leading-edge due to the heaving and pitching of the tail induced by the undulatory movement of the body. \\
\hline
$\sigma$ & Empirical parameter introduced to approximate the integral, $\sigma = 0.63$ & This is a number that approximates an otherwise complex integral expression  in thrust and is valid over a large range of relevant swimming parameters.  \\
\hline
$\mathrm{Re}_{\mathrm{cr}}$ & Critical Reynolds number, see Eq.~(\ref{eq:Recr}) & This parameter depends on kinematic as well as morphological parameters of the swimmer and it represents the swimming Reynolds number beyond which, the swimming becomes almost independent of viscous effects. \\
\hline
$\mathrm{Re}_L$ & Reynolds number defined by $L^2 f / \nu$ & Reynolds number based only on the input parameters for the swimmer. \\
\hline
$\mathrm{Re}_U$ & Swimming Reynolds number defined by $UL / \nu$ & The often used Reynolds number which is defined on the swimming velocity and body length. \\
\hline
$\mathrm{St}_{\min}$ & Minimum swimming Strouhal number, see Eq.~(\ref{eq:Stmin}) & The Strouhal number below which the fish generated net drag. This parameter depends only on the midline kinematics of the swimmer. \\
\hline
$\mathrm{St}_{\mathrm{opt}}$ & Optimal Strouhal number, see Eq.~(\ref{eq:Stopt}) & The swimming Strouhal number at which the swimmer's fin achieves the maximum Froude efficiency. This parameter depends only the midline kinematics. \\
\hline
$\mathrm{St}_A$ & Swimming Strouhal number defined by $A_F f / U$ & This is the classic Strouhal number for a swimmer defined on the peak-to-peak tail amplitude. \\
\hline
$A'^*$ & Characteristic kinematics parameter, see Eq.~(\ref{eq:Aps}) & A measure of the non-dimensional slope of the fin at the two ends of its stroke. \\
\hline
$A_F$ & Peak-to-peak tail-beat amplitude & Dimensions of length. \\
\hline
$A_F^*$ & $A_F^* = A_F R_\theta / \lambda$ & Tail-beat amplitude amplified by $A'^*$ and normalized by the body wavelength. This determines the pitching amplitude of the caudal fin.\\
\hline
$C_f$ & Skin friction factor & Dimensionless number that quantifies the friction drag on the body of the fish. It is a function of body morphology and kinematics. \\
\hline
$f$ & Tail-beat frequency & Dimensions of per second. \\
\hline
$\kappa$ & Exponent in boundary layer scaling & This is the exponent to which the Reynolds number is raised in the scaling for the thickness of the boundary layer. This is expected to depend on the laminar/turbulent nature of the boundary layer and on body kinematics. \\
\hline
$K_{\mathrm{morph}}$ & Morphological parameter, $K_{\mathrm{morph}} = 2(C_f / \beta_T)(S_b / S_f)$ & This is a morphological parameter that encapsulated information about the drag and thrust generating features of the swimmer morphology. \\
\hline
$L$ & Fish body length & Dimensions of length. \\
\hline
$R_\theta$ & Pitching amplification parameter, $R_\theta = \sqrt{1 + {A'^*}^2}$ & Amplification factor for tail-beat amplitude due to $A'^*$\\
\hline
$S_b$ & Fish body surface area & Dimensions of area \\
\hline
$S_f$ & Fish caudal fin planform area & Dimensions of area. One side only.\\
\hline
$S_x$ & Fish body frontal area in the surge direction & Dimensions of area \\
\hline
$U$ & Swimming speed & Terminal swimming speed attained by the swimmer where thrust balances drag. \\
\hline
$U_c$ & Undulatory motion wave speed, $U_c = \lambda f$ & The speed of the undulatory wave that traverses the body.\\
\hline
\end{longtable}

\renewcommand{\arraystretch}{1.0} % Keep your current row spacing

\subsection{The Force Partitioning Method} \label{FPM}
The force partitioning method (FPM) is derived by projecting the incompressible Navier-Stokes equations onto an ``influence" potential field, $\psi$, which is obtained by solving the Laplace equation with a tailored boundary condition: 
\begin{equation}
{\nabla ^2}{\psi} = 0,\,\,\,\,\,\nabla {\psi} \cdot \vec n = \left\{ {\begin{array}{*{20}{c}}
{{n_i}}&{{\rm{on}}\,B}\\
0&{{\rm{on}}\,\Sigma }
\end{array}} \right. 
\end{equation}
where $n_i$ is the component of the surface normal vector depending on the direction of the force to be partitioned, 
$i=1,2,3$ corresponds to the force in the $x,y,z$ direction, 
$B$ is the surface of the body of interest, and $\Sigma$ is for all other surfaces including the domain boundaries. 
The FPM formulation is obtained by projecting the incompressible momentum equation onto the gradient of the influence field ($\nabla\psi$), and taking a volume integral over the domain.
The formulation decomposes the total pressure force on the body ($F_B$) into the force due to the body kinematics ($F_k$), force due to viscous diffusion of momentum ($F_\nu$), and the vortex-induced force ($F_Q$):
\begin{equation}
\underbrace {\int\limits_B {P{n_i}} dS\,}_{{F_B}} = \underbrace {\int\limits_{B + \Sigma } {\left( { - \psi \rho \frac{{D\vec U}}{{Dt}} \cdot \vec n} \right)} dS}_{{F_k}} + \underbrace {\int\limits_{B + \Sigma } {\left( {\psi \mu {\nabla ^2}\vec U \cdot \vec n} \right)} dS}_{{F_\nu}} + \underbrace {\int\limits_{{V_f}} {\left( { - 2\rho \psi Q} \right)} \,dV}_{{F_Q}}, 
\end{equation}
where $Q$ is the second invariant of the velocity gradient tensor and $V_f$ is the volume of the flow domain.
At high Reynolds number and low Strouhal number, the vortex-induced force, $F_Q$ becomes the dominant component of the pressure force on the body. The integrand of $F_Q$, $f_Q=-2\rho\psi Q$ represents the vortex-induced force density. The spatial distribution of $f_Q$ then informs the contribution of local vortical structure on the total vortex-induced force, $F_Q$.

\begin{figure}
    \centering
    (a)\includegraphics[width=0.4\linewidth]{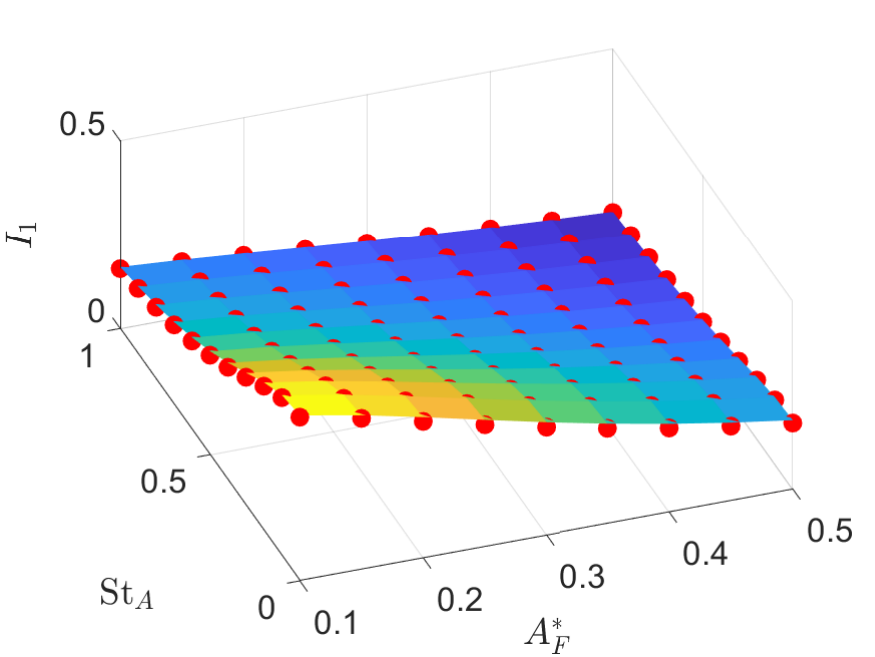}
    (b)\includegraphics[width=0.4\linewidth]{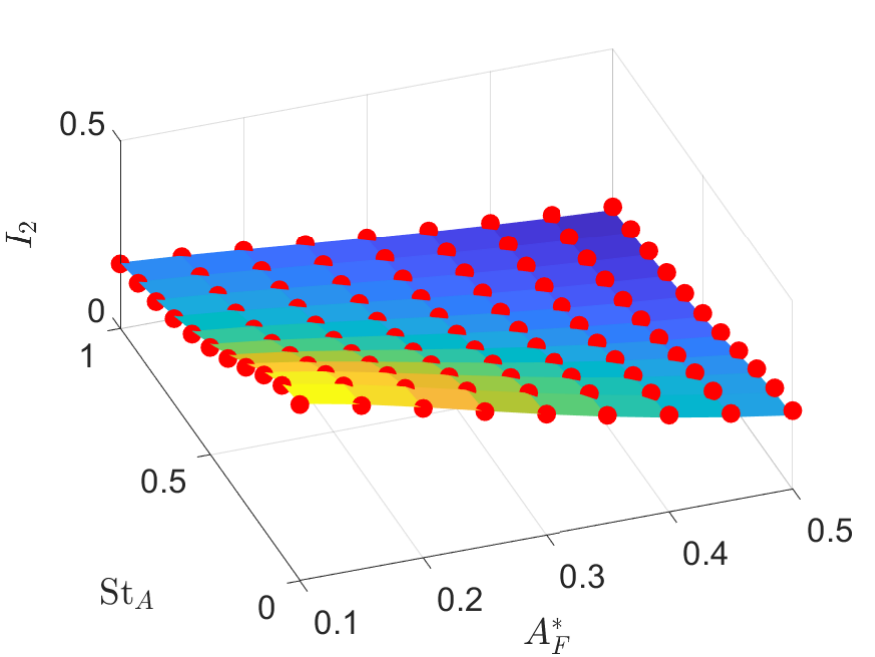}\\
    (c)\includegraphics[width=0.4\linewidth]{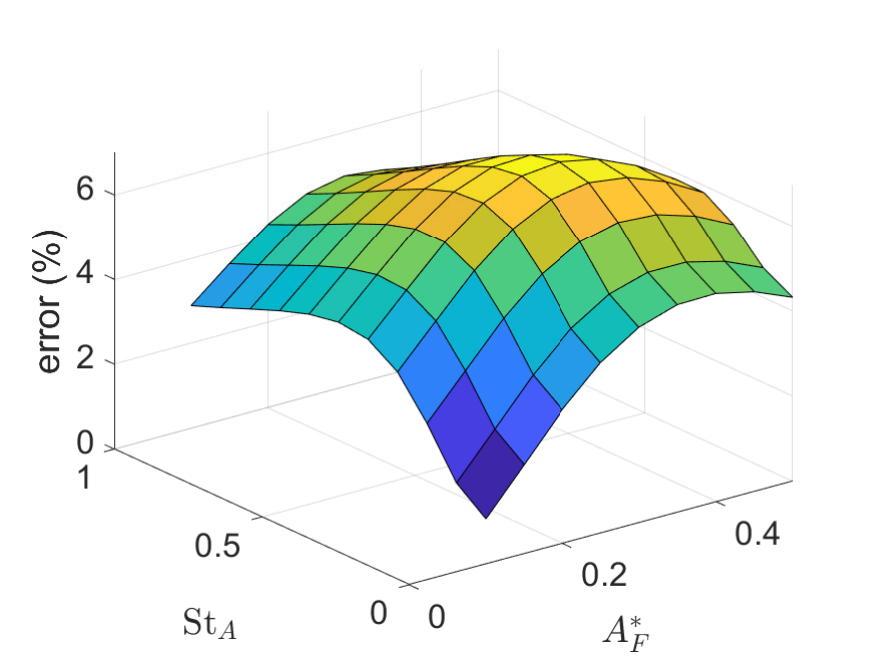}
    (d)\includegraphics[width=0.4\linewidth]{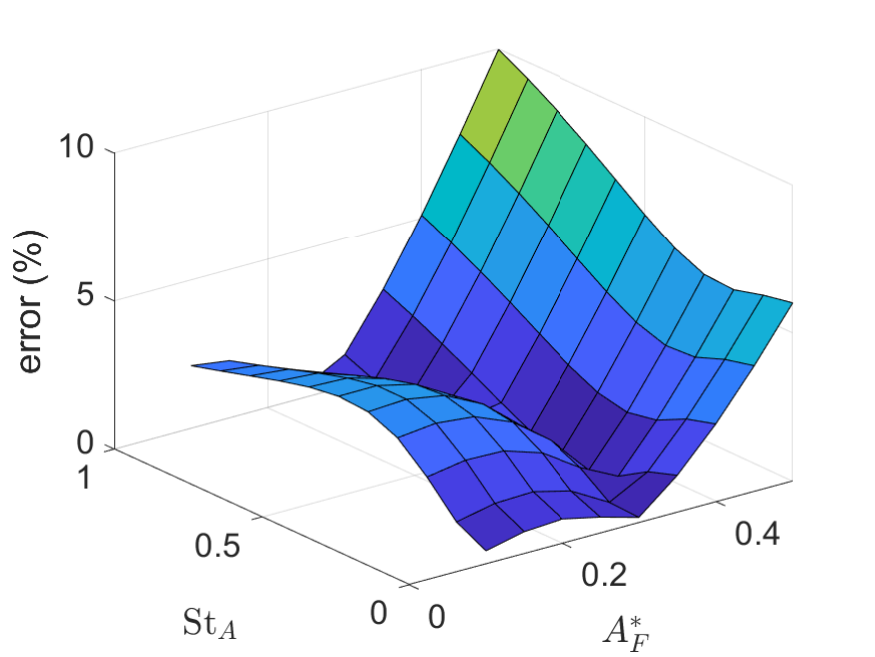}
    \caption{Comparison of the direct numerical integral to the approximated formulation. Symbols: Direct numerical integral of Eq.(\ref{eq:integrals}). Surface: Approximation given by Eq.(\ref{eq:intappx}). (a) $I_1$. (b) $I_2$. (c) Percent error between the direct integral and the approximation for $I_1$. (d) Error for $I_2$.} 
    \label{fig:Iapprox}
\end{figure}

\subsection{Approximation of Integrals}
\label{sec:appx}
Assuming that the kinematic parameters such as $\textrm{St}_A$ and $A_F^*$ do not change in one tail-beat, the mean thrust factor given by Eq.(\ref{eq:Lambda_T}) can be written with two integrals:
\begin{equation}
{\Lambda _T} = {\pi ^2}{A_F}^*\left( {\pi {\rm{S}}{{\rm{t}}_A}{I_1} - {A_F}^*{I_2}} \right),
\label{eq:lambdaT_int}
\end{equation}
where
\begin{equation}
\begin{array}{l}
{I_1} = \frac{1}{{2\pi }}\int_0^{2\pi } {\frac{{\cos ({t^*})\cos ({t^*} - {\phi _\theta })}}{{\sqrt {1 + {{(\pi {\rm{S}}{{\rm{t}}_A})}^2}{{\cos }^2}({t^*})} \left[ {1 + {{(\pi {A_F}^*)}^2}{{\cos }^2}({t^*} - {\phi _\theta })} \right]}}} \,d{t^*},\\
{I_2} = \frac{1}{{2\pi }}\int_0^{2\pi } {\frac{{{{\cos }^2}({t^*} - {\phi _\theta })}}{{\sqrt {1 + {{(\pi {\rm{S}}{{\rm{t}}_A})}^2}{{\cos }^2}({t^*})} \left[ {1 + {{(\pi {A_F}^*)}^2}{{\cos }^2}({t^*} - {\phi _\theta })} \right]}}} \,d{t^*}.
\end{array}    
\label{eq:integrals}
\end{equation}
These integrals may be challenging to perform analytically, and the approximations are proposed here.
For nominal values of $\textrm{St}_A$ and $A_F^*$, one can see that the denominator changes slower comapred to the numerator with $t^*$. Thus, we replace the denominator with a constant value by introducing an empirical parameter, $\sigma$:
\begin{equation}
\begin{array}{l}
{I_1} \approx \frac{1}{{2\pi }}\frac{{\int_0^{2\pi } {\cos ({t^*})\cos ({t^*} - {\phi _\theta })} d{t^*}}}{{\sqrt {1 + \sigma {{(\pi {\rm{S}}{{\rm{t}}_A})}^2}} \left[ {1 + \sigma {{(\pi {A_F}^*)}^2}} \right]}} = \frac{{\cos {\phi _\theta }}}{{2\sqrt {1 + \sigma {{(\pi {\rm{S}}{{\rm{t}}_A})}^2}} \left[ {1 + \sigma {{(\pi {A_F}^*)}^2}} \right]}},\\
{I_2} \approx \frac{1}{{2\pi }}\frac{{\int_0^{2\pi } {{{\cos }^2}({t^*} - {\phi _\theta })} d{t^*}}}{{\sqrt {1 + \sigma {{(\pi {\rm{S}}{{\rm{t}}_A})}^2}} \left[ {1 + \sigma {{(\pi {A_F}^*)}^2}} \right]}} = \frac{1}{{2\sqrt {1 + \sigma {{(\pi {\rm{S}}{{\rm{t}}_A})}^2}} \left[ {1 + \sigma {{(\pi {A_F}^*)}^2}} \right]}}.
\end{array}    
\label{eq:intappx}
\end{equation}
The integrals in the numerator can then be obtained easily, as shown above.
The value of the empirical parameter, $\sigma=0.63$ is obtained by fitting the approximations to the actual numerical integrals over the parameter space: $0<\textrm{St}_A\le1$ and $0.1\le A_F^* \le 0.5$. 
The comparison between the proposed approximations and the numerical integral results are plotted in Fig.\ref{fig:Iapprox} for $I_1$ and $I_2$ and the percent errors are shown in the below. The root-mean-squared error is found to be about $4\%$. The maximum error of about $10\%$ is observed in $I_2$ at the high Strouhal number ($\textrm{St}_A=1$) and high tail-beat amplitude ($A_F^*=0.5$), but this condition is rare for natural swimmers as one can see in Fig.\ref{fig:ApsQs}(a). We note that the error does not change with $A'^*$. By substituting the approximations in Eq.(\ref{eq:intappx}) into Eq.(\ref{eq:lambdaT_int}), one can get the approximate expression for the mean thrust factor, Eq.(\ref{eq:Lambda_T_app}). 
The same integrals appear in the mean power factor, $\Lambda_W$, and Eq.(\ref{eq:intappx}) is also used to get Eq.(\ref{eq:Lambda_W}).

\subsection{Effect of Tail Beat Amplitude and Frequency} \label{amp_and_freq}
\begin{figure}[!h]
    \centering
    (a)\includegraphics[width=0.45\linewidth]{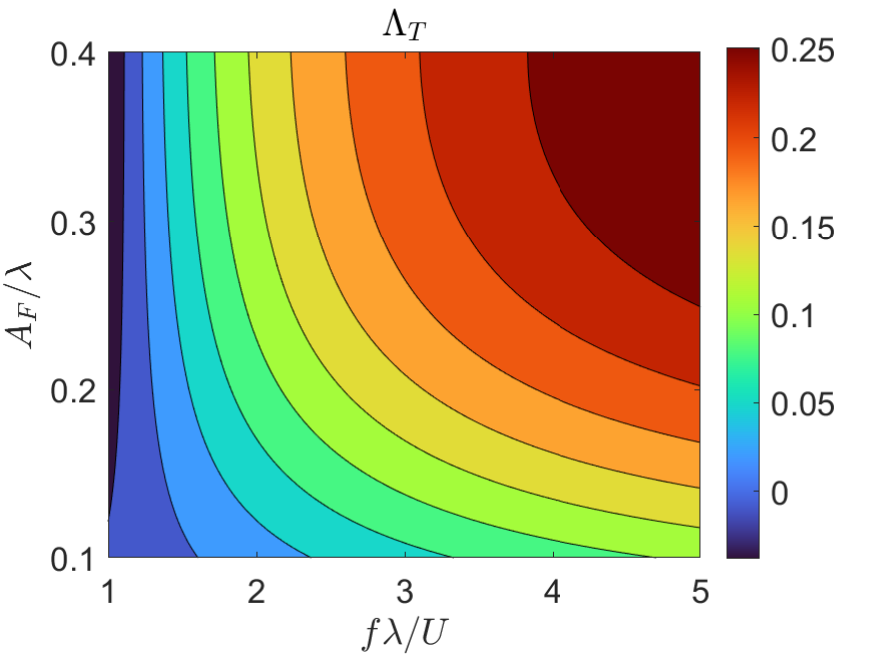}
    (b)\includegraphics[width=0.45\linewidth]{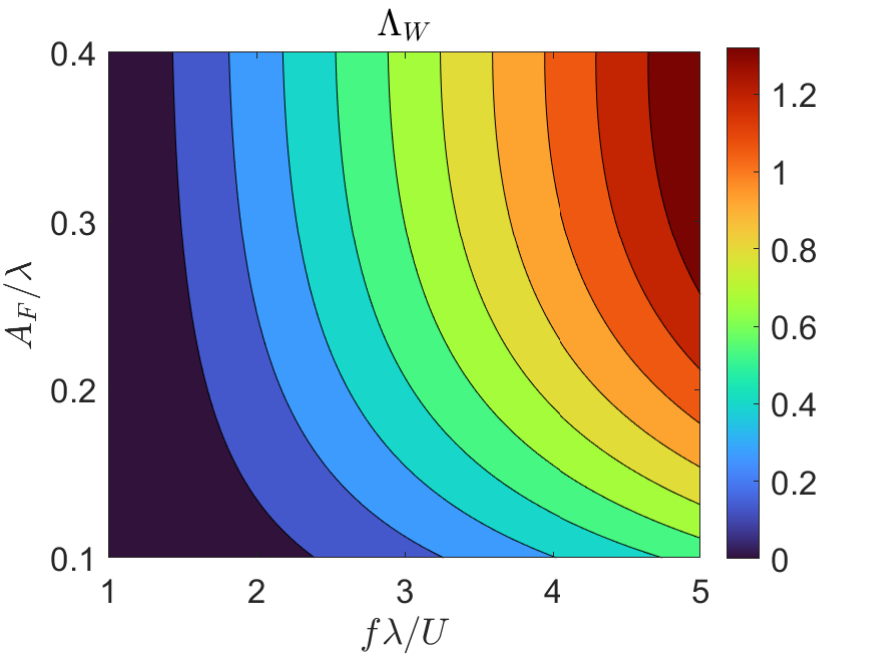}
    \caption{Effect of tail-beat amplitude and frequency on the (a) thrust factor and (b) power factor.}
    \label{fig:factors}
\end{figure}

The scaling analysis based on the LEV-based model has shown that both the thrust and mechanical power of the caudal fin depend primarily on the Strouhal number, $\textrm{St}_A=A_F{f}/U$. A fish can change this parameter by adjusting the tail-beat frequency ($f$) or the tail-beat amplitude ($A_F$). However, as evident in Eqs.(\ref{eq:CT}) and (\ref{eq:CW}), $A_F^*=A_FR_\theta/\lambda$ also appears as an independent parameter in the scaling of the thrust and power factors. Thus, the effects of tail-beat frequency and amplitude on the thrust and power are examined here. In Fig.\ref{fig:factors}, the thrust ($\Lambda_T$) and power ($\Lambda_W$) factors are plotted as functions of the normalized tail-beat amplitude, $A_F/\lambda$, and frequency $f\lambda/U$ for the present fish model. These plots show that both the thrust and the power factors are stronger functions of the tail-beat frequency than the tail-beat amplitude. We note that while at  high swimming speeds (or at low values of $f\lambda/U$), the tail-beat amplitude is not very effective in modulating the thrust, the tail-beat amplitude is highly effective in modulating thrust at low swimming speeds (high $f\lambda/U$). These observations are in line with the experimental data of Videler and Hess\cite{videler1984fast}, which showed that the variations in the amplitude of the tail-beat are smaller at high swimming speeds. A similar trend was also observed in the data from Bainbridge\cite{bainbridge1958speed}, especially for the Dace. 

\subsection{Caudal Fin Swimmer with $A'^*=0$} \label{cubicfish}
In this appendix, we provide details about the BCF kinematic model that is designed to generate  $A'^*=0$ at the caudal fin. To prescribe $A'^*=0$, we employ $(dA/dx)_{x=L}=0$ (see Eq.\ref{eq:Aps}). In order to synthesize a fish kinematic model that otherwise matches the original carangiform model (Eqs. \ref{eq:carangi}-\ref{eq:A}) in terms of head amplitude, tail amplitude and minimum body displacement, but which also satisfies this additional condition of $(dA/dx)_{x=L}=0$, we have employed the following cubic polynomial function for the amplitude envelope:
\begin{equation}
A(x)/L = 0.02 -0.16(x/L)+0.56(x/L)^2-0.32(x/L)^3.     
\label{eq:cubrcA}
\end{equation}
All other kinematic parameters and the undulatory motion equation remain the same.
This amplitude envelope function is plotted in Fig.\ref{fig:etaSt_Aps}(b) along with the original one for the carangiform swimmer.
With $A'^*=0$, $R_\theta=1$ and $\phi_\theta=0$, and thus there is no phase lag between the caudal fin pitching angle and heaving rate.
The snapshots of the fish body undulatory motion presented in Fig.\ref{fig:etaSt_Aps}(b) show that the caudal fin pitching angle at the tail is 0 when the heaving rate is 0 ($t=1/4T$ and $3/4T$). Employing the new amplitude envelope, flow simulations are performed at $\textrm{Re}_L = $5000, 10000, and 25000. The terminal swimming speed at which the mean surge force on the fish is nearly zero is found by varying the speed of the incoming current, $U$, for each Reynolds number.
The force, power, and efficiency metrics obtained from the simulations in this ``terminal speed'' condition are listed in Table \ref{tab:cubic_cases}.
\begin{table}[h]
    \begin{center}
    \begin{tabular}{|c|c|c|c|c|c|c|c|c|c|}
        \hline
        $\textrm{Re}_L$ & $U/(Lf)$ & $\textrm{St}_A$ & $F^*_{p,\textrm{body}}$ & $F^*_{s,\textrm{body}}$ & $F^*_{p,\textrm{fin}}$ & $F^*_{s,\textrm{fin}}$ & $W^*_{p,\textrm{body}}$ & $W^*_{p,\textrm{fin}}$ & $\eta_{\textrm{fin}}$  \\ \hline
        5000  & 0.46  & 0.43 & -0.95 & 3.4 & -2.9 & 0.38 & -1.4 & -2.8  & 0.47 \\ \hline
        10000 & 0.57  & 0.35 & -0.37 & 3.0 & -3.0 & 0.37 & -1.1 & -3.0  & 0.57 \\ \hline
        25000 & 0.71  & 0.28 & 0.25 & 2.3 & -2.8 & 0.35 & -0.68 & -2.8  & 0.71 \\ \hline
    \end{tabular}
    \end{center}
    \caption{The force, power, and efficiency on the free swimming fish at various Reynolds numbers simulated using a cubic amplitude envelope $A(x)$ with $A'^* = 0$. $\textrm{Re}_L = L^2f/\nu$, $F^*$ is the time-averaged force normalized by $(1/2)\rho (Lf)^2 L^2$. $W^*$ is the time-averaged power normalized by $(1/2)\rho (Lf)^3 L^2$. Negative values of $F^*$ indicate thrust. All $F^*$ and $W^*$ values in the table are to be multiplied by $\times10^{-3}$.}
    \label{tab:cubic_cases}
\end{table}

\bibliography{references}% Produces the bibliography via BibTeX.

\end{document}